# A Survey of Security in UAVs and FANETs: Issues, Threats, Analysis of Attacks, and Solutions

Ozlem Ceviz* , Sevil Sen*, Pinar Sadioglu
WISE Lab., Department of Computer Engineering, Hacettepe University, Ankara, Turkey

***Abstract*—Thanks to the rapidly developing technology, unmanned aerial vehicles (UAVs) are able to complete a number of tasks in cooperation with each other without need for human intervention. In recent years, UAVs, which are widely utilized in military missions, have begun to be deployed in civilian applications and mostly for commercial purposes. With their growing numbers and range of applications, UAVs are becoming more and more popular; on the other hand, they are also the target of various threats which can exploit various vulnerabilities of UAV systems in order to cause destructive effects. It is therefore critical that security is ensured for UAVs and the networks that provide communication between UAVs.**

**This survey seeks to provide a comprehensive perspective on security within the domain of UAVs and Flying Ad Hoc Networks (FANETs). Our approach incorporates attack surface analysis and aligns it with the identification of potential threats. Additionally, we discuss countermeasures proposed in the existing literature in two categories: preventive and detection strategies.**

**Our primary focus centers on the security challenges inherent to FANETs, acknowledging their susceptibility to insider threats due to their decentralized and dynamic nature. To provide a deeper understanding of these challenges, we simulate and analyze four distinct routing attacks on FANETs, using realistic parameters to evaluate their impact. Hence, this study transcends a standard review by integrating an attack analysis based on extensive simulations.**

**Finally, we rigorously examine open issues, and propose research directions to guide future endeavors in this field.**

## I. Introduction

UNMANNED aerial vehicles (UAVs) are aircraft capable of being flown without a human pilot or any other crew on board [1]. Commercial UAVs are aircrafts that are intended to be used for business purposes. With a CAGR of 28.58% during the projection period, it is expected that the global market for commercial UAVs will expand rapidly, increasing from $8.15 billion in 2022 to $47.38 billion by 2029 [2]. Gartner projects that there will be over one million UAVs operating by 2026, representing a staggering increase compared to the 20,000 UAVs currently in use for retail deliveries [3].

In particular, the ability to work autonomously and collaboratively without need for human intervention is pioneering the expansion of UAV applications. UAVs are now used within military missions [4], on search and rescue operations [5], [6], on target tracking assignments [7], as well as in environment protection studies [8], agricultural missions [9], [10], and many other areas where their benefits fit the need. Numerous defense ministries worldwide are investing in UAVs capabilities as a means to reducing troop casualties, and as a cost-effective alternative to the use of manned aircraft.

The high mobility of UAVs makes the network topology dynamics different from Mobile Ad hoc Networks (MANETs) and Vehicular Ad hoc Networks (VANETs). For that reason, a new type of ad hoc networks called Flying Ad hoc Networks (FANETs) has emerged and has become a popular area of research areas in recent years [11]. In addition to single UAV operations, the collaborative use of UAVs as FANETs now feature in many applications such as performing search and rescue operations within a limited or confined area [12], as well as international border surveillance [13], logistics [14], forest fire monitoring and control [15], disaster-recover scenarios [16] and agricultural remote sensing systems [17], [18].

Although FANETs is a subset of MANETs, it differs from other types of ad hoc networks by its very characteristics. One of the most obvious differences is its dynamic topology, which changes due to the high speed nature of UAVs. UAVs move in 3D, unlike the nodes in MANETs and VANETs. Their mobility patterns also differ compared to other ad hoc network types. For example, they may fly together as a group in one direction and periodically move towards the controller ground system to complete certain missions. In addition, due to the large flight area potential, the node density of FANETs is lower than other ad hoc networks. Another difference relates to platform restrictions, since UAVs allow for minimally-sized batteries, hence the problem of rapid energy depletion comes to the fore.

With the increasing use and growing interest of UAVs in both civil and military applications, UAVs have become a clear target for cyber attacks [19], [20]. UAVs can communicate with each other (UAV-to-UAV) and/or with a base station (UAV-to-ground station) via wireless links. Hence, the use of wireless connections in UAVs makes the network inherently vulnerable to eavesdropping and active interference attacks. While some attackers target communication links between UAVs, others target UAV-based features such as software, sensors, or hardware.

Attackers have focused their interest on not only the devices but also the dynamic networks that support and operate high-

Ozlem Ceviz, Sevil Sen and Pinar Sadioglu are with the WISE Lab. Deparment of Computer Engineering, Hacettepe University, Ankara, Turkey (e-mail: ozlemceviz@hacettepe.edu.tr; ssen@cs.hacettepe.edu.tr; pinar.sadioglu@gmail.com)
*Ozlem Ceviz and Sevil Sen contributed equally to this work.

speed UAVs. UAVs and their networks are exposed to attacks due to various vulnerabilities such as having nodes with limited battery power, the use of wireless links, and protocols based on the cooperativeness of nodes within in the networks. Considering the popularity of UAVs and the future of FANETs, it is important to outline the vulnerability landscape and effective attacks, and to discuss possible solutions that may be employed against them.

The primary objective of this study is to analyze the system's attack surface, identifying potential points of vulnerability. Expanding on this analysis, the study refines understanding by categorizing related attacks into a taxonomy based on the identified entry points within the attack surface. Additionally, the study discusses the proposed solutions for the prevention and detection of attacks against UAV devices and networks.

While previous studies often focused on specific aspects of UAV security, this research offers a comprehensive survey encompassing security issues and proposals for both FANETs and UAVs. Moreover, this study examines attacks against UAV communication within the routing layer through realistic network simulations. Notably, this survey paper does not solely rely on theoretical analyses but also conducts practical simulations to assess real-world implications.

### A. Contributions

The key contributions of this survey paper can be summarized as follows:

- The unique characteristics of UAVs and networks of UAVs are presented in details, and then analyzed from a security perspective.
- Attack surface analysis of UAVs and FANETs is introduced and taxonomy of attacks is presented with detailed categories based on the identified entry points within the attack surface.
- Inspired by a lack of analysis of attacks using realistic network scenarios in the literature, the current study implements and deeply analyzes four attacks against the routing of FANETs.
- Security solutions proposed for preventing and detecting such attacks are reviewed and their limitations discussed.
- Open issues and future research directions within the research domain are discussed in detail.

### B. Structure of the Survey

The organization of the survey is shown in Figure 1. Section-II discusses existing surveys for the security of FANETs and UAVs with their limitations, and emphasizes the unique contributions offered by the current study. Section-III provides a background of UAVs and FANETs and discusses their characteristics from a security point of view. Section-IV defines the possible enrty points of UAVs and FANETs. Section-V categorizes security attacks against UAVs in the light of attack surface analysis. Section-VI presents simulations and analysis of attacks against networks of UAVs using realistic simulation parameters. Section-VII then summarizes the existing security solutions in the literature, grouping them under two subsections as prevention and detection. Lastly, Section-VIII outlines the limitations of the proposed studies and discusses open research areas, which is followed by a conclusion in Section-IX. The list of acronyms used in this manuscript can be found in Table I.

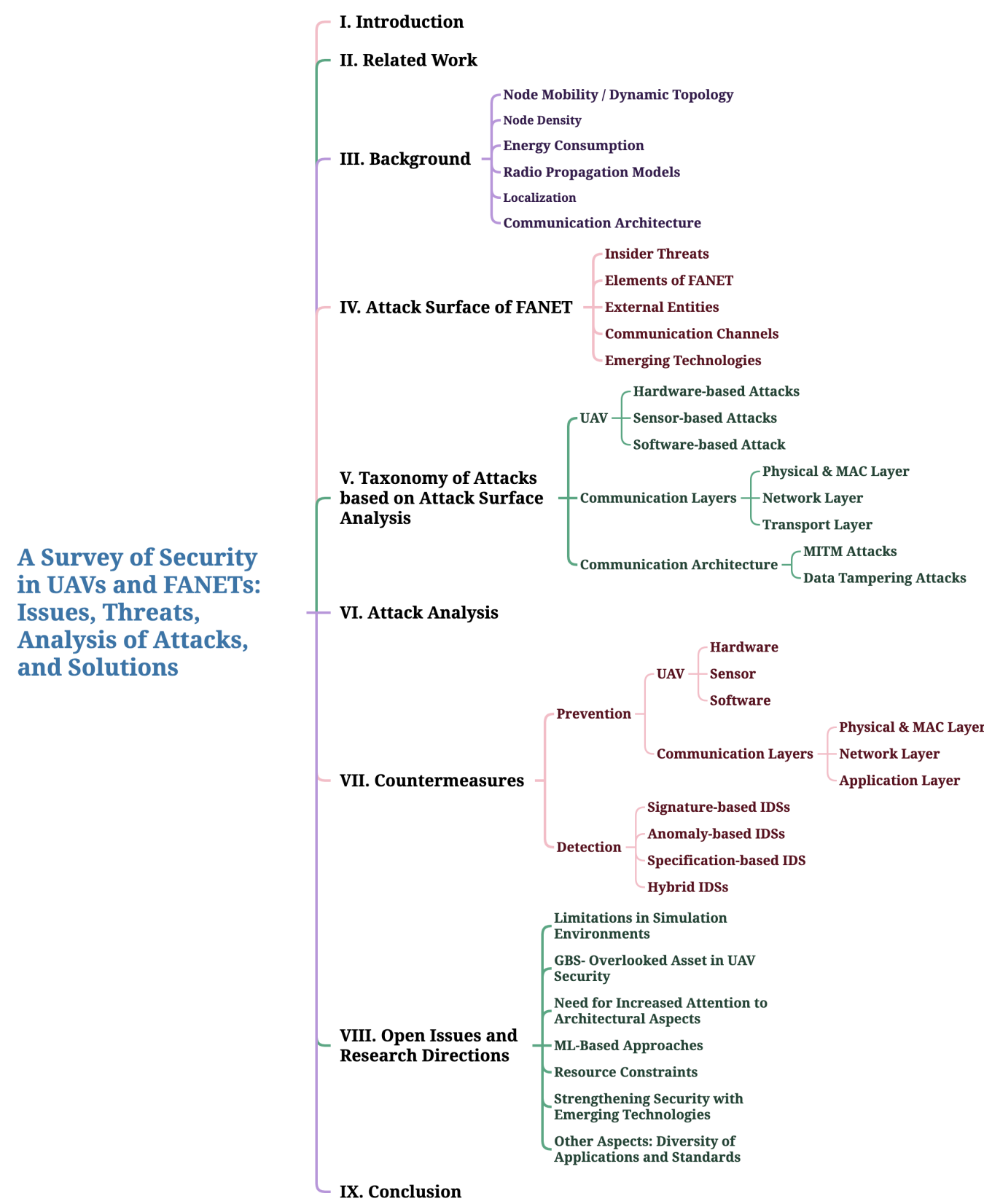

Fig. 1. Survey Organization

## II. Related Work

The first study that reviewed security issues in FANETs was [21], which not only summarized the proposed studies for secure communication in FANETs, but also gave some exemplar security solutions proposed for MANETs. However, since it was one of the first survey studies, it consisted of only a limited number of initial security solutions that had been proposed in the literature, and most of these studies were proposed for MANETs. Analysis of the existing security mechanisms proposed for MANETs and VANETs was presented as one of the open research areas [21].

Similarly, [22] focused on the security requirements of routing protocols for UAVs, and was also the first study to address this area in terms of vulnerabilities and network attacks. Attacks are examined according to three phases of routing protocols: routing discovery, route maintenance, and data forwarding. For security countermeasures, cryptography- and trust-based systems, and intrusion detection systems were discussed. Similarly, the vulnerabilities of UAV communication systems, the risks associated with data transmission and processing, and the need for secure communication protocols were discussed in [23]. Another survey [24] reviewed the security issues of FANETs, in addition to FANET communication and mobility models. However it did not give a specific classification of threats to FANETs or UAVs. A limited number of security solutions were discussed in the study, and it was emphasized that traditional security approaches are not directly applicable to FANETs due to their latency and heavy computation [24].

TABLE I
LIST OF ACRONYMS

| | | | |
|---|---|---|---|
| 2D | Two-dimensionals | MITM | Man-in-the-Middle |
| 3D | Three-dimensionals | ML | Machine Learning |
| 5G | Fifth-generation | MIMO | Multiple-Input Multiple-Output |
| ANN | Artificial Neural Network | mmWave | Millimeter-wave |
| AODV | Ad-hoc On Demand Distance Vector | NB | Naive Bayes |
| ARP | Address Resolution Protocol | NN | Neural Networks |
| AUC | Area Under the Curve | PCA | Principal Component Analysis |
| CNN | Convolutional Neural Networks | PCB | Printed Circuit Board |
| DDoS | Distributed-denial of Service | PDR | Packet Delivery Ratio |
| DNN | Deep Neural Network | PLS | Physical-Layer Security |
| DoB | Depletion of Battery | PN | Pseudorandom Noise |
| DoS | Denial of Service | ReLU | Rectified Linear Unit |
| DRL | Deep Reinforcement Learning | RERR | Route Error Packets |
| DSR | Dynamic Super Resolution | RF | Random Forest |
| DSSS | Direct Sequence Spread Spectrum | RNN | Recurrent Neural Network |
| DT | Decision Tree | RREP | Route Reply Packet |
| E2E | End-to-end | RREQ | Route Request Packet |
| FANET | Flying Ad hoc Network | RSS | Received Signal Strength Difference |
| FHSS | Frequency Hopping Spread Spectrum | RSSI | Received Signal Strength Indicator |
| FL | Federated Learning | RSUs | Road Side Units |
| FPR | False Positive Rate | RTL | Return-to-Launch |
| FRS | Fuzzy Rough Set | RTT | Round Trip Time |
| GA | Genetic Algorithm | RWP | Random Waypoint Model |
| GBS | Ground Base Station | SAODV | Secure Ad-hoc On Demand Distance Vector |
| GMM | Gauss Markov Mobility | SDN | Software Defined Networking |
| GPS | Global Positioning Stations | SDR | Software-defined Radio |
| HIS | Human Immune System | SOP | Secrecy Outage Probability |
| IDS | Intrusion Detection System | SSI | Signal Strength Intensity |
| IMU | Inertial Measurement Unit | SVM | Support Vector Machines |
| IoD | Internet-of-drone | SYN | Synchronize |
| IoT | Internet of Things | TaLU | Tanh Linear Unit |
| JBIG | Joint Bi-level Image Experts Group | TCP | Transmission Control Protocol |
| KNN | K-Nearest Neighbors | Tdoa | Time Difference of Arrival |
| LoS | Line-of-sight | TIK | Instant Key Disclosure |
| LR | Linear Regression | UAVs | Unmanned Aerial Vehicles |
| LTE | Long-Term Evolution | UDP | User Datagram Protocol |
| MAC | Media Access Control | VANETs | Vehicular Ad hoc Networks |
| MANETs | Mobile Ad hoc Networks | Wi-Fi | Wireless Fidelity |
| MEC | Mobile Edge Computing | Wi-MAX | Worldwide Interoperability for Microwave Access |
| MEMS | Micro-Electro-Mechanical Systems | XOR | eXclusive OR |

In [25]–[28], the authors presented potential threats against UAV systems, but FANET security was not covered. In [25], the UAV based-system attacks were briefly described with an overview, and then the authors focused on charging systems and battery attacks, and appropriate countermeasures. However, since UAV battery consumption attacks are new attacks, it was noted that the literature contained no fixed security solutions. Along with this issue, effective detection systems and artificial intelligence security systems were also highlighted as open research issues in [25].

Zhi et. al. [26] discussed UAV system threats by dividing them into three groups: sensor, communications, and multi-UAVs. Wi-Fi security was significantly emphasized as most UAVs require Wi-Fi connectivity for the purpose of remote control. In addition, the study revealed that sensor attacks affect the behavior of UAVs at a high level, as UAVs receive assistance from sensors such as gyroscopes to ensure balance and compass sensors to determine direction. However, compared to other surveys, the study contained only a limited number of attacks.

In [27], attacks that hinder the secure position estimation of drones are analyzed and categorized into two main classes: localization error attacks and other attacks. In addition, the authors discussed security analysis techniques, including security verification tools and methods. In [28], without giving a specific classification, some attacks against UAVs are covered such as DoS, man-in-the-middle, and de-authentication, and how these attacks exploit the vulnerabilities of different UAV applications is presented. The authors discuss applications of machine learning, blockchain, and SDN-based approaches to provide the security of UAVs only, not FANETs.

In [29]–[31], the security requirements, vulnerabilities, and privacy issues of UAVs were discussed, and included both physical threats as well as cyber threats. In [29], the authors conducted a brief review of the architecture and communication setup of UAVs, while they also summarized existing countermeasures for security issues. They provided a detailed explanation of security countermeasures for civil, government, and military UAVs. Additionally, they addressed network, communication, data, and forensic security solutions applicable to all types of UAVs. However, since they focused on countermeasures, the area of potential attacks was reviewed only briefly.

Only civil drone security and privacy issues were covered in [30], and emphasized vulnerabilities aimed at assuming flight control and landing. For this reason, the authors divided cyber-attacks into two groups: attacks on flight control and base stations, and attacks on data links. Future research topics of UAV communication were emphasized, and especially FANET security and detection systems. In another recent survey [31], attacks targeting UAV networks were categorized into physical and logical. This survey provided a broad overview by categorizing all non-physical attacks within the logical category. However, refining the classification of these attacks represents a critical step toward implementing more effective security solutions for UAV networks.

A recent survey [32] classified threats into eight groups according to attack vectors (physical, malware, sensor, communication, network, supply chain, hardening defects, miscellaneous) and presented an associated gap analysis. Another study [33] adopts the STRIDE threat model to further categorize and assess UAV security risks across various attack vectors. This approach integrates discussions on authentication techniques, physical layer security, and covert communication, providing a unified framework to understand how these elements collectively enhance UAV cybersecurity. In addition, this paper discusses relaying and trajectory optimization, which focus on optimizing communication routes and flight paths to improve both communication efficiency and network resilience.

In another recent survey [34], security issues are divided into four groups: sensor-level, hardware-level, communication-level, and software-level, and then comprehensively analyzed the countermeasures taken against each type of attack in the literature. In addition, privacy threats were also examined in three classes: individual risks, organization risks, and UAV risks. Unlike other surveys, the study also examined the basic features of UAVs under the headings of hardware, software, sensors, and communication. In this respect, this survey formed a good foundation for new researchers. However, their review of FANETs from a security perspective and potential solutions for them was very brief.

Similarly, in [35], security issues have been categorized into three groups based on the critical components of UAVs: hardware, software, and communication. However, unlike [34], fewer attacks have been discussed for each category, and these attacks have not been individually explained in detail. Emerging defense technologies have been discussed; nonetheless, this study does not handle sensor attacks nor does it outline measures against such attacks. Consequently, this study offers a general overview rather than a comprehensive examination.

In [36], threats targeting UAVs have been categorized into four groups: network, software, payload, and intelligent security. Differing from other surveys, it addresses attacks aimed at intelligent-based security solutions for UAVs. The study examines attacks to exploit machine-learning-based algorithms by manipulating data or generating malicious adversarial samples. However, it does not discuss countermeasures for these specific attacks, and limited attacks have been presented for other categories as well.

One of the comprehensive reviews in the literature was given in [37]. The study not only evaluated possible threats to UAVs according to different connection and node types, but also focused on FANET routing, characteristics, communication privacy, and security. This study presented security solutions by classifying threats and security solutions for FANETs according to the four groups of the OSI layer. However, as noted by the authors, some recent studies related to software-defined networking (SDN), machine learning, and 5G technologies were not included. Moreover, hardware-based attacks are not discussed in the study.

### A. *Positioning of Our Survey*

A summary of all current studies is presented and compared in Table II and Table III respectively. Almost all previously published surveys have maintained a focus on the general security of UAVs, but the research into the security of FANETs has been inadequate. Typically, these studies do not delve into the security implications arising from the characteristics specific to FANETs. Moreover, while all previous surveys in the literature have reviewed potential threats and solutions on a theoretical basis, there has been no research published in the current literature that has comprehensively analyzed the impact of these attacks on UAVs. The analysis of the four attacks presented in our previous study [38] has been extended in the current study with more realistic scenarios.

One important contribution that our study stand out for is the attack surface analysis. Attack surface analysis enables a more comprehensive understanding of the security landscape, not only by introducing new collaborations with emerging technologies but also by thoroughly exploring all possible entry points. As UAVs and FANETs evolve and interact with various technologies, understanding and addressing these diverse entry points are valuable for developing secure-by-design strategies.

Our study presents an attack taxonomy based on the analysis of attack surfaces. While various surveys offer diverse attack taxonomies, some of them [21] [30] [22] [26] [27] [23] [31] are not comprehensive enough as shown in Table II. Moreover, our study stands out for its comprehensive attack coverage. While many review studies present and discuss security countermeasures in UAVs and FANETs, our study provides a detailed analysis of these studies and discusses potential research directions rigorously, hence pave new ways for researchers. To the best of our knowledge, this work extensively explores research directions such as architectural aspects, multi-level security, and the role of the Ground Base Station (GBS) asset in security, contributing detailed discussions not previously elaborated upon. With the acceleration of research efforts in UAV security since 2020, we believe that such an inclusive new review study, which encompasses both a review based on the attack surface analysis and an attack analysis with simulations, will be beneficial for researchers.

TABLE II
OUTLINE OF RECENT SURVEYS ON UAV SECURITY

| Reference | Year | Taxonomy of Attacks | Description of Content |
|---|---|---|---|
| [21] | 2016 | by ad hoc networks;<br>- Eavesdropping<br>- Modification and Fabrication<br>- Selfishness | Security issues in ad hoc networks,<br>brief discusssion on FANET<br>communication security. |
| [30] | 2016 | by attacks;<br>- Flight control and base station<br>- Data link | Civilian drone security and<br>privacy requirements. |
| [22] | 2017 | by routing protocol attacks;<br>-Passive<br>-Active | Routing protocols,<br>network security and<br>countermeasures. |
| [24] | 2019 | No specific classification | Security issues in FANETs. |
| [26] | 2020 | by attacks;<br>- Sensor<br>- Communication links<br>- Multi-UAVs | Brief Overview of UAVs' Security and<br>Privacy Concerns. |
| [29] | 2020 | No specific classification | Use of UAVs, their applications,<br>potential attacks, and countermeasures. |
| [28] | 2021 | No specific classification | UAV communication security,<br>potential attacks and solutions. |
| [27] | 2021 | by attacks that hinder the drones' positions;<br>- Localization error attacks<br>- Others | Security, privacy, availability<br>authenticity, confidentiality,<br>and countermeasures. |
| [25] | 2022 | by attacks;<br>- UAVs-based systems<br>- UAVs-charging system attacks | UAV-based-system attacks,<br>focusing on battery<br>and charging system security. |
| [32] | 2022 | by classification;<br>- Confidentiality-Integrity-Availability-Privacy<br>- Threat vectors | Description of attacks and<br>countermeasures with gap analysis. |
| [37] | 2022 | by threat vectors and by OSI Layer;<br>- Physical<br>- Data link<br>- Network<br>- Transport | UAV and FANET security,<br>FANET characteristics<br>and countermeasures. |
| [23] | 2022 | by wireless communication threats;<br>- Navigational threats<br>- Data injecting and altering attacks<br>- Position altering threat<br>- Software threats | Focus on PHY and<br>cellular communication security. |
| [34] | 2023 | by classification;<br>- Hardware<br>- Software<br>- Sensor<br>- Communication | Security, privacy and<br>countermeasures. |
| [31] | 2023 | by classification;<br>- Physical and Logical | Potential attacks and<br>prevention methods. |
| [35] | 2023 | by classification;<br>- Hardware<br>- Software<br>- Communication | Security and<br>emerging defense tech.. |
| [36] | 2023 | by classification;<br>- Network<br>- Software<br>- Payload<br>- Intelligent | Attacks and<br>defense systems. |
| [33] | 2024 | by classification;<br>- Confidentiality-Integrity-Authentication<br>-Availability-Privacy-Non-repudiation | Security requirements,<br>threads and measures |
| Our Study | 2024 | by attack surface analysis;<br>- UAV<br>- Communication Layers<br>- Communication Architecture | Attacks and countermeasures,<br>attack surface analysis,<br>experimental attack analysis,<br>research directions. |

## III. BACKGROUND

Advancements in technology have led to the widespread popularity of UAVs, enabling their versatile applications across various real-world scenarios. While specific applications might favor the use of a single UAV, there are limitations to what

TABLE III
COMPARISON OF OUR SURVEY WITH EXISTING SURVEYS BASED ON DIFFERENT CRITERIA

| Study | Year | Security Impacts | Attack Surface | Taxonomy of Attacks | Attack Coverage | Focus on FANEts | Attack Analysis | Countermeasures | Open Issues |
|---|---|---|---|---|---|---|---|---|---|
| [21] | 2016 | ✗ | ✗ | ✗ | ✗ | ✔ | ✗ | ✗ | ✔ |
| [30] | 2016 | ✗ | ✗ | ✔ | ∂ | ✗ | ✗ | limited | ✔ |
| [22] | 2017 | ∂ | ✗ | ✔ | ✗ | ✔ | ✗ | ✔ | ✗ |
| [24] | 2019 | ✗ | ✗ | ✗ | ✗ | ✔ | ✗ | ✗ | ✔ |
| [26] | 2020 | ✗ | ✗ | ✔ | ✗ | ✗ | ✗ | ✗ | ✗ |
| [29] | 2020 | ✗ | ✗ | ✗ | ✗ | ✗ | ✗ | ✔ | ✔ |
| [28] | 2021 | ✗ | ✗ | ✗ | ✗ | ✗ | ✗ | ✔ | ✔ |
| [27] | 2021 | ✗ | ✗ | ✔ | ∂ | ✗ | ✗ | ✔ | ✔ |
| [25] | 2022 | ✗ | ✗ | ✔ | ∂ | ✗ | ✗ | limited | ✔ |
| [32] | 2022 | ✗ | ✗ | ✔ | ✗ | ✗ | ✗ | ✔ | ✔ |
| [37] | 2022 | ∂ | ✗ | ✔ | ∂ | ✔ | ✗ | ✔ | ✔ |
| [23] | 2022 | ✗ | ✗ | ✔ | ✗ | ✗ | ✗ | ✔ | ✔ |
| [34] | 2023 | ✗ | ✗ | ✔ | ✔ | ✗ | ✗ | ✔ | ✔ |
| [31] | 2023 | ✗ | ✗ | ✔ | ✗ | ✗ | ✗ | ✔ | ✔ |
| [35] | 2023 | ✗ | ✗ | ✔ | ✗ | ✗ | ✗ | ✔ | ✔ |
| [36] | 2023 | ✗ | ✗ | ✔ | ✗ | ✗ | ✗ | ✔ | ✔ |
| [33] | 2024 | ✗ | ✗ | ✔ | ✗ | ✗ | ✗ | ✔ | ✗ |
| Our survey | 2024 | ✔ | ✔ | ✔ | ✔ | ✔ | ✔ | ✔ | ✔ |

✔: mentioned, ✗: not mentioned, ∂:partial information operations a sole UAV can effectively execute. Single UAVs encounter challenges in completing missions when faced with rapid battery depletion, extended mission duration, potential electronic system failures due to external or internal factors, or susceptibility to targeting by attackers. These factors significantly hinder the effectiveness of single UAV operations.

In these scenarios, FANETs are recommended, as they allow multiple UAVs to join a common network and execute complex tasks in an organized manner [39]. Although FANETs inherits certain features from MANETs and its sub-classes, it also presents differences due to the very characteristics of UAVs such as their high mobility, unpredictable movements, and frequently changing network topology. Subsequently, such characteristics of UAVs are detailed along with the relevant security perspective.

### A. Node Mobility & Dynamic Topology

FANETs differ from other ad hoc networks due to UAVs' exceptional node mobility. These networks possess highly dynamic topology due to frequent changes in node positions. Mobility models differ in FANETs according to its application. UAVs might follow predetermined paths or move randomly [40]. They might exhibit independent movement or move collectively in group-based models. Unlike nodes in MANETs and VANETs, they maneuver in 3D space.

**Security Impacts:** The highly dynamic nature of the network topology poses a significant challenge in differentiating between normal and abnormal behaviour. For instance, identifying a node that is sending routing misinformation becomes intricate, as it could be an attacker or simply outdated. Furthermore, creating secure systems within dynamically changing environments poses significant architectural challenges.

Moreover, high-level mobility can impact security in both positive and negative ways. The mobility of targets, on one hand, can serve to mitigate the impact of attacks directed towards them. Conversely, the mobility also enables attackers to easily evade security measures.

### B. Node Density

Node density, which refers to the average number of UAVs per unit area [40], can vary from low to high, depending on factors such as operational areas, airspace coverage, applications, and the types of UAVs deployed. If UAVs possess high speeds and wide transmission ranges, their density tends to diminish as the distances separating them could extend across several kilometers [41]. Consequently, node density in FANETs is typically observed to be lower compared to both MANETs and VANETs.

**Security Impacts:** In scenarios with high node density, certain attacks like sinkholes can be particularly effective due to the increased connectivity. Such attacks exploit the density by attracting and redirecting network traffic, posing significant security risks. However, high node density can also positively influence distributed and collaborative security solutions [42]. Nodes in these dense networks are better equipped to detect intrusions locally and collaborate with neighbors to address insufficient local detection, enhancing overall security.

Conversely, in low-density scenarios where UAVs or nodes are sparsely distributed across vast areas, security concerns shift towards ensuring coverage, connectivity, and vulnerabilities due to larger communication ranges and reduced monitoring. In such environments, attacker nodes can evade detection more easily, especially within voting-based systems.

### C. Energy Consumption

In FANETs, energy consumption is still a critical design concern, particularly considering the utilization of mini UAVs powered by low-capacity batteries. This limited power source underscores the pressing need for innovative lightweight solutions, representing a substantial focal point for ongoing research within this domain [43], [44].

**Security Impacts:** UAVs are vulnerable to attacks due to their low energy capacity. In particular, sleep deprivation attacks [45] may be attempted right up until all available energy has been used. Battery attacks have several objectives, including battery drainage, leakage, unauthorized configuration, overcharging to overheat the battery, and draining energy [25].

Moreover, implementing security solutions for FANETs, which require extensive computation, poses a challenge. Lightweight algorithms and energy-efficient solutions are necessary for ensuring a secure network. This demands a careful balance between lightweight solutions and robust security measures to maintain network functionality while safeguarding against potential threats.

### *D. Radio Propagation Models*

Such models simulate how radio signals propagate through the airspace, directly impact the communication range and quality of wireless links between UAVs [46], [47]. FANETs, unlike MANETs and VANETs, operate at higher altitudes, affording better LoS between sender and receiver. This altitude advantage minimizes signal corruption and environmental interference, enhancing radio signal effectiveness.

**Security Impacts:**

Accurate radio propagation models are crucial for maintaining secure communication and mitigating vulnerabilities. Erroneous models can lead to connectivity misjudgments, resulting in weak or non-existent communication coverage, effectively creating blind spots. These inaccuracies can provide opportunities for attackers to exploit vulnerabilities, launching disruptive attacks or infiltrating the network undetected. Conversely, improved radio propagation in FANETs enhances security measures by enabling better monitoring of attackers through neighboring nodes, without signal disruptions between UAVs or GBS.

### *E. Localization*

Localization means determining the location of each UAV [44]. Localization techniques, such as GPS-based positioning, Inertial Measurement Unit (IMU)-based positioning, sensor fusion, and computer vision methods, play a pivotal role in many applications.

**Security Impacts:** UAV systems require accurate location information at short time intervals due to their high speed and dynamic topology. Such techniques are essential for establishing trusted communication channels, enforcing access control policies, and validating the identity of UAVs within a networked environment. Inaccurate or compromised localization data can lead to misidentified or spoofed UAV locations, resulting in various security threats. Furthermore, the rise of GPS spoofing and jamming attacks amplifies these risks, as they can maliciously alter the latitude and longitude information of UAVs or disrupt sensor data transmissions.

### *F. Communication Architecture*

The FANET communication architecture is defined by a set of rules and processes governing how UAVs communicate with each other or with GBS, exchange information, and establish connections. This architecture plays a crucial role in determining the network's resilience against security threats. Communication architectures vary based on the application areas of UAVs, yet no definitive research has proven which architecture is the most effective [39]. Communication architectures are generally classified into two main categories: centralized and decentralized [37]

In ***Centralized Communication*** architecture, each UAV communicates directly with a central controller. Since the UAVs cannot communicate with each other in this architecture, all data traffic is directed by the central controller [48].

The use of ***Decentralized Communication*** involves UAV-UAV interaction that occurs in decentralized networks, either directly or by hopping over other nodes. Without need for a centralized controller, this communication form is provided dynamically by FANETs [37]. In such systems, UAVs communicate within the group, and the base station is generally not included in these communications. Only selected UAVs are connected to the GBS, ensuring communication only between certain UAV groups [49].

**Security Impacts:** Centralized communication poses a potential single point of failure, presenting a significant security vulnerability that can impact the entire network in the event of an attack. Despite this vulnerability, central points boast higher processing capacities, enabling the implementation of complex security solutions and algorithms. Additionally, their broader network perspective allows these centralized solutions to potentially detect collaborative and distributed attacks across the network. However, decentralized security solutions enable the detection of specific attacks at a local level within each UAV, potentially accelerating attack detection and response times. Hybrid architectures, combining aspects of both centralized and decentralized approaches, try to strike a balance between control and autonomy. However, such hybrid models may inherit vulnerabilities from both centralized and decentralized systems.

Additionally, network centralization raises security issues and implies a need for trust between all nodes in the system. While the blockchain technology is a promising alternative for providing decentralization, depending on how it is used, it can also have certain limitations, such as significant latency, throughput, and block size [50], [51]. Moreover, the use and deployment of blockchain in UAVs with energy, computation, data storage resource constraints needs further examination.

### *G. Routing Protocol*

As UAVs perform their tasks, the nodes must communicate both with each other and also with the GBS. While establishing this communication, routing protocols are designed to provide real-time data transmission, to reduce processor and energy costs, and to adapt the dynamic changing topology. Due to the characteristics of FANETs, new routing protocols should be presented for such highly dynamic networks, or routing protocols developed for MANETs should be adapted to the highly dynamic structure of FANETs. Hence, in the literature, certain routing protocols proposed for MANETs have been extended and redesigned to handle issues such as broken link recovery [52], and security [53]. Routing protocols for FANETs can be examined under five classes [54] as shown in Figure 2. For a comprehensive overview of recent advancements in FANETs routing, interested readers may refer to the recent surveys [52], [55].

TABLE IV
COMPARISON OF FANETS WITH MANETS AND VANETS

| | **MANETs** | **VANETs** | **FANETs** |
|---|---|---|---|
| **Devices** | laptops, cell phones, etc. | vehicles, RSUs | UAVs, GBSs |
| **Node density** | medium/high | medium/high (city centers) | low |
| **Mobility and speed** | - 2D mobility | - 2D mobility | -3D mobility |
| | - low speed (7-20 km/h) | - medium speed (20-130 km/h) | - high speed (above 720 km/h) |
| **Topology changes** | medium | high | high |
| **Energy constraints** | high | low | high for small UAVs |
| **Security requirements** | low or high | high | high |
| | (application dependent) | | |
| **Radio propagation** | low | low | high |
| **Localization** | GPS | GPS | GPS and IMU |

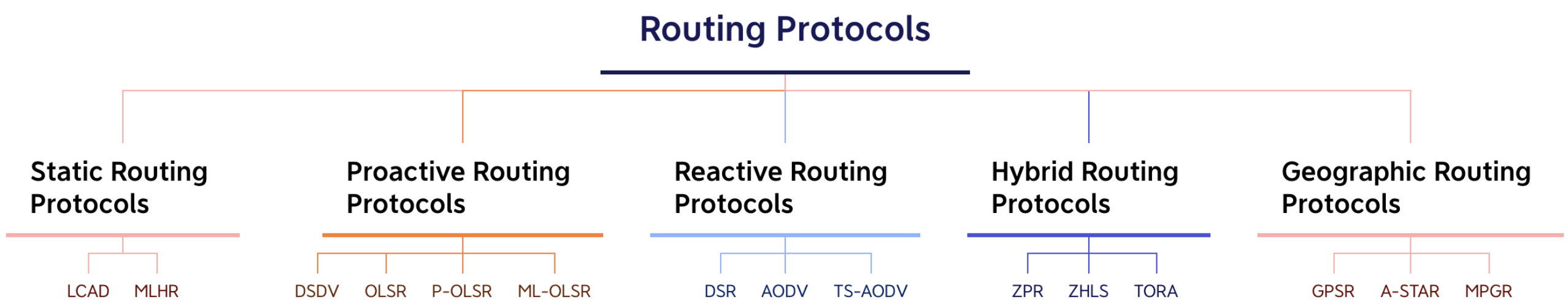

Fig. 2. Classification of routing protocols in FANETs

**Security Impacts:** Many protocols in FANETs assume that nodes are cooperative and non-malicious. In such a cooperative and dynamic environment, distinguishing well-behaved nodes from malicious ones becomes challenging. For instance, the frequent broadcast of messages to discover new routes due to broken links-resulting from the high mobility of UAVs-can raise security concerns. This behavior may resemble a denial-of-service (DoS) attack, where an attacker intentionally floods the network with excessive traffic to disrupt communication. However, it could also be a legitimate response to network dynamics. Consequently, the ambiguity makes it difficult to identify malicious activities.

To sum up, as outlined in Table IV, MANETs, VANETs, and FANETs have different characteristics, hence are faced with different security challenges. Building on these unique characteristics, the following sections will examine the attack surface of FANETs and the specific attacks they face, emphasizing how their operational environments contribute to these vulnerabilities.

## IV. ATTACK SURFACE OF FANET

The term "attack surface" refers to all the potential entry points on a system, system component, or environment where an attacker could attempt to breach it, have an impact there [56]. Attack surfaces constantly fluctuate as a system incorporates new components or interacts with existing ones. The categories of vulnerability, however, often stay the same. An attack surface diagram offers a comprehensive perspective on all possible flaws within a system. Therefore, in this study, we firstly classify the potential entry points and the landscape of threats against UAVs and FANETs in Figure 3, and then correlate them with the potential attacks targeting these categories depicted in Figure 4. Each entry point is discussed below.

### A. *Insider Threats*

Insiders are legitimate system users with a high attack potential. They often posses authenticated access to sensitive information. They may also be aware of the weaknesses in the implemented operation systems and processes. Unlike external attackers, whose attack traces are difficult to conceal, malicious insiders' activities are often harder to detect. Consequently, threat actors have a strong incentive to use the insider threat vector. These threats can come in various forms, ranging from unintentional actions, such as accidental information disclosure or data alteration, to deliberate misuse or neglect of safety measures.

Within the dynamic and decentralized nature of FANETs, insider threats pose additional challenges, especially given the crucial reliance on trust. Distinguishing between malicious intent and legitimate actions becomes more challenging in such naturally dynamic and collaborative environments. Moreover, incidents involving insider threats could significantly impact FANET operations, where tasks are highly sensitive and essential.

### B. Elements of FANET

In the context of FANETs, the inclusion of multiple UAVs and GBSs significantly expands the attack surface. UAVs are susceptible to various threats like signal jamming, physical tampering, and communication interception. Similarly, GBSs, serving as central data collection and command centers, are prone to cyber attacks leading to data tampering, unauthorized access, system interruptions, and potential takeovers. Moreover, the effects of some attacks such as DoS are more accentuated at GBSs. Due to their pivotal role in FANETs, GBSs represent an attractive target for attackers seeking to exploit vulnerabilities and gain network control. Furthermore, attackers can exploit out-of-date components, unsafe default configurations, and weak update mechanisms in both UAVs and GBSs. Securing the entire FANET ecosystem and ensuring reliable operation depend on addressing vulnerabilities in UAVs and GBSs, which are constant components.

### C. External Entities

The attack surface of FANETs extends further to include various interconnected components and systems such as cloud services, mobile devices, VANETs, IoT devices, edge computing, and other connected devices. When utilizing cloud environments for data processing and storage to overcome UAVs' memory constraints, it's essential to ensure the security and privacy of the data [57]. While edge computing improves real-time processing capabilities by processing data closer to the source of data generation, it also adds more attack points that adversaries might exploit [58]. IoT and mobile devices, offer possible points of exploitation and increase the risk of data manipulation or network compromise [59]. To conclude, additional interconnected components would increase the attack surface overall.

### D. Communication channels

Due to the mobility of UAVs, wireless communication serves as the primary medium for data transmission among them. Within the framework of FANETs and UAV deployment, the wireless channel constitutes a critical attack surface intricately linked to the unique operational characteristics of these systems and the inherent vulnerabilities of wireless channels. The wireless channel facilitates direct paths for potential eavesdropping and interception by malicious entities, making the confidentiality and integrity of transmitted data inherently more challenging to secure. Moreover, the constant and high mobility of UAVs across varying altitudes leads to continuously changing network topologies and fluctuating signal quality on the wireless channel, which complicates secure channel management. These communication challenges further complicate the implementation of distributed security mechanisms, which are generally better adapted to the distributed network utilized by UAVs. Additionally, the reliance on wireless communication makes UAVs particularly susceptible to jamming attacks, where adversaries can disrupt communication by broadcasting interference on the communication channels.

Different connectivity channels introduces specific points of interaction that contribute to the overall attack surface. These channels include connections to the cloud, IoT devices, and mobile devices, enhancing the network's versatility and connectivity. Moreover, FANET allows UAVs to communicate directly with each other and with GBS, satellites, and cellular networks, ensuring robust communication links in diverse operational environments. However, this array of communication channels highlights the diversity of protocols and technologies employed, presenting specific points of vulnerability to potential attacks. Security flaws in the communication protocols or weak security controls in their implementation may expose FANETs to vulnerabilities. Additionally, connected external entities could have their own unique vulnerabilities, necessitating customized security and communication strategies to ensure the overall integrity and resilience of FANETs.

### E. Emerging Technologies

Emerging technologies such as 5G, 6G [60], blockchain [61], digital twin [62] are essential for improving communication, trust mechanisms, and decision-making in FANETs. 5G and beyond represent a significant advancement in wireless communication technology, capable of fulfilling various objectives through the use of existing technologies. Operating as both a user device and a relay, a flying base station might cater to different requirements of various 5G use cases [63]. The purpose of a flying base station is to extend its connectivity coverage and improve its communication channel quality to achieve a greater communication rate and latency performance. This is particularly useful in emergency scenarios where traditional communication infrastructure may be disrupted or overloaded.

The innovative use of drones for communication during disasters presents an intelligent and creative solution, facilitating effective coordination among first responders and remote cities in crisis zones or conflict areas. For example, in the aftermath of a natural disaster such as an earthquake or hurricane, ground-based communication infrastructure may be severely damaged, making it difficult for emergency responders to coordinate and communicate effectively. In such situations, a flying base station mounted on a UAV can be deployed to establish a temporary communication network, enabling emergency personnel to coordinate their efforts, gather information, and provide assistance to affected populations.

Integrating emerging technologies presents potential security concerns, including novel attacks and vulnerabilities. These issues might have drastic results in certain applications, such as crisis management. For example, while digital twins enhance security, they may also introduce new vulnerabilities. Similarly, although AI aids in discovering complex characteristics of FANETs, adversarial attacks against AI-based systems [64] should be considered. In summary, due to the unique characteristics and challenges of FANETs, further exploration is necessary to understand and address the complex issues that arise from integrating emerging technologies into FANETs and UAVs.

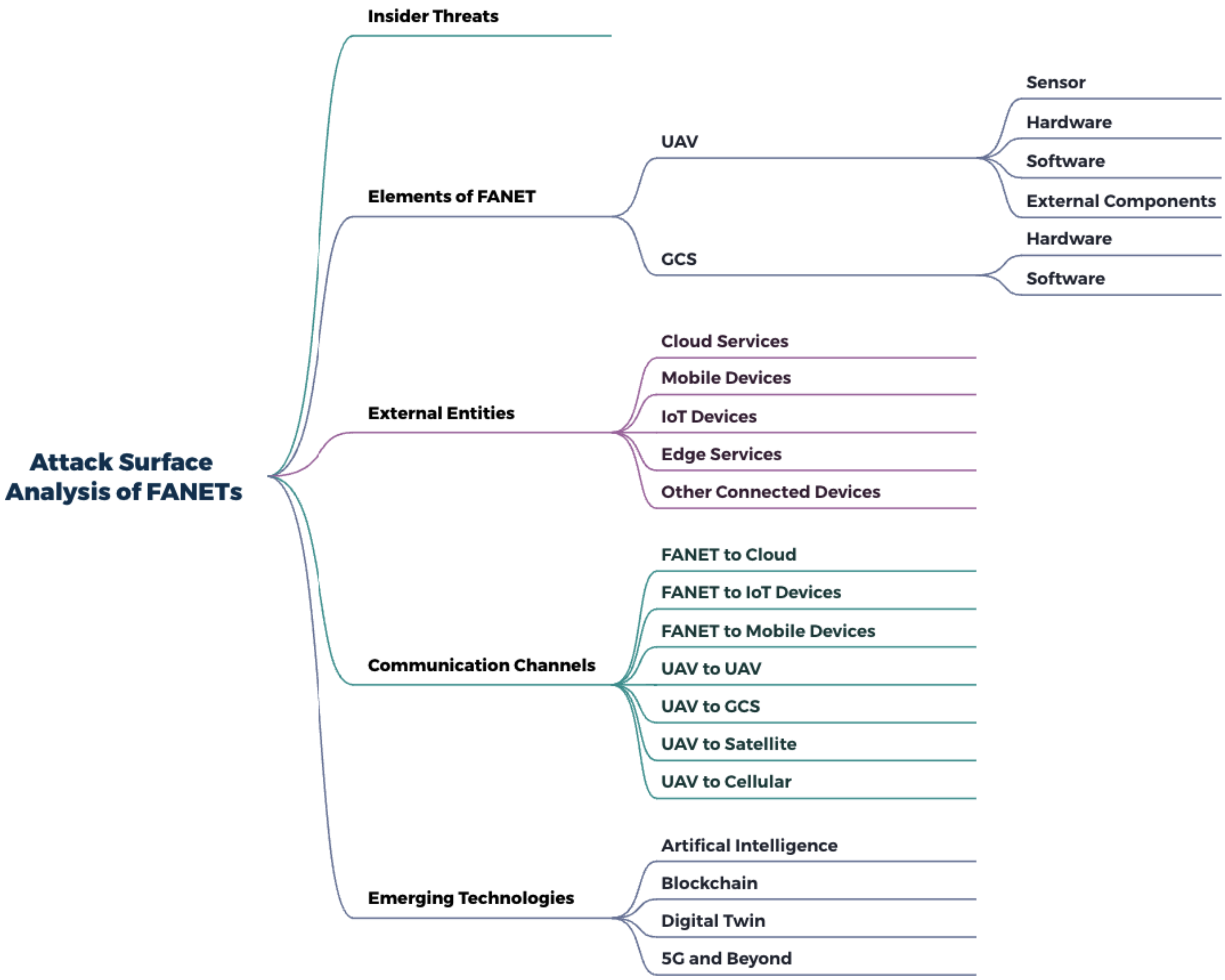

Fig. 3. Attack Surface Analysis of UAVs and FANETs

## V. Taxonomy of Attacks based on attack surface analysis

Attack surface analysis involves identifying and understanding potential points of vulnerability within UAVs and FANETs. The taxonomy of attacks complements this by categorizing related attacks based on the entry points identified in the attack surface. In this section, we prioritize categories based on their potential impacts, emphasizing the entry points that could have the most significant consequences in UAVs and FANETs. While there is overlap between the attack surface and the taxonomy of attacks, our focus is on areas with the highest potential impact, avoiding redundancy in categorization. This targeted approach aims to provide a comprehensive exploration of vulnerabilities within FANETs. With this in mind, we present a detailed review of attacks targeting UAVs, communication layers, and communication architecture in Figure 4.

### *A. UAV*

In this section, we primarily focus on attacks targeting UAVs, which are core components of FANETs. While GBSs are also integral to FANETs and share some vulnerabilities with UAVs, concentrating on UAV vulnerabilities covers a significant portion of the threats against GBS in both hardware and software domains. However, certain attacks can have a particularly severe impact on GBS, as they often represent a single point of failure. Additionally, it is important to recognize that GBS typically have heightened security measures and greater computational, storage, and processing capabilities compared to UAVs. These factors influence the nature of the attacks they might encounter. Furthermore, whether GBS are static or dynamic affects their ability to evade certain types of attacks. With this in mind, our focus will remain on thoroughly exploring UAV-specific threats, as they not only overlap significantly with GBS-related attacks but also highlight critical vulnerabilities impacting the entire system.

Attacks targeting UAVs are categorized into three groups: hardware-based, sensor-based, and software-based attacks. Each group is covered in detail in the subsequent subsections.

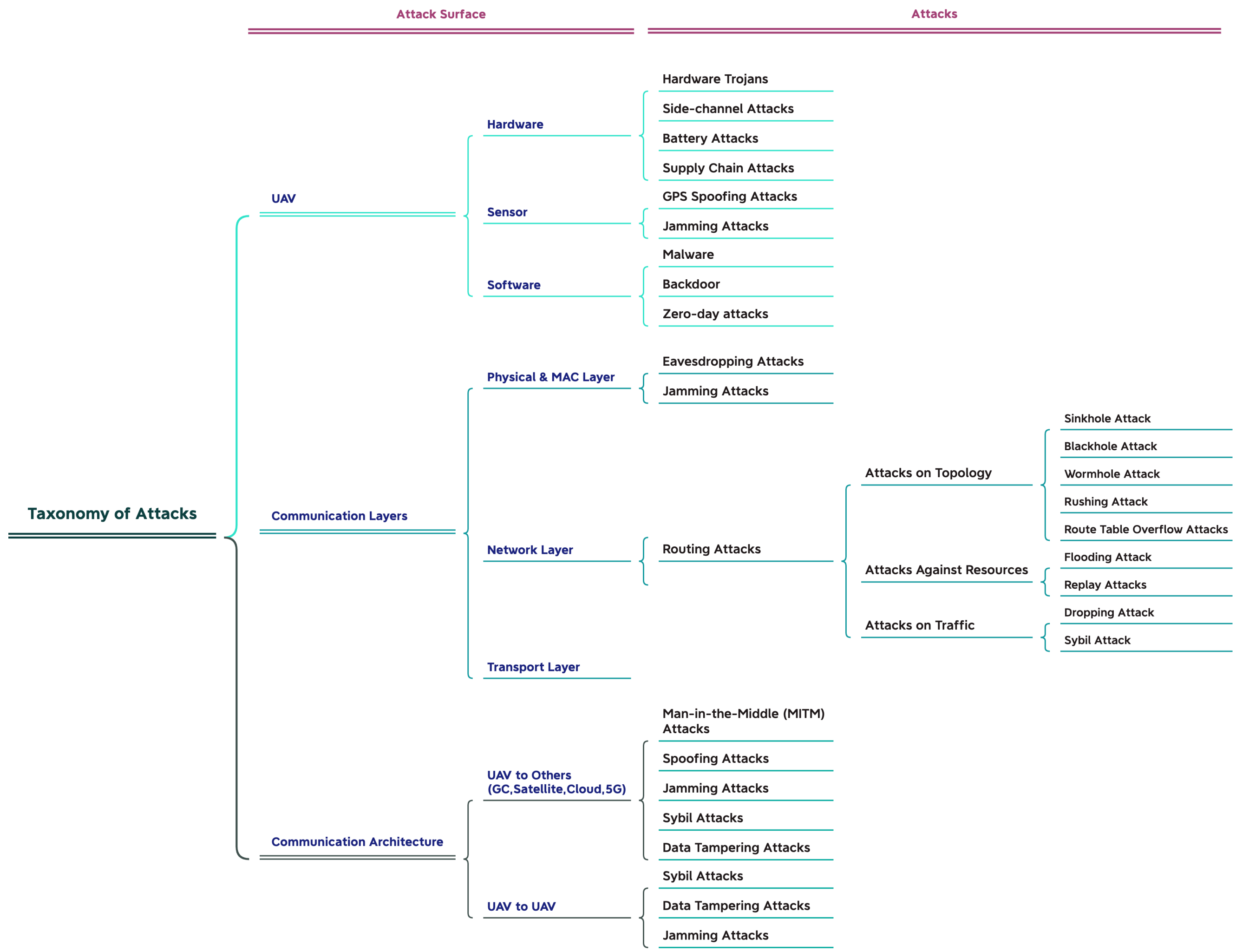

Fig. 4. Taxonomy of Attacks Based on Attack Surface Analysis

*1) Hardware-based Attacks:* Hardware attacks are aimed at accessing UAV components during the manufacturing process, or later during maintenance or usage [65] [66]. Components can be tampered with by an attacker installing harmful software or interrupting the flow of data. In addition, an adversary can add external components that could later assume remote control or capture data. Hardware attacks can not only result in the loss of control of UAVs, but also enable attackers to collect critical data.

In this section, we summarize the most significant attacks in this category from the literature, including hardware trojans, side-channel attacks, battery attacks, and supply chain attacks. Please note that attackers aim not only to disrupt the UAV's hardware through cyber-attacks but also to cause actual physical damage to the hardware through direct attacks. There have been several news reports of UAVs being physically shot down as a form of counteraction [67]–[69]. The scope of vulnerabilities broadens significantly with additional risks such as terrorist attacks, mid-air collisions, and illegal surveillance, which pose socio-political and physical threats beyond hardware damage [70].

Hardware Trojans: Hardware Trojans are described as malicious modifications embedded within electronic hardware, potentially causing detrimental effects when activated [71]. These modifications are often concealed and designed to be triggered under specific conditions, making them difficult to detect using standard testing procedures. Notably, hardware Trojans have been identified as a risk in the supply chain of defense-related technologies, including F-16 fighter jets [72]. This risk arises from the integration of counterfeit chips-equipped with these Trojans into critical military hardware, which could compromise functionality and leak sensitive information to unauthorized parties.

Side-channel Attacks: These types of attacks aim to exploit information leaked through physically observable phenomena resulting from the execution of tasks in microelectronic components [73]. Most side-channel attacks can be conducted without the need for specialized equipment, making them difficult to mitigate without impacting the performance of the UAV system. In [74], side-channel attacks were classified into three categories: time-based attacks, power consumption attacks, and electromagnetic radiation attacks. Time-based attacks exploit the device's operating time [75], while power consumption attacks utilize information related to the device's battery consumption. Electromagnetic radiation attacks measure the magnetic field around the device while it processes information.

Battery Attacks: Small drones, which are in high demand for many kinds of tasks, typically use compact batteries. While this creates resource constraints for certain operations, attackers can also target nodes with small batteries to deactivate them. Such attacks can lead to premature battery depletion, interrupting connections, triggering mission failures, and even causing UAVs to crash land. These types of attacks can significantly disrupt essential UAV operations, such as controlling flight [25]. Moreover, adversaries can also target the charging system used by UAVs. In [76], a battery attack was demonstrated in which attackers sent fake requests to the charging system to induce voltage fluctuation problems. It was shown that a *Depletion of Battery (DoB)* attack can quickly consume a UAV's battery power [77], [78], with energy levels of nodes depleting 18.5% faster when under attack. Another type of battery attack is a *denial of sleep* attack, which prevents UAVs from entering into sleep mode [79], and thereby continuing to use power unnecessarily.

Supply Chain Attacks: Attacks that target the supply chain aim to exploit various security vulnerabilities that can arise during the procurement phase of UAV components. In [80], a supply chain attack was presented in which attackers remotely modified 3D design files to produce faulty UAV components. Another study [81] described a supply chain attack that attempted to integrate components not specifically designed for the printed circuit board (PCB) connecting various electronic circuit components. However, ensuring supply chain security is challenging due to the large number of manufacturing companies operating within the sector.

*2) Sensor-based Attacks:* UAVs are equipped with a variety of sensors that monitor events or changes in the environment and collect data to perform various tasks, ensuring optimal service levels. These sensors are particularly attractive to attackers due to the sensitive and confidential nature of the information they handle. Such attacks can jeopardize the flight missions of targeted UAVs [34] and hinder their efficient operation [32]. A sensor attack on gyroscopes was introduced in [82], where audible and ultrasonic noise was applied to 15 different gyroscope sensors in both simulated and real-world experiments. These types of attacks can affect multiple sensors, leading to disturbances in flight balance. In all 20 trials conducted in a real-world environment, the target drone lost balance and crashed. To elaborate on sensor-based attacks, we categorize them into two types: spoofing attacks and jamming attacks.

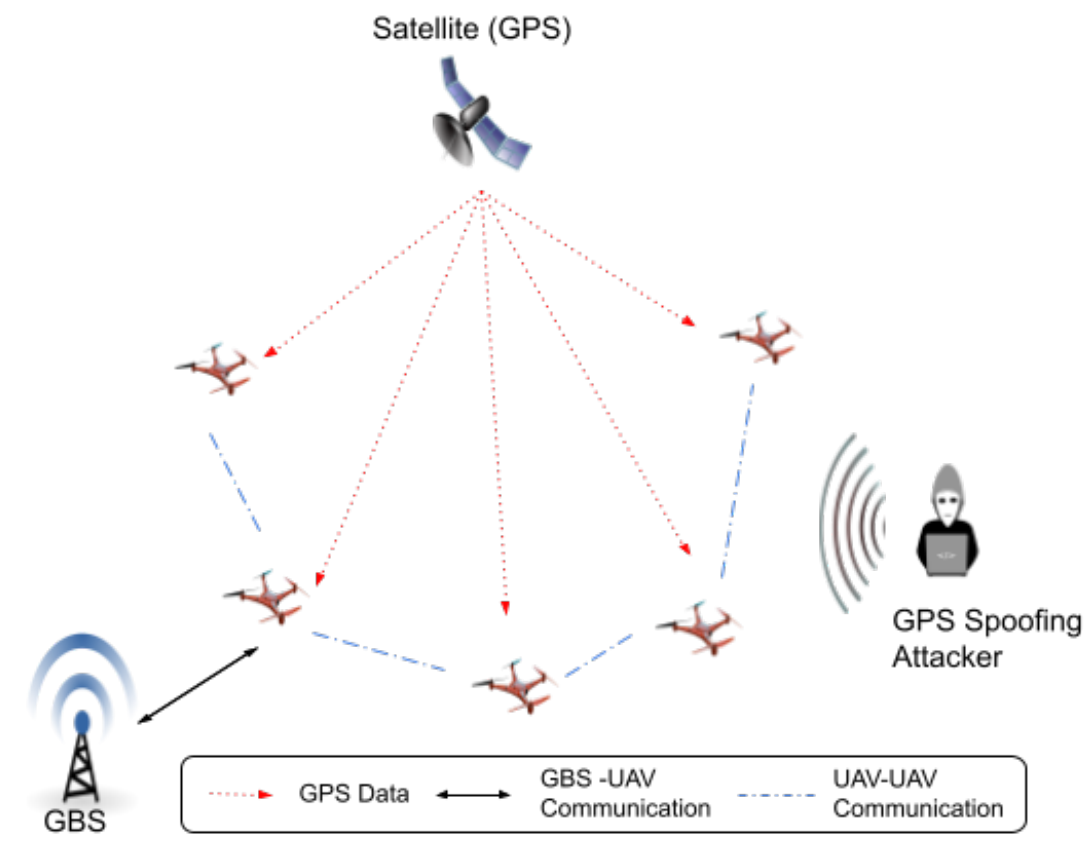

Fig. 5. An exemplar GPS spoofing attack

Spoofing Attacks: In a passive spoofing attack, data is captured and eavesdropped without affecting the UAV system, whilst in an active attack, sensor data is modified or falsified data is introduced. These attacks are generally used to alter previously planned UAV behaviors in order to assume control of the device. The most commonly used spoofing attacks involve GPS spoofing, in which fake GPS data is sent from a malicious device in order to fool the GPS sensor of an operating UAV as shown in Figure 5. This attack was first presented by the University of Texas in 2012 [83]. In [84], fake latitude and longitude information was sent to a GPS unit without disturbing the signals of the original GPS sensor. In another study [85], a GPS spoofing attack was conducted by generating fake GPS data using a device that generates GPS signals and integrated using the SimGen simulation tool [86].

Since sensors cannot detect differences between real and fake data, they transmit all the information they have without question, which may result in UAVs flying off to unintended locations where they could be either damaged or captured. Another study [87] presented GPS attacks with real-world tests shows that the DJI Phantom 3 Standard drone is vulnerable to GPS attacks, which endanger its functioning and control. Attacks on the drone can cause it to depart from its intended flight route, display unpredictable behavior, or even interrupt communication from the remote control.

Jamming Attacks: Jamming attacks utilize jamming equipment to interrupt sensor signals, hence completely preventing the target UAV from receiving valid sensor information. Since the flight control and stabilization of UAVs rely upon stable sensor information, their stabilization systems can be disturbed or even damaged. Although jamming is not effective in all conditions, such as where the jamming signal frequency is inadequate, or where the attacker is excessively distanced from the target, jamming equipment is considered generally available and inexpensive to acquire. An attacker, who was thought to be located near to the test flight area in South Korea, jammed the GPS sensor signals using a jammer which resulted in the UAV crashing into the ground base system, killing an engineer and injuring two remote pilots [65].

*3) Software-based Attacks:* Software-based atacks target the integrity, confidentiality, and availability of the UAV system. Various UAV components can suffer harm in such attacks, including flight displays, navigation systems, and any vital system functions utilized to control and operate the UAV during flights [88]. To elaborate on software-based attacks, we cover them in two general groups: malware and zero-day attacks.

Malware: Malware refers to malicious software designed to disrupt, damage, or gain unauthorized access to computer systems or networks. In the context of UAVs, malware can specifically target the onboard systems and software that control the drone's operation. This allows attackers to manipulate the drone's movements, potentially leading to data theft or physical damage. For instance, attackers can coerce UAVs to navigate towards specific locations as per their instructions [89]. The initial occurrence of malware infiltrating a UAV without causing direct damage was reported in [90].

A backdoor provides unauthorized access to a system or encrypted data by circumventing the system's security measures. This can exploit software vulnerabilities or utilize hardware backdoors present in encryption algorithms or networking protocols. In the realm of UAVs, backdoor attacks pose significant risks, including the theft of sensitive information, unauthorized manipulation of drone control systems, and even the potential hijacking of multiple drones. The first documented UAV backdoor, named Maldrone [91], employed a Transmission Control Protocol (TCP) connection to gather sensor and driver data, potentially enabling the hijacking of UAV control.

Zero-day Attacks: Protecting networks and systems from unauthorized access or potential threats from unknown attacks is a challenging task. The period in which software developers have to fix a publicly disclosed vulnerability is referred to as a zero-day. Vendors attempt to deliver a patch or update during this period to address the recently discovered vulnerability. If a patch is not made available as soon as possible, attackers might take advantage of the vulnerability and launch a zero-day attack or leverage it.

### *B. Communication Layers*

Secure communication is one of the significant requirements for UAVs to ensure stable, reliable, and secure data transmission and flight control. In this subsection, essential communication attacks are categorized into three groups: physical & MAC layer, network layer, and transport layer attacks.

*1) Physical & MAC Layer:* Radio signals and wireless networks serve as fundamental communication channels between UAVs and GBS, as well as within multi-UAV setups. In [26], the authors demonstrated the disruption of the connection between UAVs and terminals by attacking specific UAV components. Subsequently, they managed to crack the acquired password from a simulated multi-UAV attack. In [92], well-known number of attacks against physical and MAC layers were discussed, including the ARP injection attack, the dictionary attack, and the PTW attack. In another study [93], it was described how an attacker can sniff the signals emitting from UAV devices using Bluetooth.

Physical layer attacks are generally classified into two groups: eavesdropping attacks and jamming attacks [94].

Eavesdropping Attacks: Eavesdropping attacks are categorized into two: passive and active. As a type of passive attack, eavesdropping is where a message is captured by an unauthorized attacker. The attacker then violates the network privacy to listen in on the communication without interrupting the transmission [25]. In addition, attackers can elevate an attack by introducing fake messages, delete or modify the intercepted message.

In active eavesdropping, the primary objective is to disrupt the main communication channel by reducing its capacity [95]. This is typically achieved by transmitting jamming signals to the legitimate receiver, thereby degrading the channel's performance. Furthermore, active eavesdroppers may attempt to enhance the capacity of their eavesdropping channel. As a result, active eavesdropping poses a greater threat compared to passive eavesdropping. The LoS channel characteristic plays a pivotal role in this context [95]. LoS communication provides a direct and unobstructed transmission path between the UAV and the legitimate receiver, ensuring reliable communication. However, this characteristic also presents vulnerabilities. If a third-party eavesdropper positions themselves within the LoS communication range, they can intercept the transmitted data intended for the legitimate receiver. Furthermore, they may exploit the optimized channel characteristics to achieve higher data rates and better reception.

In the context of duplexing modes, both full-duplex (FD) and cooperative half-duplex (HD) eavesdropping present challenges. In FD scenarios, the eavesdropper transmits jamming noise and intercepts confidential signals simultaneously and independently. In response, the victim UAV may increase its transmission power to mitigate channel degradation, despite the associated higher energy consumption. Since this also strengthens the signal received by the eavesdropper, it increases the possibility of eavesdropping. Conversely, HD eavesdroppers divide tasks, with some transmitting jamming signals to the legitimate receiver while others intercept confidential data. In these scenarios, HD eavesdroppers collaborate to mimic FD behavior. Regardless of the communication mode used, if an unauthorized party obtains Channel State Information (CSI) about the legitimate receiver, it significantly increases the vulnerability of UAV systems. This is because it provides detailed insights into communication conditions, enabling attackers to exploit weaknesses in security measures and launch more targeted attacks.

Jamming Attacks: Various jamming techniques are discussed in the literature [96]–[98]. A simple form of jamming attack is the *noise attack*, which is designed to disrupt the radio signals utilized by UAVs through the introduction of pulses and disruptive noise, consequently impeding their operational functionalities. Attackers can utilize powerful transmitters to generate strong signals that disrupt not only the victim's communication, but also all elements of network communication. They can even be be attempted from a remote distance [94]. Additionally, when a continuous wave signal, frequently at a specific frequency or tone, is employed to disrupt communication or navigation systems, it is referred to as *tone jamming*.

Another form of jamming attack is *swept jamming*, which combines elements of both noise and tone jamming techniques. Similar to noise jamming, it focuses on a singular tone, yet it also simulates the effects of tone jamming through sweeping. Unlike noise or tone jamming, Swept jamming ensures comprehensive coverage across the entire spectrum, including all hop frequencies of the data signal [98]. In contrast, *follower jamming* relies on identifying and jamming specific frequency hops within an FHSS system, requiring spectrum analysis to detect energy fluctuations indicating new signal presence or departure from the band.

Traditionally, methods like Frequency Hopping Spread Spectrum (FHSS) and Direct Sequence Spread Spectrum (DSSS) have been proposed as robust defenses against jamming attacks [21]. These techniques alter the sent frequency values, making it challenging for adversaries to disrupt communication by jamming a narrow frequency band. FHSS is a technique in which signals are quickly switched among various frequency values. Similarly, DSSS changes the frequency value of an original signal by adding noise into a normal frequency signal [99].

FHSS and DSSS have proven effective against conventional jamming methods. However, the emergence of smart jamming techniques underscores the necessity for new security solutions. *Smart jamming* attacks selectively disrupt essential signals to disrupt successful communication, optimizing power efficiency and efficacy through transmitted data analysis and critical point identification. Protocol-aware jamming techniques, developed using Software Defined Radios (SDR), generate jamming signals that mimic target signals, including hopping patterns for FHSS and Pseudorandom Noise (PN) code data rates for DSSS. This approach yields energy-efficient jamming performance with reduced bit error rates.

*2) Network Layer:* UAVs communicate with each other or a GBS via an ad hoc network. One of the biggest advancements in ad hoc networks has been the development of new protocols and the improvement of existing ones in terms of energy efficiency, packet delivery ratio, and other metrics [22], [100]. However, most of the proposed protocols rely on the cooperativeness of nodes in the network and fail to incorporate mechanisms for enhancing security.

As a result, attackers primarily target routing protocols. A malicious node that aims to disrupt a routing mechanism can significantly decrease network performance by easily integrating itself into the network and subsequently obtaining critical information [22]. Here, we categorize routing attacks into three groups: attacks on topology, attacks against resources, and attacks on traffic. It should be noted, however, that certain attacks could be classified as belonging to more than one group.

Attacks on Topology: Many routing protocols incorporate a route discovery mechanism. In reactive protocols, this mechanism is initiated when a node requests a new route to send packets to a destination. Malicious nodes could exploit this mechanism to create non-optimal routes between endpoints, thereby capturing data packets. Some of the most significant topology attacks are summarized as follows.

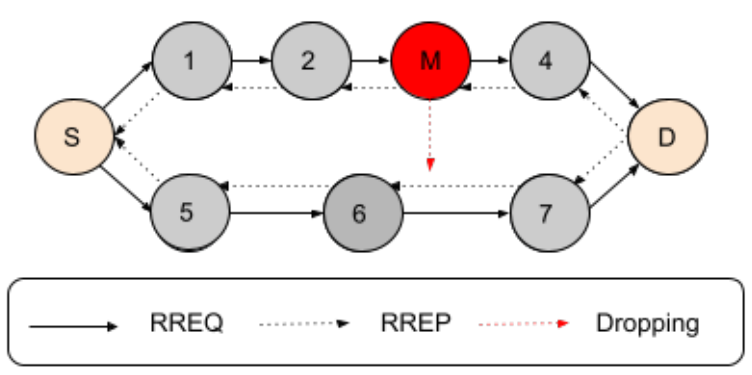

Fig. 6. An exemplar blackhole attack on AODV

- Sinkhole Attack: Here, attackers advertise as if they have a better route to a certain destination. If the route is then selected by the source node, all network communication between the source and destination nodes can be eavesdropped by the attacker node, therefore it is referred to as a sinkhole attack [38]. This attack type is generally employed as a first step prior to launching further attacks that will aim to drop and modify data packets.
- Blackhole Attack: This attack is a combination of sinkhole and dropping attacks. First, the attacker advertises that it has the best route to the required destination, as in a sinkhole attack, and subsequently directs the network traffic to itself. This is followed by other attacks such as modification and packet dropping attacks. Figure 6 illustrates an exemplar blackhole attack launched against the AODV protocol. The source node (S) broadcasts an RREQ message to discover a route to the destination node (D). When the malicious node (M) receives the RREQ message, it sends a fake RREP message to the source node. As can be seen in the Figure 6, M is not located on the shortest path to the destination. However, even if it does not have the shortest or freshest route to the destination, it continues to receive the data packets sent from S to D, and then drops the packets, hence the blackhole effect. A blackhole attack can cause disconnections due to increased network overhead, and the attack can also increase the UAV system's overall energy consumption due to route re-discovery and shorten the lifetime of the network.
- Wormhole Attack: This attack introduced in [101] sends

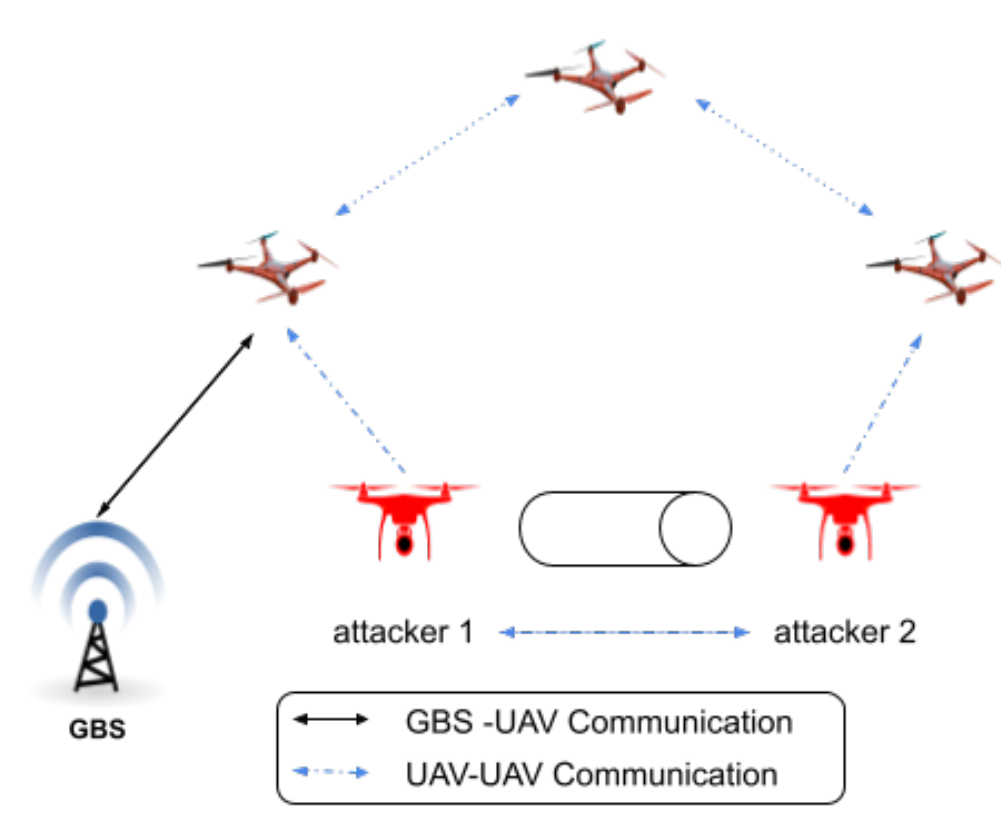

Fig. 7. An exemplar wormhole attack on AODV

information as if an attacker is in the neighborhood of other nodes in the network. Hence, genuine nodes start to send their data through the attackers. As shown in Figure 7, in a wormhole attack, attackers create a tunnel between themselves, and then forward the every packets they receive on to other attackers. When attackers gain access to the network, they can perform further attacks such as dropping, altering, and false data injection. In addition, these attacks can damage the proper working of routing protocols as they can prevent discovery of other nodes two hops or further away, resulting in packet loss and network performance reduction.
- Rushing Attack: Introduced by Hu et. al. [102], the rushing attack is a type of DoS attack. It was claimed in [102] that this type of attack can be effective against all existing reactive protocols proposed for ad hoc networks such as AODV [103], DSR [104], and even secure protocols such as SAODV [105] and SUCV [106]. The rushing attack exploits the vulnerability of the route discovery mechanism, in which the destination node receives the first RREQ packet and discards the others. When a request packet is received by the malicious node, it is immediately forwarded to the destination node. Since the packet from the malicious node will reach the destination node faster, the destination node will discard other legitimate request packets. As a result, the source node will be forced to use routes containing the attacker node. Many techniques can be employed in order to conduct rushing attacks such as creating wormholes, disregarding MAC or routing layer delays, holding other transfer node queues as full, and sending data at a higher wireless transmission rate [107].
- Route Table Overflow Attacks: Even though this attack is not specific to ad hoc networks, it might be more effective on resource-constrained nodes of such networks. In this instance, an attacker sends a large number of route advertisements with the aim of causing an overflow of the routing tables and hereby preventing new routes from being established [107]. Hence, this attack type can be more effective in proactive rather than reactive routing protocols [94].

Attacks Against Resources: Attacks in this group aim to increase network traffic and overhead, and thereby slow down data transmission and reduce the network's overall performance. By congesting accessible links, this attack limits the availability of the network and effectively reduce its lifetime. Another effect is to consume nodes and network resources, hence eliminating nodes from further communication, and even creating partitions in the network. Resource-constrained characteristics of nodes make such nodes very attractive for attackers. Some of the most significant routing attacks targeting resources are summarized as follows.

- Flooding Attack: This attack can manifest in various implementations, such as by sending large numbers of control and data packets. Some routing protocols send Hello packets to one-hop away nodes in order to determine their neighbor nodes. Neighbor nodes receive these messages but do not forward them. The attacker takes advantage of this stage and sends a large number of Hello packets to neighboring nodes. Another type of flooding attack is carried out using the route discovery phase of the AODV routing protocol. This attack, in which a large number of RREQ messages are broadcast at regular intervals, is referred to as the RREQ flooding attacks [38]. These packets can be directed to nodes in the network or to node addresses that do not exist within the network. This attack generally exploits the route discovery mechanism, since the control packets are broadcast in this phase in many reactive routing protocols.
- Replay Attacks: This is a DoS attack that re-sends outdated but legitimate data in order to slow down and/or interrupt communication.

Attacks on Traffic: These attacks target the dropping, modifying, forging or replaying of data packets. A malicious node may also then perform additional attacks using the data captured in a prior attack. Some of the more well-known attacks in this group are summarized as follows.

- Dropping Attack: In this attack, an attacker may drop all of the packets that they receive or selectively drop packets intended for a specific destination. Such attacks, where attackers solely drop certain data packets and forward others, are called *Grayhole Attacks* and are more difficult to detect [108]. Attackers may even randomly drop a few packets to evade detection; however, in that scenario, the effect of the attack on the network might also be limited. Generally, the attacker node intervenes in the routing protocol during route discovery, similar to a sinkhole attack, positioning itself within a valid route to initiate a dropping attack. Although dropping attacks fall under the category of traffic attacks, attackers might also drop routing control packets to disrupt the establishment of valid routes. In such cases, restarting the routing discovery mechanism will consume additional network resources.
- Sybil Attack: This attack is considered a form of impersonation attack. Nodes must have original IP addresses in order to join the routing process [22]. However, if the network does not have a central authority node in place to check the identities of nodes in the network, as in many scenarios, attackers can use the address of other nodes or even generate addresses not present in the network. In other words, attacker nodes generate stolen or fabricated identities in attacks referred to as sybil attacks [109]. These sybil nodes perform further attacks, such as placing themselves within a route and modifying data packets.

*3) Transport Layer:* Attacks in this layer are well-known for targeting transport layer protocols, including Transmission Control Protocol (TCP) and User Datagram Protocol (UDP), through methods such as SYN flooding, UDP flooding, and session hijacking. In [110], well-known DoS attacks such as SYN flooding attacks were implemented by using some tools such as Hping3 [111], LOIC [112], Netwox [113] against a particular UAV (AR.Drone 2.0). It has been observed that the attacks disrupt the communication channels of the drone, resulting in reduced responsiveness, reduced video stream quality, and even a complete loss of control.

### C. Communication Architecture

Communication architecture is fundamental for establishing a secure and efficient network for FANETs and UAVs. Understanding and addressing these attacks is essential to prevent unauthorized access, data manipulation, and disruptions to the communication flow. To achieve this, the following attacks are discussed.

*1) Man-in-the-Middle (MITM) Attacks:* A Man-in-the-Middle (MITM) attack occurs when an attacker covertly intercepts and alters communication between two parties, thereby gaining unauthorized access to confidential data [114]. MITM attacks jeopardize the security and integrity of data transfer and can take many different forms, including packet sniffing, DNS spoofing, and session hijacking. In [115], the impact of unauthorized nodes entering the VANET is illustrated, highlighting how MITM attackers seek to spread and exchange malicious content with vehicles.

*2) Data Tampering Attacks:* : Attackers could tamper with various data, leading to inaccuracies in critical information such as the UAVs' position, altitude, or direction shared with other parties. Tampering with navigation data can misguide UAVs, compromising their situational awareness. Additionally, manipulating sensor data could distort their perception of the environment, further undermining their operational capabilities. Furthermore, altering information related to alarms, emergency circumstances, or safety procedures could impede appropriate responses or result in false alarms.

## VI. Attack Analysis

Routing attacks pose significant threats to FANETs, originating from nodes that bypass prevention methods and can cause dramatic damage to the network. Therefore, it is imperative to analyze these attacks to develop effective countermeasures. Despite the importance of routing security, there is a notable lack of studies focusing on the analysis of routing attacks in FANETs. Consequently, in our study, we aim to address this gap by conducting a comprehensive analysis of routing attacks.

This study covers four attacks against the widely used AODV protocol, each with different goals. Initially, a concise overview of AODV, 3D Gauss Markov Mobility (GMM), and the specific attacks is presented. Following this, the simulation results obtained from networks with diverse topologies are demonstrated and deeply analyzed. Details of the experimentation process are provided in the subsequent subsections.

### A. Routing Protocol: AODV

AODV is one of the most widely used reactive routing protocols in ad hoc networks. In this study, AODV was chosen over alternative protocols due to its widespread adoption, simplicity in implementation, and minimal operational overhead [116].

In AODV, when a node needs to send data to a destination, it first checks for an existing rout. If no route is found, a route request (RREQ) is broadcasted, prompting nodes with a valid route to reply with a route reply (RREP). The source node then selects the most recent and shortest route based on maximum sequence number and minimum hop count. Once established, data transfer commences. To handle link breakages caused by mobility, AODV utilizes route error packets (RERR) to notify nodes, allowing affected nodes to trigger the route discovery mechanism for alternative routes, thereby ensuring robustness in dynamic ad hoc networks.

### B. Mobility Models: 3D GMM

3D GMM is a memory-based model with a single parameter that can potentially address the need for a realistic mobility model capable of adjusting various degrees of randomness [117]. The model is applied as a time-based mobility model specifically to avoid abrupt changes in the direction or velocity of UAVs. By utilizing this model, it becomes feasible to simulate various real applications while accommodating 3D movement in UAVs [52]. To maintain meaningful consecutive positions, the model stores prior node movements in memory and utilizes the parameter $\alpha$ (ranging between 0 and 1) to govern subsequent node mobility behaviors [118].

### C. Implementation of Attacks

In the analysis conducted in this study, four attacks against AODV were implemented in realistic simulation scenarios: sinkhole, dropping, blackhole, and flooding attacks.

*1) Sinkhole Attacks:* In this implementation, when the attacker node receives a RREQ packet, it sends a fake RREP packet in return. The fake RREP packet contains a destination sequence number that is higher than the current one. Hence, if the RREP packet reaches the source node, it is guaranteed to be selected as the route to the destination. Moreover, it claims to be one hop away from the destination node.

*2) Dropping Attacks:* In this attack implementation, if the attacker node is deemed to be located on the active route between the source node and the destination node, it drops every data packet it receives, and thereby prevents communication between these endpoints. However, since malicious nodes are selected randomly, there is no guarantee that they are located on active routes. In such cases, the effect of an attack on the network would be limited.

*3) Blackhole Attacks:* This attack is a composite attack that includes both sinkhole and dropping attacks. In other words, the attacker node first places itself in a route, then drops data packets that are transmitted through that route.

*4) Flooding Attacks:* This exploits a vulnerability in the route discovery mechanism. An attacker node sends a significant number of RREQ packets to randomly selected nodes in the network. In the simulations, a destination node is selected randomly, and then 10 sequential RREQ messages are broadcast for this destination node. The attack is repeated every 3 seconds for the duration of the simulation.

### D. Simulation Parameters

In this study, attacks were simulated using the ns-3 simulation tool [119]. The simulations use 25 and 50 nodes in order to assess the effects of node density on the network performance. A specific immobile node designated as the GBS is located at the center of the simulation area. Ten network communications are built by randomly assigning 10 source and 10 destination nodes. The remaining nodes may function as relay nodes. The destination nodes are responsible for collecting data from other nodes and sending it to the GBS. The communication between destination nodes and the GBS begins at the 10th second and continues until the end of the simulation. Attacker nodes are selected randomly from nodes excluding the source and destination, and nd remain fixed for each attacker ratio (ranging from 5% to 25%), regardless of the type of attack being carried out.

TABLE V
SIMULATION PARAMETERS

| Parameters | Values |
|---|---|
| Routing protocol | AODV |
| MAC protocol | IEEE 802.11b |
| Simulation time | 1800 seconds |
| Area | 12000 m x 12000 m x 300 m |
| Number of nodes | 25, 50 |
| Avg. speed | 100 m/s |
| Transmission range | 250 m |
| Traffic type | UDP |
| Packet size | 512 bytes |
| Packet count | 1/s |
| Bandwidth | 11 Mbps |
| Ratio of malicious node | No attack, 5%, 10%, 15%, 20%, 25% |
| Mobility model | 3D GMM Model |
| Bounds for GMM | X: [0; 12000], Y: [0; 12000], Z: [0; 300] |
| $\alpha$ for GMM | [0.25-0.7] |

Several different network topologies are run, and results are obtained without any attacks. Next, various attack types (blackhole, sinkhole, dropping, and flooding) are individually simulated on these selected topologies. Specifically, 10 simulations are run without attacks, and for each attack type, 10 simulations are performed for each of the 5 attacker ratios (ranging from 5% to 25%), resulting in a total of 210 simulations. The average performance result was then used for the subsequent attack analysis.

3D GMM was employed to simulate the natural 3D flight of UAVs in a realistic manner, as demonstrated in [117]. The alpha parameter value of the 3D GMM, which provides a balance of randomness and predictability in the UAV's mobility, was initially set at 0.25 and then incrementally increased by 0.05 to create different network topologies. The simulation parameters used are summarized in Table V. Our previous study [38] was extended to include additional simulations where UAVs operate in a larger area with more nodes over longer simulation durations. Moreover, more realistic scenarios were implemented. For instance, whereas all nodes sent their data to a single mobile server in the previous study, in this study, nodes can communicate with each other and send the collected to a stationary GBS located at the center of the simulation area.

## *E. Experimental Results*

In this subsection, we present and examine our experimental findings, focusing on the network's performance under diverse attack scenarios. The following performance metrics were used to evaluate the performance of networks under attack: packet delivery ratio (PDR), end-to-end (E2E), and overhead (OVH). PDR is the average ratio of the total number of packets received by all nodes in the network to the total number of packets sent to the same nodes. E2E latency is the measurement, in seconds, of the average of all delays that occur on the network during data transmission between end communication points. Overhead is the ratio of the total control packets generated by the routing protocol and received by the nodes to the total number of data packets received. Under simulated blackhole, sinkhole, dropping, and flooding attacks across different attacker ratios (ranging from 5% to 25%), the network's performance was observed. The average performance metrics derived from simulations involving 25 and 50 nodes are presented in Table VI to Table X.

TABLE VI
EFFECTS OF SINKHOLE ATTACK

| | Attacker ratios | PDR | E2E (s) | OVH |
|---|---|---|---|---|
| Low Density | 0% | 93.70% | 0.084 | 7.40 |
| | 5% | 93.70% | 0.097 | 7.69 |
| | 10% | 93.60% | 0.102 | 8.16 |
| | 15% | 93.50% | 0.107 | 8.42 |
| | 20% | 93.60% | 0.106 | 8.68 |
| | 25% | 93.60 % | 0.109 | 9.29 |
| High Density | 0% | 94.00% | 0.100 | 3.77 |
| | 5% | 93.40% | 0.117 | 4.36 |
| | 10% | 94.00% | 0.139 | 4.93 |
| | 15% | 94.00% | 0.149 | 4.90 |
| | 20% | 94.00% | 0.162 | 5.96 |
| | 25% | 94.00% | 0.174 | 6.23 |

*Sinkhole Attack:* The attacker node attempts to divert or attract network traffic towards itself. As shown in Table VI, while the PDR value is high in a network without any attack present, the PDR value decreases when the attackers join the network. However there are several cases which might decrease the impact of the attack. For instance, if the attacker node is already located on a route between the source and destination node, and the target node is physically only one hop away from the attacker which is mostly possible in a high dynamic topology, the impact of the attack might be somewhat restricted. Here, the attack does not alter even the length of the existing active route.

In another scenario, attackers could prompt the establishment of an alternative, albeit longer, route that still provide packet forwarding to the destination. Therefore, the noticeable increase in delay times during this attack highlights that packets are being forwarded along a longer alternative route instead of the shortest available path. Moreover, randomly selected malicious nodes may not attract data packets and it may not have as much impact as thought due to FANETs' dynamic topology. Although PDR may appear unchanged during the ongoing attack, the attack still modifies the network's behavior, leading to indirect impacts. When this attack is initiated initially and combined with another attack, it has the potential to amplify its impact on the network, as demonstrated in consecutive analyses.

With an increase in the attacker ratio, the E2E latency also rises, as previously noted, potentially due to the establishment of longer alternative routes. During network attacks, the presence of invalid active routes and disrupted node connections resulted in the broadcasting of control messages or initiated route discovery mechanisms. Consequently, as the network experienced a higher volume of control packets, the observed overhead increases.

TABLE VII
EFFECTS OF DROPPING ATTACK

| | Attacker ratios | PDR | E2E (s) | OVH |
|---|---|---|---|---|
| Low Density | 0% | 93.70% | 0.084 | 7.49 |
| | 5% | 93.68% | 0.083 | 7.41 |
| | 10% | 93.00% | 0.082 | 7.43 |
| | 15% | 93.00% | 0.080 | 7.43 |
| | 20% | 92.80% | 0.080 | 7.49 |
| | 25% | 92.42% | 0,080 | 7,49 |
| High Density | 0% | 94.00% | 0.100 | 3.77 |
| | 5% | 93.60% | 0.090 | 3.67 |
| | 10% | 93.50% | 0.088 | 3.67 |
| | 15% | 93.40% | 0.087 | 3.63 |
| | 20% | 92.68% | 0.086 | 3.59 |
| | 25% | 93.80% | 0.094 | 3.62 |

*Dropping Attacks:* In this type of attack, the attacker node drops the data packets it received during simulation time. Two possible scenarios are exist for this attack: one involving randomly selected attackers which might not be on any route and another focusing on attackers specifically selected among nodes on the active route. Since the attacker node can only receive data packets if it is located on an active route, we expect that the attack will be more effective in the second scenario. As seen in Table VII, for the first scenario, the attack's impact is limited to certain communication routes, resulting in localized disruption rather than significantly impairing the overall functionality or availability of the entire network. Moreover, an increase in the total number of attacker nodes doesn't necessarily significantly raise the probability of these nodes being located on active routes in a high-density network. Therefore, the impact of an attack with a higher attacker ratio might still be limited on PDR, potentially resulting in slight decreases in latency or minor fluctuations in overhead.

The average metrics after simulating the second scenario are as shown in Table VIII. When a dropping attack occurs on active routes within a network, the consequences for network performance can be profound. The attacker disrupts the data flow, causing a noticeable decline in the PDR by approximately 4% in low-density networks and up to 10% in high-density networks. In high-density networks, as the number of attackers increases, their probability of settling on active routes also increases, consequently leading to the observed lower PDR. Please note that while attackers are initially chosen on active routes, this selection might change throughout the simulation due to mobility. As shown and expected, the position of attacker nodes is highly critical for the impact of this attack. Moreover, failure to transmit packets from the source node to the destination node increased the overhead by increasing the number of control messages in the network. On the other hand, since the number of data packets that reach the destination under attack decreases, E2E decreases.

TABLE VIII
EFFECTS OF DROPPING ATTACK WITH SELECTED ATTACKERS ON ACTIVE ROUTES

| | Attacker ratios | PDR | E2E (s) | OVH |
|---|---|---|---|---|
| Low Density | 0% | 93.70% | 0.084 | 7.49 |
| | 5% | 92.80% | 0.084 | 7.49 |
| | 10% | 92.60% | 0.083 | 7.49 |
| | 15% | 91.70% | 0.078 | 7.53 |
| | 20% | 90.74% | 0.075 | 7.66 |
| | 25% | 89.60% | 0.073 | 7.71 |
| High Density | 0% | 94.00% | 0.100 | 3.77 |
| | 5% | 92.80% | 0.087 | 3.78 |
| | 10% | 91.60% | 0.089 | 3.79 |
| | 15% | 90.10% | 0.089 | 3.84 |
| | 20% | 88.30% | 0.089 | 4.09 |
| | 25% | 84.60% | 0.083 | 4.21 |

TABLE IX
EFFECTS OF BLACKHOLE ATTACK

| | Attacker ratios | PDR | E2E (s) | OVH |
|---|---|---|---|---|
| Low Density | 0% | 93.70% | 0.084 | 7.49 |
| | 5% | 87.00% | 0.073 | 8.34 |
| | 10% | 83.50% | 0.070 | 9.07 |
| | 15% | 81.50% | 0.066 | 9.63 |
| | 20% | 80.70% | 0.061 | 10.14 |
| | 25% | 79.10% | 0.056 | 10.89 |
| High Density | 0% | 94.00% | 0.100 | 3.77 |
| | 5% | 83.50% | 0.117 | 4.37 |
| | 10% | 79.00% | 0.066 | 5.66 |
| | 15% | 78.71% | 0.068 | 6.26 |
| | 20% | 77.00% | 0.073 | 7.18 |
| | 25% | 76.00% | 0.070 | 7.69 |

*Blackhole Attack:* Blackhole attack poses a major threat to the performance and reliability of FANETs, as evidenced by the results presented in Table IX. This composite attack, comprising both sinkhole and dropping attack respectively, deceitfully attract and then drop data packets, causing substantial disruptions in network functionality. Resulting in a sharp decline in essential performance metrics such as PDR, leading to a decrease of up to 15% in low-density networks and approximately 18% in high-density networks.

As the number of attackers increased on the network, the blackhole attack proved more effective than solely applying either the sinkhole or dropping attacks. Furthermore, in networks with a high volume of malicious nodes, increased overhead intensified network congestion and depleted vital resources. The consequent decrease in the number of packets reaching their destinations resulted in an overall reduction in end-to-end efficiency.

*Flooding Attack:* In a flooding attack, a form of DoS attack, the RREQ control packets are incessantly broadcasted, aiming to overwhelm the network by transmitting 10 RREQ messages every 3 seconds. This repeated transmission is intended to exhaust network resources and deliberately induce network congestion, disrupting normal operations and impeding the network's ability to efficiently process legitimate data packets. Legitimate nodes, bombarded with excessive number of RREQ messages, encounter significant decline in the PDR as presented in Table X. The rise in the number of attackers in both high-density and low-density networks correlates with a substantial decrease in the PDR.

TABLE X
EFFECTS OF FLOODING ATTACK

| | Attacker ratios | PDR | E2E (s) | OVH |
|---|---|---|---|---|
| Low Density | 0% | 93.70% | 0.084 | 7.49 |
| | 5% | 92.66% | 0.141 | 15.48 |
| | 10% | 91.40% | 0.134 | 15.94 |
| | 15% | 90.47% | 0.138 | 16.02 |
| | 20% | 86.30% | 0.138 | 16.09 |
| | 25% | 84.45% | 0.139 | 16.27 |
| High Density | 0% | 94.00% | 0.100 | 3.77 |
| | 5% | 92.33% | 0.099 | 4.91 |
| | 10% | 89.00% | 0.183 | 8.14 |
| | 15% | 85.60% | 0.184 | 8.12 |
| | 20% | 76.30% | 0.175 | 8.14 |
| | 25% | 62.70% | 0.172 | 8.11 |

In low-density networks, while the attack has an impact on performance, the effects might be comparatively less severe due to the sparser node distribution. However, flooding attacks exert a pronounced impact on high-density networks, exacerbating congestion and severely compromising network performance, resulting in a reduction of up to 31%. Moreover, the excessive transmission of RREQ messages intensifying increased the network overhead to almost more than double, thereby creating a bottleneck. This bottleneck leads to significant increases in E2E metrics, differentiating it from other attacks and hindering the timely delivery of remaining data across the network.

*Summary of Lessons Learned from Attack Analysis:* The effects of all attacks on PDR are comparatively illustrated in Figure 8. Sinkhole and dropping attacks, conducted by randomly selected attackers, seem to exert minimal impact on network performance. Indeed, in the case of a sinkhole attack, the primary objective is to deceive network nodes by providing false routing information and redirecting traffic through malicious nodes. These attacks aim to mislead rather than directly interfere with data packets, potentially resulting in their impact being less pronounced compared to other attacks. Similarly, dropping attacks attempted by randomly chosen attackers might indeed have a limited impact on network performance due to their positional constraints within the network. On the other hand, deliberate placement of attackers on active routes in dropping attacks allows them to strategically receive and drop packets passing through these active routes. This interference significantly disrupts the transmission process, leading to a reduction in PDR. The contrast between these two scenarios underscores the pivotal role of attackers' placement within a highly dynamic network.

The blackhole attack combines characteristics from sinkhole and dropping attacks, posing a significant threat to network security. This unique combination enables the blackhole attack to profoundly impact network performance, especially evident when the attacker ratio reaches 25%. This scenario notably decreases PDR, signifying the attack's substantial hindrance to successful data transmission within the network.

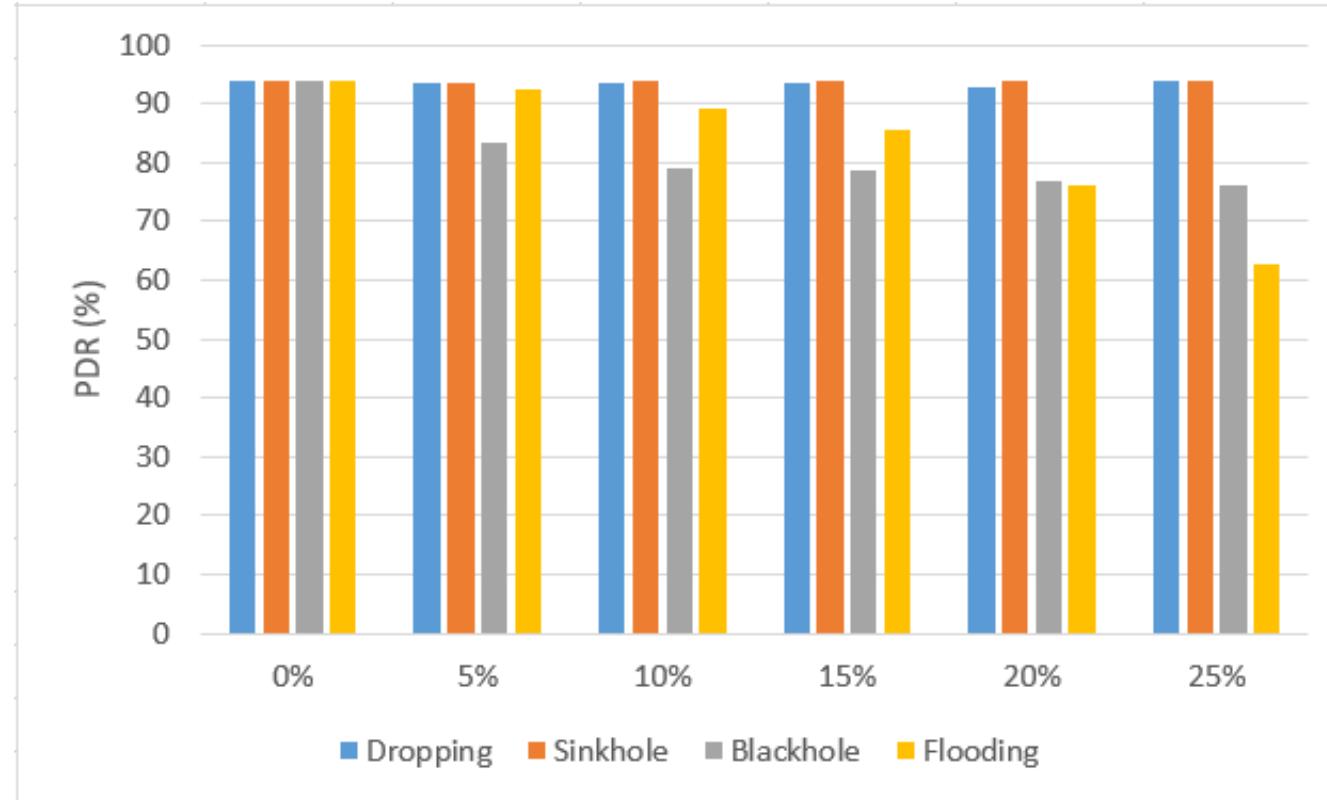

Fig. 8. Comparison of PDR on networks under different attack types

Conversely, flooding attacks create severe traffic congestion by flooding the network with frequent RREQ control messages, leading to dropped data packets before reaching their destinations. Notorious for consuming substantial network resources, flooding attacks prove to be the most impactful, resulting in a staggering reduction in PDR of up to 62%. It is also important to note that the effect of all attacks is directly related to their specific parameters.

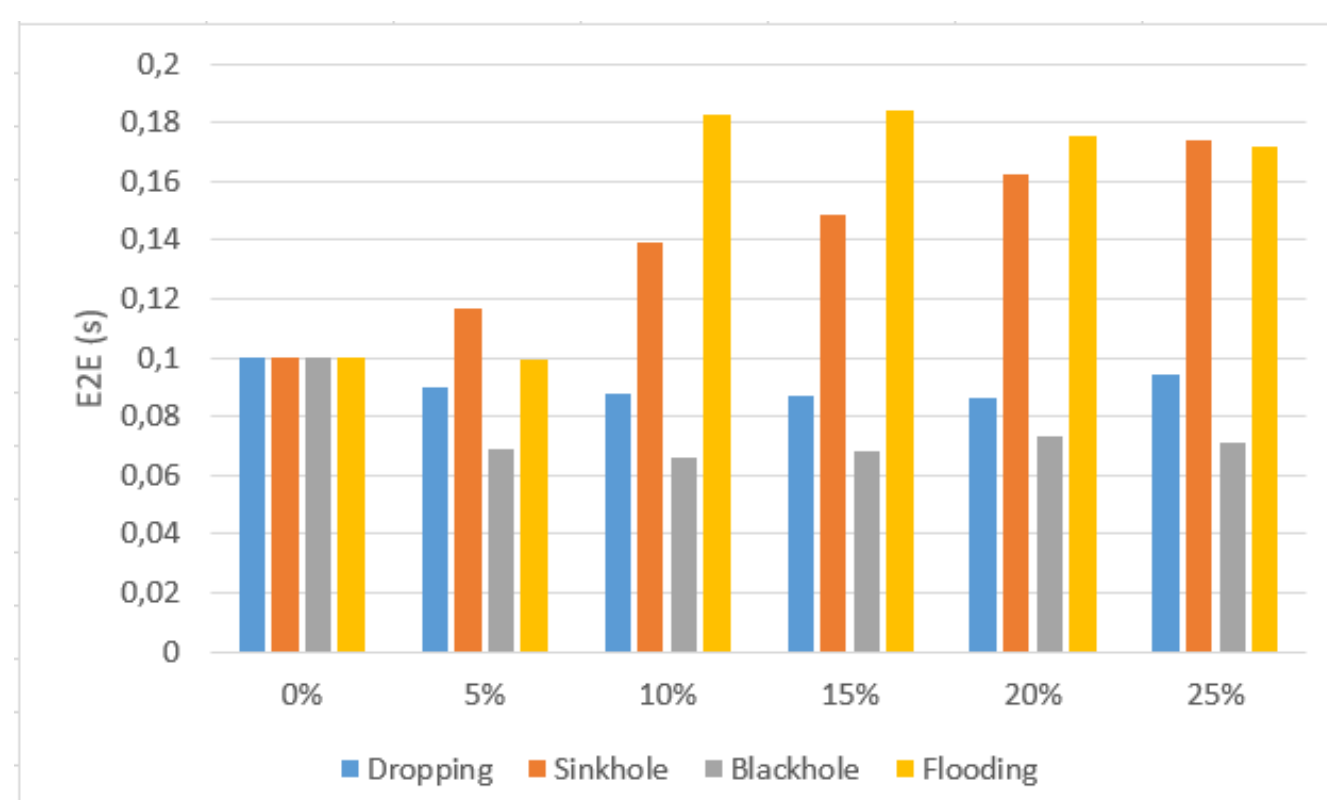

Fig. 9. Comparison of E2E on networks under different attack types

Figure 9 demonstrates a notable increase in the E2E delay when the network is not under attack, compared to instances with active sinkhole and flooding attacks. The sinkhole attack prolonged packet delivery times by rerouting packets through attackers, establishing alternative routes that deviated from the shortest paths. These detours caused delays in package delivery as the alternative routes were less efficient, consequently prolonging the overall delivery time. A distinctly different scenario was observed in flooding attacks, where an escalation in attacker ratio notably increased the volume of control packets. This surge in control packets, alongside an increase in dropped packets, led to traffic congestion, contributing to delays in delivering packets intended for their destinations.

The E2E delay is directly calculated based on the number of successfully delivered packets, hence closely connected with PDR. Consequently, dropping attacks executed by attackers positioned on active routes and blackhole attacks significantly reduced the number of delivered packets, resulting in notably lower the E2E delays. However, dropping attacks employing randomly chosen attackers exhibited a limited effect on delay, causing only slight fluctuations in the E2E delay. The inefficiency of these attacks, particularly in terms of dropping packets, had a negligible impact on the overall packet transmission time, demonstrating their limited effectiveness as an associated factor.

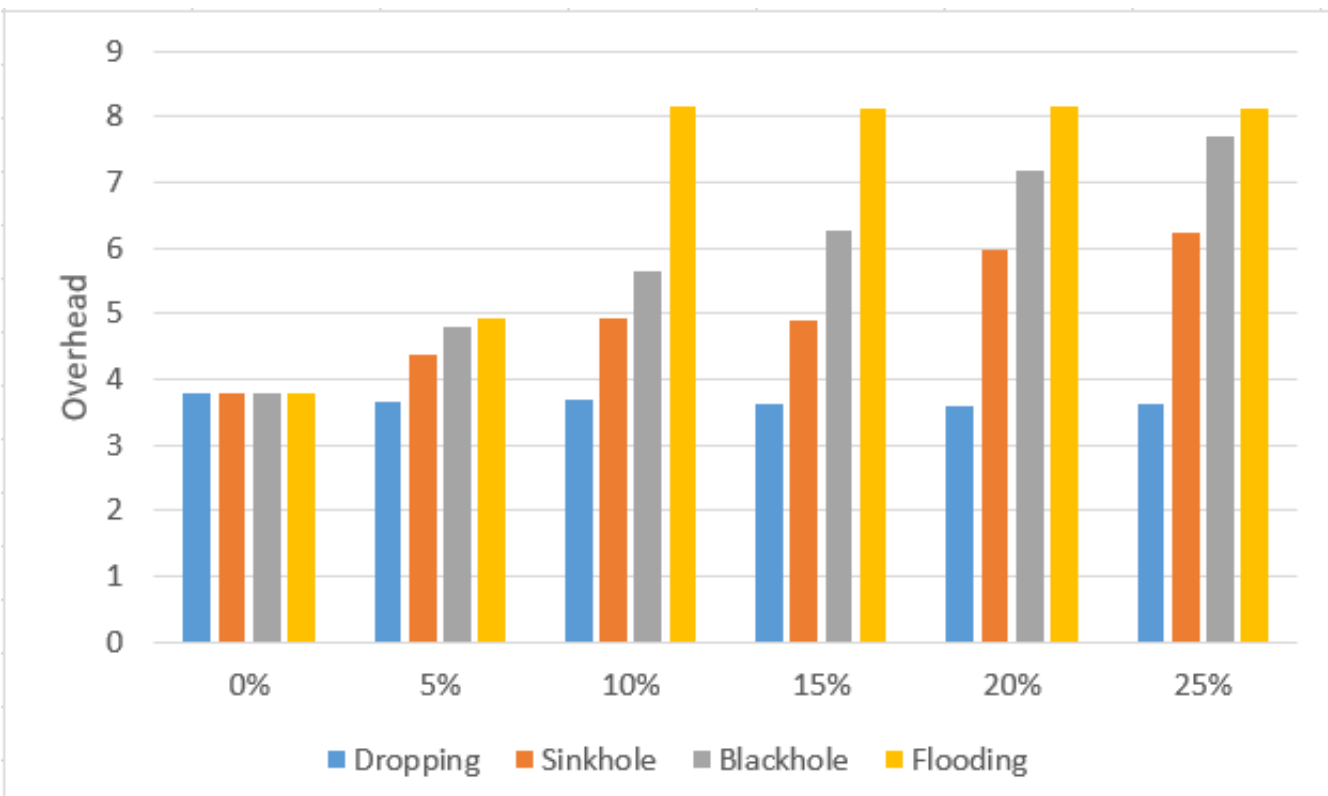

Fig. 10. Comparison of overhead on networks under different attack types

As the number of attackers on the network increased, the overhead also escalated due to the re-initiated route discovery mechanism and error messages, as depicted in Figure 10. This rise was observed as highly significant for all attack types, as anticipated. Notably, in the case of flooding attacks, the overhead was considerably higher compared to other attacks. This was primarily due to the periodic broadcasting of RREQ packets by the malicious node, alongside numerous other control packets, resulting in an excessive volume of network traffic.

To sum up, this study pioneers a comprehensive analysis of attacks in FANETs, using realistic parameters. Simulations with 25 and 50 nodes demonstrated that in high-density networks with attacker ratios of 10% and 20%, PDR dropped below 80% for blackhole and flooding attacks respectively, with relatively lower impact on low-density networks. High density fosters node cooperation, aiding attackers in disrupting the network. Furthermore, as the node intensity decreased in the network, the results of different attack types were seen to converge. Blackhole and flooding attacks in high-density networks reduced PDR by over 15%, with flooding attacks proving notably more effective.

## VII. Countermeasures

The previous section highlighted how attackers can significantly hinder network performance by disrupting routing mechanisms, causing packet loss, and creating congestion, and hence adversely impact UAV missions. Consequently, researchers are actively developing solutions focusing on prevention and detection. We believe this analysis will expedite research efforts aimed at enhancing the security of FANETs.

The first line of defense for UAV systems is referred to as prevention, with the aim being to prevent an attack from entering the system at all. These preventative measures are typically provided through traditional methods such as authentication, encryption, and secure routing protocols. However there is often a trade-off to be managed between the added security and availability of the system being protected. Moreover, insider threats are always a possibility. Therefore, detection, as the other form of countermeasure, aims to monitor the system and to detect anomalies and attacks. This second line of defense is an unavoidable component part of the security structure. In the subsequent subsections, an overview of the countermeasures proposed in the literature for FANETs and UAV networks is presented.

### A. *Prevention*

The solutions presented in this section are categorized in alignment with the attack surface analysis, focusing on two key groups: UAV and Communication Layers.

*1) UAV:* This section is classified under hardware, sensor, and software-level prevention methods, which mirror the layers of security implementation levels within the system.

*Hardware:* In the realm of hardware security, extensive research primarily focuses on combating Hardware Trojans (HTs). Alongside HT-specific strategies, comprehensive studies also explore advanced, generalized measures to protect firmware updates and leverage Physically Unclonable Functions (PUFs) with Field Programmable Gate Arrays (FPGAs) to fortify security against a variety of threats.

HTs denote stealthy alterations within circuits that pose a challenge for detection, holding the capability to induce malfunction in integrated circuits (ICs) [120]. In [121], an HT prevention technique that uses memristor technology integrated with CMOS is proposed to protect integrated circuits during fabrication. This technique mainly hides the functional behaviors and connections of the circuit by obscuring its network IP list through the use of a hybrid design. This methodology, in contrast to conventional methods like split manufacturing [122], eliminates the need for numerous fabrication phases, enabling the production of the whole circuit in unreliable foundries without sacrificing security. This is made easier by memristors, which offer programmable switches that randomize net connections with little effect on the circuit's size, power requirements, or delay limitations.

In [123], a novel strategy aimed at mitigating HT risks by eliminating available routing resources for malicious use is proposed. In the proposed method, following the principal circuit implementation, all residual metal and polysilicon layers are exhaustively employed to construct on-chip magnetic probes. This comprehensive utilization leaves no available space for the integration of Trojan circuits. Furthermore, these on-chip magnetic probes are strategically utilized to capture the magnetic signatures of the chip, which facilitates the detection of any performance anomalies indicative of HT insertions. This approach not only prevents the placement of HTs but also enhances the detectability of subtle manipulations typically overlooked by conventional testing methods.

In [124], a multi-layer security model utilizing Physically Unclonable Functions (PUFs) and Field Programmable Gate Arrays (FPGAs) is introduced to safeguard complex and high-speed transactions among IIoT devices without relying on centralized authentication services. This model leverages the unique, unclonable properties of PUFs to authenticate devices, providing a hardware-based security layer that is nearly impervious to spoofing. By embedding security protocols directly into the hardware, the system efficiently handles high volumes of device communications with minimal latency, crucial for IIoT operations.

A blockchain-based framework for *securely updating firmware* on IoT devices is presented in [125]. Centralized systems pose risks like DoS attacks which can be mitigated by blockchain's decentralized nature, enhancing the security and efficiency of firmware distributions without compromising the privacy and integrity of device operations. The authors propose a PUSH-based update method, where the verification of firmware uses a hash chain to link and authenticate firmware versions via smart contracts, ensuring the integrity and authenticity of each update. This method not only reduces the risks associated with centralized architectures but also handles the high demands of IoT ecosystems effectively, preventing common vulnerabilities in IoT devices through secure, verified updates.

*Sensor:* There have also been proposals [82] [126] aimed at preventing attacks against UAV sensors. In real-world applications, optical flow sensors initially require a feature detection algorithm in order to pinpoint areas of the ground plane image that are especially conducive to tracking. The adversary can simply exploit environmental settings such as covering the flight area to interrupt the vision of the optical flow camera, alter the plausible inputs to affect the sensors input, or create valuable input for the sensor system by utilizing knowledge of the optical flow algorithm.

In [126], a more robust optical flow algorithm was proposed to prevent spoofing attacks on the downward-facing optical flow camera sensors, which provide stabilization to UAVs during flight. In [82], a noise attack against a gyroscope was performed and the possibility of an attacker utilizing deliberate sound noise to damage UAVs with Micro-Electro-Mechanical Systems (MEMS) gyroscopes was explored. Real-world attack tests revealed that in each of the 20 attack test trials, one of the two target UAVs with weak gyroscopes was shown to malfunction, and crash-land soon after the attack had been initiated. Recommended prevention methods include hardware modifications such as physical isolation from the attacker sound noise, differential comparator, and resonance tuning.

*Software:* To the best of our knowledge, while there is no specific research dedicated solely to software security prevention methods in the domain of UAVs, the general principles of cybersecurity still apply and are crucial for prevention to software attacks. Nevertheless, existing studies on software security for IoT devices offer valuable insights that could be adapted to enhance the security of UAV networks.

In [127], a blockchain-based protocol for privacy-preserving software updates for IoT devices is proposed, which also includes a proof-of-delivery mechanism. The protocol uses a smart contract on the blockchain to manage transactions and ensure that updates are securely and reliably delivered without compromising the privacy of the device's users. This is achieved through the use of double authentication preventing signatures (DAPS) and outsourced attribute-based signatures (OABS), which help to minimize the computational load on constrained IoT devices and protect user anonymity. The protocol not only secures the transmission of updates but also provides a financial incentive for transmission nodes that successfully deliver updates to the IoT devices.

The study [128] introduces Secure Asynchronous Remote Attestation (SARA), designed to enhance the security of IoT devices through asynchronous attestation to guarantee the integrity of a software running on devices. This protocol allows IoT devices to verify integrity and operation without pausing their functions, enabling continuous operation and reducing downtime. SARA incorporates selective attestation and uses historical interaction data to maintain device integrity over time, making the process efficient and less resource-intensive.

*2) Communication Layers:* In this section, security measures are categorized to encompass communication layers such as the Physical & MAC Layer, Network Layer, and Application Layer which serve as the umbrella for physical-layer security, secure routing protocols, authentication mechanisms and blockchain-based solutions respectively.

Physical & MAC Layer: Securing UAV communication encounters challenges due to inherent resource limitations, rendering traditional cryptography impractical. Employing Physical-Layer Security (PLS) in UAVs offers secure information-theoretic transmissions [129] with minimal computational complexity, addressing energy, computational, and memory constraints [130].

In [23], prevention methods for PLS are extensively discussed. For instance, the *Noise-Aided PHY Security* method intentionally degrades an eavesdropper's channel by introducing fabricated noise into information transmissions [131]. Its objectives include reducing the Secrecy Outage Probability (SOP), increasing system throughput, and enhancing ergodic secrecy capacity [132] to bolster secrecy functionality.

Another approach, *Cooperative Jamming-Aided PHY Security* [133] [134], involves a UAV transmitting both data and a jamming signal to deter eavesdroppers. Self-interference cancellation helps the intended receiver filter out the jamming noise. Objectives here include minimizing SOP [135], improving system throughput, and enhancing ergodic secrecy capacity. A proposal for the secure communication between GBS and UAVs in the presence of multiple eavesdroppers within mmWave relaying networks is discussed in [136]. Two relay schemes, with and without jamming, are proposed: the first scheme examines secrecy performance without jamming by splitting transmissions into two time slots, while the second scheme involves cooperative jamming using both the destination and a jamming UAV to thwart eavesdropping in both time slots.

Incorporating LoS links and utilizing multiple UAVs strategically for cooperative jamming [137] necessitates meticulous trajectory planning to prevent collisions. The *Legitimate Eavesdropping Aided PHY-Security* approach leverages the receiver's null space to interfere with the eavesdropper's link. Here, a UAV mimicking an attacker emits jamming signals to disrupt dubious users. Authorized receivers utilize self-interference cancellation to filter unwanted signals. However there are some issues to be considered [138] [139] such as aerial to ground or aerial to aerial channel planning, precise Channel State Information estimation, and cooperative tactics using UAVs for enhanced covertness and covert data transfer.

The recent advancements in *Intelligent Reflecting Surfaces (IRS)* [140] have introduced new opportunities for PLS due to its advantages including low power requirements, cost-effectiveness, and versatile deployment options to enhance spectrum efficiency and secrecy performance in the 5G and beyond communication systems. IRS are advanced structures with many controllable units that modify electromagnetic waves for directed signal enhancement or suppression, serving as a promising tool for boosting wireless network performance and security efficiently with minimal energy use.

In [141], the concept of dynamically adjusting the IRS's surface elements is introduced. The approach counteracts interference from smart jammers, preserving the integrity and quality of legitimate transmissions. The study presents a strategic formulation of an optimization problem aimed at bolstering communication resilience against jamming, through the simultaneous optimization of power distribution at the base station and beamforming strategies at the IRS. This addresses the volatility and unpredictability associated with jamming activities. To adeptly manage the intricacies associated with undetermined jamming behaviors, it introduces an innovative learning mechanism based on fuzzy win or learn fast-policy hill-climbing (WoLFCPHC). Simulation results show that this approach significantly improves both the IRS-assisted system's rate and the security of transmissions. It effectively strengthens PLS, proving to be highly effective against complex jamming attacks.

Authors introduces a frequency hopping strategy based on a federated deep Q-network (DQN) in [142], which combines federated learning with deep reinforcement learning techniques enabling UAVs to adaptively learn and update their frequency hopping strategies without compromising data privacy among the network. Every UAV client in the network gathers channel data from its surroundings and uses it to train a local DQN model. The Multi-access Edge Computing (MEC) server receives the local model weights from the UAVs on a regular basis and aggregates them using a federated averaging approach. After then, the UAVs receive this combined global model weight, which helps them choose the best communication channels and avoid jamming.

In [143], the study explores the potential of countering remotely controlled UAVs through RF jamming. It focuses on commercially available UAVs that often use proprietary protocols. These UAVs operate across varied frequency bands, which poses challenges for detection and neutralization. To address this, the authors propose a flexible jamming technique called *protocol-aware jamming*. This technique is tailored to the characteristics of the targeted communication system to optimize power efficiency and minimize the impact on other communication systems. A key contribution of this study is the development of a software-defined radio (SDR) implementation that can adapt in real-time to different UAV systems' communication protocols and RF parameters. The proposed jammer demonstrates superior efficiency and reduced interference with WLANs compared to less adaptable alternatives.

In [144], a secure transmission scheme for wiretap channels, addressing scenarios where a source communicates with a legitimate UAV amidst eavesdropping activity is proposed. A full-duplex active eavesdropper engages in both eavesdropping and malicious jamming concurrently. To counter this threat, the source employs artificial noise (AN) signals to confuse the eavesdropper. Utilizing a ground-to-UAV channel model, the hybrid outage probability is analyzed, considering both transmission outage and secrecy outage probabilities. Optimal power allocation between information signals and AN signals is determined, alongside identifying the optimal operating height of the UAV to minimize the hybrid outage probability. This paper offers a framework for designing confidential UAV communication systems.

To address the challenge of secure remote control in multi-UAV systems, [145] propose a PLS-based solution, incorporating cooperative jamming. The method leverages the Nakagami-m fading model [146] and a mixed LoS/NLoS probability model to assess the likelihood of wireless attacks such as eavesdropping, jamming, and spoofing. This hybrid approach quantifies the risks of multiple attack types, using SOP to measure cooperative jamming's effectiveness. The authors design three levels of control schemes-centralized, partially centralized, and distributed-to balance energy efficiency and security needs, applying them to both UAVs and GCSs. Simulation results validate the approach's ability to defend against various attacks and secure communications, offering flexibility in adapting to different operational scenarios and security demands.

By utilizing effective channel modeling and the spatial correlation characteristics of channel sparsity, Teng et al. [147] present a PHY-layer authentication approach tailored for transmitter authentication in mmWave MIMO (Multiple-Input Multiple-Output) UAV-ground systems. This scheme is specifically designed to counter identity-based impersonation attacks. Utilizing a Laplace prior approach, the authors accurately characterize the statistical model of angular-domain mmWave MIMO channels, revealing their spatial correlation properties. The framework employs joint feature extraction via Expectation Maximization and Generalized Approximate Message Passing algorithms, combined with hypothesis testing, to effectively improve authentication performance. Theoretical performance analysis is conducted, providing closed-form expressions for false alarm and detection probabilities. Simulations confirm the proposed model's feasibility and the framework's resilience against identity-based impersonation attacks, especially showcasing the benefits of channel sparsity in high SNR and large antenna array settings.

Network Layer: While the proposed routing protocols for UAVs aim to address the inherent challenges posed by their high mobility, dynamic topologies, and limited resources, they are typically not designed with security as a primary consideration [55]. Therefore many solutions in this layer focus on development of secure routing protocols.

In [148], a secure communication architecture for UAV Ad hoc Networks (UAANETs) is proposed in [148]. This secure routing protocol aims to provide authentication, confidentiality, and integrity. The protocol minimizes signaling overhead to optimize resource usage. Building upon this research, another study [149] introduced the SUAP (Secure UAV Ad hoc routing Protocol) to provide a tailored solution for UAANETs. SUAP integrates AODV-based routing with asymmetric cryptography and hash chain mechanisms for message authentication and integrity. Additionally, it includes measures to detect and prevent wormhole attacks. Experimental evaluations include hybrid simulation and real-world tests with UAVs and Ground Control Stations (GCS).

FANETs demand robust algorithms capable of managing node mobility and ensuring secure data transmission without compromising Quality of Service (QoS) metrics. In this context, a secure routing protocol designed for FANETs is SecRIP [150], which employs chaotic algae and dragonfly algorithms to assign Cluster Heads (CHs). CHs have responsibilities that include intra-cluster and inter-cluster routing, as well as aggregating information about other nodes within the cluster. The involvement of CHs also aids in reducing energy consumption during data transmission. Additionally, SecRIP incorporates encryption mechanisms to bolster data security, specifically between CHs. However, the use of meta-heuristic algorithms may increase delay and communication overhead, impacting network performance, particularly in dynamic FANET environments with fast-moving UAVs. Moreover, SecRIP's implementation in a two-dimensional environment limits its compatibility with FANETs

Another secure routing protocol, a new protocol called SEEDRP [151] introduces optimization techniques aimed at enhancing network efficiency. By strategically selecting neighbor nodes for data forwarding and mitigating control packet flooding, SEEDRP proves effective in scenarios characterized by high mobility and network density. Moreover, when compared to SecRIP, SEEDRP exhibits enhanced performance in factors such as end-to-end delay and packet delivery ratio.

Hosseinzadeh et al. [152] introduces a novel Q-learning-based secure routing scheme (QSR) for FANETs, designed to defend against wormhole attacks, including those involving encapsulation and packet relay. The system is divided into two phases: secure neighbor discovery and Q-learning-based routing. In the first phase, UAVs securely identify neighboring UAVs, with a local monitoring system in place to detect and mitigate packet relay wormhole attacks. This system monitors data exchanges and enforces rules to isolate suspicious nodes. In the second phase, a distributed Q-learning algorithm helps UAVs determine the most secure routes, focusing on preventing encapsulation-based attacks. A reward function evaluates factors like one-hop delay, hop count, data loss, packet transmission frequency, and packet reception frequency to choose optimal paths. Simulation results show that QSR outperforms methods like TOPCM [153], MNRiRIP [153], and MNDA [154] in terms of accuracy, malicious node detection, and data delivery, albeit with slightly higher latency than TOPCM.

In response to the security challenges of large-scale UAV communication, [155] present RLPC-SIT, a reinforcement learning and location confusion-based protocol that addresses threats such as eavesdropping, data tampering, replay attacks, and MITM. The approach uses reinforcement learning to establish optimal and stable communication paths between UAVs by calculating Q-values based on parameters like sensitivity, stability, and distance. In parallel, the location confusion method ensures that UAV positions are masked, preventing malicious actors from accurately identifying their locations. Additionally, the system incorporates an encryption mechanism for message authentication, which encrypts the transmitted data at each hop to safeguard it against unauthorized access or tampering. Simulation results show that RLPC-SIT not only improves data transmission speed and stability but also outperforms existing protocols like AODV, PBQR and GPSR [156] in terms of data delivery success and traceability, while maintaining robust protection of location privacy.

Application Layer: This section reflects the diverse strategies employed to reinforce the integrity and security of communication protocols within the application layer. This includes traditional authentication mechanisms alongside emerging blockchain-based approaches.

*Authentication Mechanisms:* The authentication protocol stands as a fundamental security measure within distributed systems, aiming to uphold the integrity and trustworthiness of nodes during communication. A recent study [157] suggests that securing UAV communication requires implementing encryption and authentication mechanisms for protocols such as MAVLink [158], which facilitates the communication between the UAV and the GBS, as well as the inter-communication among systems on board the aircraft. These protocols are often left vulnerable in standard configurations. The research demonstrates that secure key exchange protocols and message authentication codes could reduce the risk of interception and modification of UAV control data.

In another study [159], a lightweight mutual authentication mechanism is presented, aiming to ensure secure communication between UAVs and the base station. The fundamental principle of the proposed mechanism is that UAVs and GBS employ a challenge-response combination of a physical unclonable function as the initial condition of a chaotic system. This system is used to randomly mix the message, which carries a seed to generate a secret session key. The simulation consisted of three nodes (UAV, server, and user) and was conducted using OMNeT++ [160]. The results showed that the mechanism outperformed the well-known cryptographic proposal [161] in terms of computational cost, communication overhead, and energy consumption.

Mallikarachchi et al. [162] delves into the condensation of data frame payloads through compression and data hiding, aiming to authenticate the payload of data frames received by nodes in a FANET. The primary objective is to verify the integrity of each received packet along the communication path. The key contribution of this research lies in the design of a payload authentication scheme that combines masking (XOR), a lossless compression technique, and data hiding. In order to ensure the lightweight and energy-efficient nature of the scheme, a straightforward XOR operation is employed, coupled with image generation and JBIG2 compression to generate the bitstream embedded into the CP. Upon reception, the hidden data is extracted and decoded, allowing for a comparison against the payload of the received data frame. The evaluation of the proposed scheme, in terms of bit error rate, revealed BERs of less than $0.7 \times 10^{-4}$, aided by the implementation of a 7-bit Hamming code. Furthermore, experiment results affirm the proposed scheme's capability to localize tampered data frames. The validation of the scheme is conducted using a simulated FANET model implemented in MATLAB 2018b, featuring a 3D Random Waypoint (RWP) mobility model with the AODV routing protocol.

In [163] the design of a robust and lightweight authentication and key agreement scheme for cloud-assisted unmanned aerial vehicles using blockchain in FANET (LAKA-UAV) is introduced. The primary aim in this study is to ensure integrity and decentralization functionalities for data sharing for cloud-assisted UAVs in FANETs. LAKA-UAV leverages cloud technology to attain ample storage resources and computing capabilities. Within each block, only metadata is stored to enhance block construction and minimize distributed storage waste. Additionally, LAKA-UAV employs blockchain technology, specifically Hyperledger Fabric, to guarantee efficient access control, data integrity, and decentralization through log transactions. Through testbed experiments and blockchain implementation, LAKA-UAV demonstrates efficient computation cost and a high-security level. While LAKA-UAV incurs a higher communication cost than comparable schemes, it ensures lightweight computation and storage costs, along with superior security features compared to existing schemes.

Wu et al. [164] proposed an improved version of the three-party authentication protocol given in [165] in order to protect an Internet-of-drones (IoD) environment from known security threats. The improved protocol consisted of three phases: drone registration, user registration, and login authentication. While an adversary could collect stored data from the server, intercept the messages in the public channel, and draw out data from a captured UAV, as shown in [165], the improved version introduced in [164] addressed these vulnerabilities. In [166], an improved Temporal Credential-based Anonymous Lightweight Authentication Scheme (iTCALAS) is proposed for the Internet of Drones (IoD), with the aim of mitigating traceability and preventing stolen verifier attacks. By employing lightweight symmetric hash functions and temporal credentials, it ensures compatibility with resource-constrained drones.

In [167], PRLAP-IoD is proposed as a lightweight authentication protocol designed for UAV-to-UAV and UAV-gateway communication in IoD environments. It integrates PUF technology to ensure secure and efficient authentication while minimizing memory and computational costs. The protocol uses a user-gateway server framework, ensuring session privacy by preventing third-party involvement in session key generation. Formal validation using the AVISPA [168] tool and the ROR oracle model, along with informal security analysis, demonstrates its resilience against attacks such as replay, impersonation, session key disclosure, desynchronization, privileged insider attack, and MITM, offering a balance between security and efficiency compared to existing authentication schemes.

In [169], PARTH, a PUF-based authentication protocol tailored for Software-Defined UAV Networks (SDUAVNs) is proposed. It generates secret keys in real-time using PUFs, thereby removing the need to store secret keys in the physical memory of UAVs. PARTH achieves mutual authentication across multiple layers of entities, ensuring identity protection, message integrity, and data confidentiality. Furthermore, it addresses security concerns such as drone tampering and various attack scenarios. Formal security proofs and comparisons with state-of-the-art protocols are provided to validate PARTH's effectiveness.

In [170], a lightweight digital signature protocol is proposed in order to prevent MITM attacks, in which malicious nodes eavesdrop on communications between UAVs and GBS by posing as the GBS and sending falsified commands in order to jeopardize a UAV mission. The chaotic complex system employed by the GBS in the proposed protocol allows it to construct a digital signature based on the command message, which it then appends to the command message. The UAV then verifies the digital signature prior to executing the command it received by comparing it to the digital signature produced from the command message itself. If the verification of the digital signature is not proven, the request is instantly denied, and the Return-to-Launch (RTL) mode is initiated, forcing the UAV to return to its takeoff position.

In [171], an improved and secure access control system that utilized certificates initially registered by a trusted authority was presented for IoT-enabled drone environments. The mechanism provided key agreement and mutual authenticity among drones, and also between drones and the ground station. The proposed system completed the access control procedure by exchanging just two messages. The proposal was claimed to be robust to known threats such as MITM attacks, physical attacks, forgery attacks, privileged insider attacks, replay attacks, and session-specific temporary information threats.

*Blockchain-based Solutions:* Another point to consider is to secure drone communication during data collection and transmission while preserving the integrity of collected data. Blockchain is considered as a solution to data integrity and privacy problem ensuring a safe and trusted communication and implemented to ad-hoc networks [172]–[176]. Blockchain uses cryptographic algorithms to secure transactions and prevent unauthorized access. This can help to protect sensitive data in FANETs, such as mission-critical information or personal data of individuals involved in the operation. Through blockchain, transparent record of all transactions on the network can help to increase accountability and trust among participants in FANETs.

The use of blockchain technology, coupled with private key cryptography, to enhance security is proposed in [176].

In the study, the benefits of blockchain are emphasized, including its distributed nature and immutability, which offer a safe way for controllers and drones to communicate. In order to increase security, timestamping and GPS are used in conjunction with data encryption between the UAV and control panel and data hashing for the cloud. In addition to UAVs, the study proposes a decentralized blockchain-based solution with potential applications across other industries.

In [177], a novel blockchain-based secure data delivery and collection scheme is proposed for maintaining data integrity and confidentiality within the IoD networks. The proposed scheme aims to provide access control between the drones and their respective GBS in each flying zone, establish session keys for secure communication, and record all transactions among the drones, GBSs, and Control Rooms (CR) to form private blocks. A consensus-based algorithm facilitates transaction block verification within the network. Security analysis, formal verification, and performance evaluation demonstrate the method's robustness and suitability for resource-constrained drone environments.

Another blockchain-based distributed key management scheme for secure communication in FANET is proposed in [178]. Drones can autonomously manage keys without centralized authority, leveraging blockchain for tamper-resistant and traceable transactions. The scheme is designed for a heterogeneous FANET, which typically comprises drones with different capabilities, including computation and battery power. CHs have a central role in key management within the network, being tasked with generating, distributing, updating, and revoking cryptographic keys essential for securing communication and data transmission among drones. Notably, CHs oversee the management of the blockchain infrastructure, leading to reduced costs and enhanced scalability of the system.

In [179], a blockchain-based framework for secure collaborative computing in UAANETs is proposed. The framework incorporates an upgraded Practical Byzantine Fault Tolerance (PBFT) consensus algorithm, which reduces overhead by employing trust evaluation among UAVs. Additionally, a smart contract-driven task allocation mechanism is also developed to improve task efficiency and offloading security. Simulations demonstrate a 47% reduction in consensus latency, a 76% decrease in message count, and a 48% reduction in task costs. The effectiveness of the framework in supporting secure and efficient collaboration within UAANETs is further confirmed through real-world UAV swarm experiments.

*3) Summary of Lessons Learned from Prevention Studies:* The proposed prevention approaches are summarized in Table XI. In conclusion, it is imperative to design systems equipped with defenses to thwart attacks that could potentially harm UAVs and FANETs. Attack prevention often involves enhancing UAV or FANET components to render attacks impractical or dysfunctional, as well as implementing effective preventative measures capable of withstanding potential threats.

In the domain of UAV and FANET security, preventive measures at the hardware level are often less specific and somewhat fragmented, focusing predominantly on the physical components themselves rather than holistic hardware security strategies. The majority of existing research tends to apply broad software security solutions, which do not fully address the real-time operational demands and highly dynamic network topologies of UAVs. While the current preventive measures for UAV hardware and software provide a baseline level of security, they fall short in offering comprehensive protection against more advanced and sophisticated threats.

A secure UAV architecture framework should be proposed, seamlessly integrating hardware, sensor, and software-level security measures to ensure robust protection against emerging and dynamic threats. This comprehensive approach should encompass the protection of the physical components of UAVs through improved hardware security mechanisms tailored to UAV specifications, securing sensor-level vulnerabilities, and implementing robust software-level defenses to proactively mitigate potential attacks. Additionally, attention should be given to the software implementation process and the update life cycle to ensure that security measures are continuously maintained and updated to address evolving threats.

In FANETs, routing protocols are essential; however, many of these protocols were originally designed with insufficient security consideration. Additionally, several studied protocols for FANETs are primarily designed for ad hoc networks in general, without specifically addressing FANETs' unique requirements. Limited research has been conducted on the security aspects of FANET routing protocols, posing a considerable danger that attackers could potentially take control of the network or interfere with its normal operations due to the vulnerability of these protocols to attacks.

Additionally, it is crucial to delve into privacy and integrity concerns associated with UAV-collected data in the light of addressed FANET-UAV challenges. Adopting cryptographic techniques for UAV systems or FANETs, either for authentication, access control, privacy, confidentiality or trust establishment, require some form of trade-off in terms of computational cost and energy limitations, network bandwidth consumption, and potential latency on the chip. Leveraging blockchain-based solutions within FANETs can strengthen prevention systems through the establishment of transparent and tamper-proof records for transactions and communications.

UAVs have become a crucial component of the communication networks that connect cellular clusters to various infrastructures including IoT and VANET. UAVs have the potential to function as aerial ground stations, relay nodes, and infrastructure in remote regions within the realm of wireless communication systems. Despite considerable research progress, numerous challenges persist on physical layer security such as data interception and jamming which pose more formidable challenges than conventional terrestrial eavesdropping. Addressing these issues requires innovative approaches like Noise-Aided and Cooperative Jamming-Aided PHY Security, and the deployment of advanced technologies like Intelligent Reflecting Surfaces. However, adapting these approaches from theory to practice poses significant challenges. The air-to-ground (A2G) and air-to-air (A2A) communication channels differ significantly from terrestrial channels and require specific adaptations. These channels are affected by the 3D arrangement of aerial nodes and the unpredictable movements of UAVs, which can complicate security measures and may also provide advantages to attacks in this layer. Therefore, future efforts can focus refining these techniques and overcoming practical deployment challenges to ensure secure and reliable UAV communications. Additionally, advanced techniques within the realm of physical layer security could be explored. To sum up, exploring distributed and collaborative prevention mechanisms within the FANET framework stands out as critical research areas within the prevention paradigm. Last but not the least, evaluations of the proposals are very limited as shown in Table XI.

### *B. Detection*

Prevention techniques are effective against known attacks; however, they may not always prevent new types of attacks or insider threats. Therefore, detection systems are essential complements, aiming to identify attacks that evade existing prevention mechanisms. Intrusion detection systems (IDSs) play a crucial role in detecting threats before compromising a system's integrity, confidentiality, or availability, striving to minimize inflicted damage. This section categorizes proposed studies on intrusion detection based on their detection methods, encompassing signature-based, anomaly-based, and specification-based approaches.

*1) Signature-based IDSs:* These systems rely on employing signatures, rules, or patterns that define known attacks. While highly effective and efficient against known threats, they are primarily favored in commercial systems. However, their limitation lies in their inability to detect novel, unknown attacks or newly evolved variations of known attack patterns. Additionally, these systems require regular updates to their signature databases to remain effective.

A rule-based study inspired from the human immune system (HIS) is given in [187]. The study consists of three phases. In phase 1, safe routes between source and destination are identified by using consecutive *Hello* packets. These selected secure routes progress to Stage 2. In Phase 2, they use reverse test packets sent from the destination to the source to find potential malicious UAVs among the intermediary nodes. This method relies on spotting differences in the packets for detection. Routes free from flagged malicious nodes move to Phase 3 for thorough robustness checks. This phase assesses hop count, Round Trip Time (RTT), and Signal Strength Intensity (SSI) as evaluation criteria. Routes with fewer hops, shorter RTT, and stronger signal take precedence for safety. The proposed approach effectively identifies blackhole, sinkhole, wormhole, and fake information dissemination attacks. However, the increased communication between source and destination endpoints might lead to added overhead. The authors expand on their study by introducing decision-making agent defense agents in [188]. The UAV designated as the defense agent replicates itself near UAVs on suspicious routes and transmits test packets. Another study that use Hello packets to identify secure routes is given in [189].

A recent study [190] uses both a rule-based approach and a mobile agent-based negotiation process. In the initial phase, the system employs specific rules and principles, analyzing various aspects such as node behavior, data transmission patterns, route response messages, sequence numbers, and hop counts to detect potentially malicious UAVs. This phase also involves the investigation of node activities and interactions among neighboring nodes to identify anomalies or suspicious behavior. In the second phase, certain designated nodes function as agents, selected randomly to facilitate data transmission between source and destination UAVs. These agents play a crucial role in discovering neighbour UAVs within a one-hop distance. Furthermore, they employ a hash function to secure information from potential adversarial UAVs within the communication network. To ensure security, these agents generate digital signatures for any information exchanged between source and destination UAVs. The system demonstrates heightened residual energy and packet delivery ratio while maintaining lower levels of false positives.

A traditional approach for detecting jamming attacks was proposed by [191], in which a detection framework monitors the signal power density of each device, compares it with the signal strength of that device, and executes intrusion detection and prevention mechanisms. Another study [192] presented a novel lightweight distributed rule-based detection approach called Lids in order to detect flooding attacks. Lids limits the packets drones can send within set time frames, verified by shared transmission data among drones. This method efficiently prevents flooding attacks by swiftly neutralizing adversary drones. But it poses challenges of potential network congestion and increased drone energy consumption, notably impacting small-sized UAVs.

*2) Anomaly-based IDSs:* The system creates a profile of normal behavior and flags any activities deviating from this pattern as anomalies, potentially indicating an attack. These methods are good at detecting new attack types. However, defining what constitutes normal behavior poses a challenge, especially as it can evolve over time. Consequently, this approach might generate a significant number of false positives. Anomaly detection employs various techniques, including statistical-based and machine learning-based approaches.

*Traditional Machine Learning-based Studies:* A recent study [193] proposes a machine learning-based (ML) approach to detect various attacks like hijacking, GPS signal jamming, and DoS attacks targeting drones in smart cities. The study applies several classification algorithms, including Support Vector Machines (SVM), Naive Bayes (NB), Linear Regression (LR), and Random Forest (RF), using data from the DJI Phantom 4 drone dataset [194]. This dataset comprises GPS, gyro, and power data collected from a single drone. While the aim is to bolster drone system security against potential threats, forming a comprehensive security strategy solely based on data from a single drone presents challenges.

In [195], a study on detecting GPS spoofing attacks is presented. The study introduces the UAV attack dataset, utilizing logs that collect sensor data including GPS, accelerometers, and gyroscopes during flight. However, it is limited by the use of only a few drones and a restricted sensor set. Similarly, in [195], the authors introduced an anomaly-based approach for detecting GPS spoofing attack. This approach allows for the use of logs that collect sensor data, such as GPS data, accelerometers, and gyroscopes during flight, to create a dataset for training. The results of the approach show that it achieves high F1 scores of 99.56% and 99.73% for benign and malicious sensor readings, respectively, indicating its effectiveness in detecting GPS spoofing attacks.

TABLE XI
OUTLINE OF THE PREVENTION SOLUTIONS ON UAVS' SECURITY

| Reference | Year | Classification | Method | Attacks | Simulations |
|---|---|---|---|---|---|
| [125] | 2019 | Hardware | PUSH-based method with blockchain | DoS | No sim. |
| [124] | 2020 | Hardware | Multi-layer security model PUFs and FPGAs | Hardware trojan | Python/SimPy |
| [123] | 2021 | Hardware | A new tamper resilient solution | Hardware trojan | High-frequency structural simulator |
| [121] | 2023 | Hardware | New method utilizing memristor technology | Hardware trojan | ISCAS 85 benchmark circuit C432 27-channel interrupt controller. |
| [82] | 2015 | Sensor | Physical isolation differential comparator resonance tuning | Sensor attacks MEMS gyroscopes | Multiwii DIY drones |
| [126] | 2016 | Sensor | Novel algorithm proposed | Sensor input spoofing | No sim., Burrows-Abadi-Needham logic analysis [180] |
| [127] | 2019 | Software | Blockchain | Chosen message selective-attribute | Smart contact in solidity |
| [128] | 2020 | Software | Novel secure asynchronous remote attestation protocol | Replay software adversary mobile adversary | Contiki |
| [181] | 2016 | Phy. & MAC Layer | Watchdog timer | DoS, buffer overflow, ARP cache poisoning | Parrot Bebop Drone |
| [143] | 2018 | Phy. & MAC Layer | SDN | Jamming | Simulink |
| [144] | 2019 | Phy. & MAC Layer | Artificial noise | Eavesdropping | Numerical result |
| [131] | 2020 | Phy. & MAC Layer | Artificial noise | Jamming | Numerical result |
| [140] | 2020 | Phy. & MAC Layer | IRS | Jamming | Numerical result |
| [141] | 2021 | Phy. & MAC Layer | IRS | Jamming | Base station single-antennas |
| [136] | 2021 | Phy. & MAC Layer | Time slots based transmissions | Eavesdropping | Numerical result |
| [142] | 2021 | Phy. & MAC Layer | Federated deep Q-network | Jamming | Tensorflow |
| [135] | 2022 | Phy. & MAC Layer | Power splitting based secure non-orthogonal transmission | Eavesdropping | Numerical result |
| [145] | 2024 | Phy. & MAC Layer | Cooperative-jamming-based secure remote control mechanism | Jamming, spoofing, passive/active eavesdropping | Numerical result |
| [147] | 2024 | Phy. & MAC Layer | A novel PHY-layer auth framework | Identity-based impersonation attacks | Numerical result |
| [148] | 2015 | Network Layer | Certification compliance via model driven development | Warmhole | MATLAB Simulink |
| [149] | 2019 | Network Layer | Asymmetric cryptography and hash chain mechanisms | Warmhole | OMNET++, Delair-Tech DT18 drones. |
| [150] | 2021 | Network Layer | Chaotic algorithm, dragonfly algorithms, encryption mechanisms | Attacks targeting data integrity and confidentiality | NS2 |
| [151] | 2021 | Network Layer | Distinct dynamic key generation | Attacks targeting data integrity and confidentiality | NS3 |
| [152] | 2024 | Network Layer | Q-learning-based secure routing | Wormhole attack | NS2 |
| [155] | 2024 | Network Layer | RL-based secure data transmission | Eavesdropping, replay, MITM, data tampering | MOP and EPPS schemes [182] |
| [176] | 2019 | Application Layer | Blockchain-based solution | Jamming | No Sim. |
| [159] | 2020 | Application Layer | Mutual auth. protocol | Cloning, MITM, replay, tampering | OMNeT++ |
| [177] | 2020 | Application Layer | Blockchain-based solution | Replay, MITM, impersonation, privileged insider , physical drone capture, Ephemeral Secret Leakage (ESL), | AVISPA [168] |
| [166] | 2020 | Application Layer | Auth. scheme | Stolen verifier attacks | ProVerif [183] |
| [178] | 2020 | Application Layer | Distributed key management scheme | Eavesdropping, tampering, replay, impersonation, hijack, cloning | No Sim. |
| [169] | 2020 | Application Layer | Lightweight mutual auth. protocol | MITM, replay impersonation | Mao and boyd logic [184] |
| [170] | 2020 | Application Layer | Digital signature protocol | MITM | Customized sim. framework |
| [171] | 2020 | Application Layer | GCACS-IoD | Drone impersonation, session key disclosure, physical capture | No sim., ROR model analysis [185] |
| [164] | 2022 | Application Layer | Enhanced auth. protocol | Privileged insider, MITM, replay, physical capture | No sim., ROR model analysis [185] |
| [165] | 2022 | Application Layer | Mutual auth. protocol | Replay, impersonation, brute force, DoS , MITM | No sim., Burrows-Abadi-Needham logic analysis [180] |
| [162] | 2023 | Application Layer | Auth. scheme | Eavesdropping, MITM | MATLAB 2018b, bit error rate analysis |
| [163] | 2023 | Application Layer | Lightweight auth. and key agreement scheme | Replay, impersonation, session key disclosure, desynchronization, Ephemeral secret leakage (ESL), Off-line password guessing, MITM | ROR oracle model [186], AVISPA [168] |
| [167] | 2024 | Application Layer | Lightweight auth. and key agreement scheme | Replay, impersonation, session key disclosure, desynchronization, privileged insider attack, MITM | ROR oracle model [186], AVISPA [168] |
| [179] | 2024 | Application Layer | Blockchain-based solution | False task attack, active malicious nodes | AerialBC [179] |

Another recent study [196] employed ML to detect DoS attacks within UAV networks. Utilizing the AWID2 dataset [197], the study implemented gradient boosting techniques such as XGBoost, CatBoost, and LightGBM to train a model. Real drone testing followed the model's training. Notably, LightGBM demonstrated superior performance in Area Under the Curve (AUC) metrics and training time among these algorithms. However, the AWID2 dataset, derived from a typical Small Office/Home Office (SOHO) network infrastructure, lacks the capacity to effectively model the complex behaviors and mobility inherent in UAV operations. To further refine the model's effectiveness, the study applied Bayesian optimization specifically to LightGBM for hyperparameter tuning.

Similarly, other efforts to fortify cellular-connected UAV networks against DDoS attacks, highlighted in [198], also rely on ML-based techniques. Yet again, the utilization of the CSE-CIC IDS-2018 dataset [199] for attack detection in this context lacks alignment with FANETs and the intricate infrastructure of 5G networks. This makes the dataset less suitable for such UAV-based studies.

A very recent study based on ML [200] focuses on real-time GPS spoofing detection for UAVs. It explores static and dynamic attack scenarios through flights using both authentic GPS signals and simulated spoofing attacks via a software-defined radio (SDR) transceiver module. The study's significant contribution lies in introducing a real-time GPS spoofing detection solution compatible with standard receivers and common modules, eliminating the need for hardware modifications. The study recorded GPS signal attributes during normal and spoofed encounters, and employed various ML algorithms (RF, K-Nearest Neighbors (KNN), SVM, Decision Tree (DT), and Neural Networks (NN)).

As stated above, several ML-based studies have previously relied on public datasets gathered from various environments, potentially unsuitable for FANETs. Nonetheless, in the most recent literature, two studies have presented their datasets specifically collected from simulations mimicking FANET environments to address this gap. In [201], an ML-based approach for Sybil attack detection in FANETs is introduced. The study created a FANET dataset by considering the 3D movement and low-density characteristics using the OMNET++ simulation tool. This dataset encompasses two radio signals: Received Signal Strength Difference (RSS) and Time Difference of Arrival (TDoA), extracted from the physical layer. Reported experimental results showcase a detection rate exceeding 91%, with a claimed false positive rate (FPR) of less than 9%. However, to ensure real-world applicability, further evaluation of the model across diverse attack scenarios becomes essential, as relying solely on a single scenario might not guarantee its overall security.

The other study presented in [202] introduces an attack dataset designed for detecting time delay attacks in FANETs. This dataset gathers latency-related information from pre-planned routes established by different routing protocols within simulations conducted via the ONE simulator [203]. Employing ML algorithms, the study detects attacks and utilizes K-means clustering to identify malicious nodes. The study achieves an accuracy exceeding 80% with less than 2.5% overhead across various network configurations. However, its emphasis on pre-planned flight paths, slow speeds (6 m/s), and 2D movements of UAVs limit its suitability to a variety of FANET applications.

Several studies have employed distinct methods such as artificial neural network-based (ANN) and fuzzy-based algorithms to detect specific types of attacks. In [204], an ANN-based approach is proposed for detecting false data injection attacks. It utilized the Thor Flight 111 dataset [205] for model training and evaluated the model's performance based on detection time and false positive rate. However, with increasing network density, a decline in detection rate and an increase in false positive rates were observed. Another study [206] explores the use of neural networks and fuzzy-rule-based IDS for detecting DDoS attacks. The attacks were conducted in real-time against the Parrot AR. Drone. Hence the tests were limited to a single small-scale drone, prompting the need for observation of their effects in a larger network setting. [207] proposed another fuzzy-based IDS for detection various attacks such as wormhole, sinkhole, selective forwarding. Each node computes the trust value of its one-hop neighbors based on its experience and recommendations from neighboring nodes using a fuzzy method.

Recent research [208] has introduced an innovative IDS that utilizes a hybrid model combining ANN and genetic algorithms (GA) to further enhance system performance. Additionally, the study employs a novel dimensional reduction technique, which integrates correlation coefficient analysis, information gain, and principal component analysis (PCA) to reduce the dimensionality of the UAV attack dataset, significantly lowering computational and memory requirements. The proposed model has demonstrated superior prediction accuracy and time efficiency compared to traditional classifiers, positioning it as a more effective solution for UAV network security.

*Deep Learning-based Studies:* Various deep learning (DL) approaches have been explored in recent studies for intrusion detection in UAV networks. These methods are proposed in many studies due to their inherent advantages in discovering complex patterns from data, demonstrating adaptability, and achieving high accuracy in classification tasks. A DL-based approach showcasing the superior performance of Convolutional Neural Networks (CNN) over traditional machine learning algorithms is presented in [209]. The system used encrypted Wi-Fi traffic data records from the UAV-IDS-2020 dataset [210], achieving an accuracy of 99.50% with a prediction time of 2.77 ms.

The implementation of a recurrent neural network (RNN) algorithm is explored in [211]. However, due to the absence of a tailored intrusion dataset for FANETs, training was performed using datasets such as KDDCUP99 [212] and NSL-KDD [213] collected from different network types. The proposed IDS deployed both in each UAV and within GBS. In [214], a Deep Reinforcement Learning (DRL) approach was proposed for training IDS models on the central system. Regular model updates between UAVs and the central station resulted in a higher detection rate but led to significant energy consumption. Hence, an offline learning strategy, wherein model updates occur when UAVs return to the charging station, was suggested to conserve UAV energy while maintaining IDS effectiveness.

In another study [215], DL combined with hierarchical SVM was employed to detect GPS spoofing and jamming attacks in UAVs. Upon detecting an attack, UAVs trigger a Q-learning-based adaptive route learning algorithm to navigate back to a secure zone. However, while the study claimed that DL algorithms provide a lightweight solution for this purpose, no experiments were conducted to validate this assertion. Building on the theme of lightweight solutions, another study [216] proposes a feature selection technique based on an improved fuzzy rough set (FRS) to identify the most relevant features, aiming to reduce computational costs and achieve a more lightweight IDS. The study also presents a model that integrates DL with RF to ensure high detection accuracy. Although the initial evaluations indicate the method's effectiveness, further investigation is required to fully verify its efficiency and potential for real-world implementation. Another recent study [217] suggests a Greedy-based GA for feature selection, followed by the implementation of a DL-based IDS. The IDS is trained using UAV data generated from a MATLAB Simulink model, which incorporates various features such as drone velocity, height, and width.

A recent study proposes the Exhaustive Distributed Intrusion Detection System (E-DIDS) [218]. This system decentralizes intrusion detection by distributing multiple IDS units (Sensor, System-based, and Network) across various UAV subsystems, enabling each component to independently monitor for anomalies. The Sensor and System-based IDS units are linked to key UAV components to gather data, including sensors and the core-central system. Additionally, the Network IDS unit is specifically dedicated to detecting communication-based attacks, collecting data from both the GBS and the UAV. E-DIDS has demonstrated high detection accuracy, achieving an average of 98.6%.

In [219], the study proposes the Collaborative Intrusion Detection System (CIDS), which leverages cooperation between multiple UAV nodes equipped with IDS units. These units are initially trained centrally, ensuring a coordinated and comprehensive detection strategy. A key innovation in CIDS is the use of a hybrid activation function that combines the advantages of both ReLU (Rectified Linear Unit) and TaLU (Tanh Linear Unit) activation functions, enhancing the model's ability to process complex data patterns. This paper also presents a real-world application of CIDS using actual drones, demonstrating its ability to perform online, offline, and hybrid predictions.

*Federated Learning-based Studies:* In [220], a federated learning-based approach is proposed to detect jamming attacks using two datasets. The first dataset, generated using the ns-3 simulation tool, contained 3,000 samples with eight features like Packet Delivery Ratio, throughput, and Received Signal Strength Indicator. The second dataset adapted the CRAWDAD VANET dataset [221] for FANETs' unbalanced data. Each UAV trained a local model and sent weights to a central system, which aggregated them to form a global model. A selective approach utilizing the dumper-shaper method effectively reduced communication costs, achieving approximately 82% accuracy on the CRAWDAD dataset and 89.5% accuracy on the FANET dataset. Traditional solutions were shown to yield notably lower accuracy on these datasets. In their extended study [222], a reinforcement federated learning-based method identified a defense strategy in new environments. This approach devises alternative routes avoiding jamming attack areas through spatial retreat.

In [223], an IDS detects GPS jamming and spoofing attacks using an unsupervised federated learning approach with the UAV Attack Dataset [224]. Various federated learning aggregation methods like FedAvg [225], FedAdagrad [226], FedAdam [226], and FedYogi [226], were tested, with FedAvg notably displaying robustness and achieving an F1-score of 0.887.

In a very recent study [227], FL using CNN and DNN algorithms was introduced for detecting FANET routing attacks (blackhole, sinkhole, flooding). A significant contribution was the creation of a comprehensive FANET dataset with 50 nodes featuring 3D movements and essential FANET-specific characteristics. Moreover, they compared IDSs developed via FL, traditional central, and local methods. FL closely approached central IDS performance in most experiments, demonstrating its potential for FANET's distributed architecture. Additionally, they employed the Bias Towards Specific Clients (BTSC) approach to enhance detection performance. In their extended study [228], the authors propose a Few-shot FL-based IDS (FSFL-IDS), which integrates few-shot learning (FSL) techniques with federated learning (FL) to detect routing attacks. This approach significantly reduces the data required for training while ensuring privacy.

*3) Specification-based IDSs:* These techniques aim to combine the advantages of both signature-based and anomaly-based systems. In such systems, any deviation from system specifications is flagged as a potential attack, enabling the detection of new attacks that do not adhere to these specifications. However, they inherently struggle to detect DoS attacks, as they align closely with the system's specifications. Additionally, defining specifications for all system components is a time-consuming task.

Since routing protocols represent a significant advancement in MANETs, numerous specification-based IDSs focus on detecting routing attacks in this context [42]. To our knowledge, there have not been specific specification-based proposals in FANETs. Adapting MANET proposals to FANETs is plausible, yet detecting more evasive DoS attacks in such dynamic systems should be a key consideration.

*4) Hybrid IDSs:* A hybrid approach merging signature-based and statistical-based anomaly detection techniques is proposed in [229] for identifying various types of DDoS attacks targeting FANETs. The signature-based method was tested against two known attack variants and a new type. However, this approach displayed reduced robustness against these new attack types and even variants of existing ones. To address this limitation, a statistical anomaly-based approach was introduced, aiming to enhance detection robustness. In another study [230], a hybrid approach was introduced to detect blackhole, grayhole, GPS spoofing, and jamming attacks in FANETs. Each UAV was equipped with a rule-based local IDS, and an intrusion response system, developed using SVM, was implemented in the GBS. The results highlighted that this hybrid method offered a high level of accuracy and provided a lightweight security solution with minimal overhead.

A trust-based approach was proposed in [231] to detect wormhole and data integrity attacks. In [232], an IDS was described using the *belief* approach to monitor the behavior of each UAV and create a threat level. The IDS was located on each UAVs. The experimentation results showed that, despite a high number of attackers, a low false positive ratio (≈3%) and a high detection ratio (≈93%) were obtained.

*5) Summary of Lessons Learned from Detection Studies:* The proposed intrusion detection approaches are summarized in Table XII. While research on UAVs and FANETs is rapidly growing, their security exploration remains in its early developmental stages. Although literature proposes numerous approaches for MANETs and VANETs [237], these solutions might not readily adapt to FANETs due to their higher dynamic topology and distinct mobility patterns and architectures. Nonetheless, the research community might utilize certain solutions from the literature, such as specification-based IDSs developed for specific routing protocols, to develop hybrid solutions.

In our review, we have presented significant studies on intrusion detection in UAVs and FANETs, categorized by intrusion detection methods. Notably, artificial intelligence-based approaches stand out for their ability to uncover complex properties. However, retraining these models within resource-constrained environments requires careful consideration. Exploring trade-offs between security and resource consumption for different applications with distinct requirements remains an open area for investigation. Additionally, many of these studies rely on datasets not collected from UAVs, raising concerns about their real-world applicability.

Examining IDS architecture and the deployment of proposed solutions is pivotal. Federated learning-based IDS holds promise in ensuring communication privacy among IDS agents. Nevertheless, ensuring the security of these proposed solutions, particularly safeguarding against adversarial attacks on AI-based solutions in highly mobile systems, remains an area requiring further research.

## VIII. Open issues and Research Directions

Studies published for preventing and detecting attacks against UAVs and networks of UAVs are listed in Table XI and Table XII. As can be seen, the research on FANETs and UAV communication is still at an early stage, hence there are currently only limited studies on the securing of such networks. UAVs are used in many applications including those that are considered mission-critical, which naturally make them prime targets for attacks. While a good number of security solutions have been proposed for MANETs and VANETs, FANETs have different requirements to other ad hoc networks. Very high node mobility in 3D, dynamic topology, low density networks, and nodes with small batteries are some of the biggest differences resulting in challenges faced by FANETs from the security perspective. Hence, there is a need to explore novel prevention and detection techniques tailored specifically for FANETs. These methods should account for the network's unique features, whether by devising new solutions or adapting existing security approaches.

The studies on the security of UAVs and FANETs have accelerated in the very few years, reflecting the growing recognition of the vulnerabilities inherent in these systems. However, a critical analysis reveals several gaps and challenges that warrant further investigation. In the subsequent sections, we delve into a detailed examination of the identified shortcomings in the existing literature, shedding light on specific areas where research efforts could be directed to fortify the security state of UAVs and FANETs.

**Limitations in Simulation Environments:**

The published studies have generally been either not evaluated or tested within simulation environments as shown in Table XII. However, these simulations should reflect real-world flight dynamics and 3D movements of UAVs. Unfortunately, however, this has not been the case in the literature. To the best of the authors knowledge, approaches for securing FANETs have been generally simulated on networks where nodes move only in 2D, and only a few studies [38], [201], [222] have implemented the 3D movement of UAVs in their simulations. Furthermore, some studies have utilized parameters such as a small number of nodes [215], [220] or low speeds that are more suited to MANETs [116], [202], [230]. Therefore, we believe that the analysis of attacks carried out in more realistic real-world scenarios, as exampled in the current study, is an important initial study which aims to accelerate the research in this important area.

While proposing and assessing security solutions for UAV communication, it is pivotal to assess their adaptability across diverse tasks and applications. Factors like the presence of central nodes and the mobility patterns of these nodes can significantly influence the performance of proposed security solutions. For instance, certain missions may require coordinated movement of UAVs in a specific direction, followed by periodic reorientation towards the controller ground system. As a result, security solutions are only practical and effective if they are aligned with these mission-specific network configurations. The absence of mission-specific simulations in current literature is a notable gap. This limits our ability to assess security proposals within real-world mission contexts. Incorporating realistic network settings aligned with specific missions is an unexplored area that hampers the development of tailored security solutions optimized for diverse UAV operations.

**GBS- Overlooked Asset in UAV Security:**

Unlike typical ad hoc networks, which lack central points and distribute data across nodes, GBS play a vital role in various UAV applications. GBS serve essential functions such as data aggregation, decision-making, and more. Security proposals in these scenarios can leverage the presence of central nodes within UAV operations. Given their superior compu-tational powers and ample resources compared to UAVs, these central stations can execute more robust algorithms, thereby enhancing the security measures implemented within the network. Additionally, the deployment of these nodes, whether in static or mobile capacities, opens avenues for innovative security solutions. As static central nodes, much like returning to the trusted comfort of a mother's embrace, GBS provide a reliable hub for tasks like updating models or databases, integrating signatures, and establishing rules during UAV battery charging at these stations [214]. Similarly, in their mobile capacity, they can undertake similar tasks. Exploring various alternatives becomes crucial, considering network density, mobility characteristics, and diverse applications.

TABLE XII
OUTLINE OF THE DETECTION STUDIES ON UAVS' SECURITY

| Reference | Year | Method | Dataset | Attacks | Simulations |
|---|---|---|---|---|---|
| [231] | 2016 | Trust-based (Belief approach) | No dataset | Wormhole<br>Data integrity | No sim. |
| [206] | 2016 | Fuzzy | No dataset | DDoS | Real UAV was used |
| [230] | 2018 | Rule-based<br>ML | a FANET dataset | Blackhole<br>Grayhole<br>GPS spoofing and jamming | ns-3<br>2D movement<br>50-250 nodes |
| [191] | 2018 | Rule-based | No dataset | DoS<br>False information<br>injection | ns-3<br>100-400 nodes |
| [215] | 2019 | DL+SVM | a FANET dataset | GPS spoofing and jamming | ONE simulator<br>2D movement<br>20 nodes |
| [229] | 2019 | Signature-based<br>Statistical Anomaly-based | No dataset | CFC<br>PFC | OMNeT++<br>2D movement |
| [187] | 2020 | Rule-based | No dataset | Blackhole, Grayhole<br>Wormhole and Fake Information Dissemination | Ns-3<br>100-400 nodes, 2D movement |
| [220] | 2020 | FL | CRAWDAD VANET dataset [221]<br>a FANET dataset | Jamming | ns-3<br>3D movement<br>4 nodes |
| [222] | 2020 | FL | CRAWDAD VANET dataset [221]<br>a FANET dataset | Jamming | ns-3<br>3D movement<br>6 nodes |
| [207] | 2020 | Trust-based<br>Fuzzy Classification | No dataset | Dropping | Omnet++<br>2D movement<br>100 nodes |
| [195] | 2020 | ML<br>AI | a UAV dataset | GPS Spoofing | PX4 and Gazebo<br>3D movement<br>6 nodes |
| [211] | 2021 | DL | KDDCup 99<br>NSL-KDD<br>UNSW-NB15<br>Kyoto<br>CICIDS2017 [199]<br>TON_IoT | Backdoor<br>DoS<br>Injection<br>Mitm<br>Password<br>Scanning<br>XSS<br>Benign | No sim. |
| [198] | 2021 | ML | CSE-CIC-IDS2018 [199] | DoS, DDoS<br>Brute Force, BotNet<br>Web Attack, Infiltration | No sim. |
| [214] | 2021 | DL | CICIDS2017 | Brute force<br>DoS<br>Botnet<br>Port scanning<br>SQL injections<br>XSS | No. sim. |
| [192] | 2021 | Rule-based | No dataset | Flooding | OMNeT++<br>2D movement<br>40 nodes |
| [193] | 2022 | ML | DJI Phantom 4 drone datase [194] | DoS attacks | No sim. |
| [196] | 2022 | ML | AWID2 dataset [197] | DoS attacks | No sim. |
| [200] | 2023 | ML | UAV dataset | GPS Spoofing | a real UAV was used |
| [201] | 2023 | ML | a FANET dataset | Sybil | OMNET++<br>3D movement<br>6 nodes |
| [202] | 2023 | ML | FANET dataset | Time Delay | ONE<br>2D movement<br>13,24 nodes |
| [223] | 2023 | FL | UAV Attack dataset [224] | GPS Jamming and Spoofing | No sim. |
| [227] | 2023 | FL | a FANET dataset | Sinkhole<br>Blackhole<br>Flooding | ns-3<br>3D movement<br>50 nodes |
| [218] | 2024 | DL | UAV Attack dataset [224]<br>UAV UAVCAN dataset [233]<br>Kitsune [234] | GPS Jamming and Spoofing<br>Flooding, Replay, MiTM<br>DoS, Injection | No sim. |
| [219] | 2024 | DL | UAVIDS [235] | anomalous networked UAV traffic | No Sim. |
| [208] | 2024 | ANN+GA | UAV Attack dataset [224] | GPS Jamming and Spoofing | No sim. |
| [216] | 2024 | DL+RF | CIC-DDOS2019 [236]<br>CSE-CIC-IDS2018 [199] | DDoS<br>DoS, DDoS<br>Brute Force, BotNet<br>Web Attack, Infiltration | No sim.<br>. |
| [228] | 2023 | FL+FSL | a FANET dataset | Sinkhole<br>Blackhole<br>Flooding | ns-3<br>3D movement<br>50 nodes |
| [217] | 2024 | DL | a UAV dataset | anomalous location | MATLAB Simulink |

In addition, the vulnerability of GBS remains insufficiently analyzed, despite their critical role in UAV operations. Existing studies predominantly concentrate on attacks against UAVs, overlooking potential threats to GBS. Attacks targeting these central nodes could have significant consequence, given their status as potential single point-of-failure in certain tasks. Moreover, the impact of such attacks can vary based on whether these central nodes are static or dynamic, underscoring the need for comprehensive assessment and mitigation strategies.

**Need for Increased Attention to Architectural Aspects:**

While studies predominantly concentrate on their methods, they often neglect to address the architectural aspects in their proposals. For instance, in certain applications, UAVs might operate collectively, moving together in a specific direction to accomplish tasks. In such densely coordinated systems, a distributed and cooperative architecture is likely to yield superior performance compared to a network where UAVs move randomly and autonomously, resulting in sporadic connectivity. Moreover, as pointed out above, the positive inclusion of GBS in a hybrid or hierarchical architecture should be further explored.

In a distributed and cooperative architecture, prioritizing privacy and secure communication among agents is critical. Blockchain technology has shown promise in addressing these concerns, as evidenced by several studies exploring its applications in this research domain [238] [239]. However, the use and deployment of blockchain in UAVs with energy, computation, data storage resource constraints needs further examination.

Another impactful approach is the utilization of federated learning, particularly in bolstering the performance of machine learning methods for security purposes. Federated learning involves transferring local models' parameters, instead of large data volumes, to train a global model and redistribute its parameters to local models. This approach holds significant potential for highly dynamic networks prone to frequent link disruptions, where a distributed and cooperative solution is likely to outperform traditional methods. Notably, federated learning addresses privacy concerns inherent in such systems.

Existing literature showcases federated learning-based approaches for intrusion detection. Studies, such as those aimed at detecting jamming attacks [220] [222], and recent research [227] comparing FL-based proposals with central and local architectures, demonstrate the potential of federated learning in terms of accuracy, security, and communication cost. Future research should explore federated learning-based systems for various attack types. Further investigation is warranted in the realm of blockchain-based federated learning [174]. Greater emphasis is needed to understand and mitigate attacks against federated learning systems, such as fake parameter updates or model poisoning attacks in general, backdoor attacks. In addition, convergence challenges due to device heterogeneity, and communication delays during parameter collection and redistribution among local nodes should be taken into account [174].

Connectivity among nodes is another critical architectural consideration in UAV networks. The high speeds of UAVs can lead to intermittent connectivity, causing potential communication disruptions. Moreover, relying on wireless links heightens susceptibility to packet drops, impacting network reliability. These challenges complicate real-time data monitoring and must be carefully addressed in architectural design to ensure robust communication for security solutions. An alternative solution could be an Internet of Digital Twin UAVs in the cloud, capable of gathering traffic information from other twin UAVs [240]. The integration of Digital Twin technology for enhancing the security of UAVs represents an emerging area of research that warrants further exploration.

**ML-based Approaches:**

UAVs high mobility and energy constraints make them very challenging for manual proposals. Hence, artificial intelligence-based studies offer promising solutions with their ability to automatically discover the complex characteristics of a system. Therefore, researches investigate the use of machine learning techniques for UAV's security. While a few security proposals have been based on machine learning [198], [201], [215] and deep learning [211], [214], [235] for UAVs and FANETs, most of these studies have utilized public datasets proposed for environments other than FANETs. Moreover, even in the simulations applied in [202], [215], the nodes used moved only in 2D. Thus, there is a clear need for additional research to thoroughly evaluate the practical implementation of these methods in real-world scenarios.

A multi-level security system that incorporates anomaly detection across various aspects can be quite robust. For instance, anomalies detected in sensor inputs, routing packets, and communication between UAVs and the Ground Base Station (GBS) could collectively indicate a potential intrusion. This layered approach enhances the overall security posture by monitoring multiple levels for potential threats, hence an open research area.

Moreover, techniques that adapt to dynamic environments warrant further investigation. FANETs can be considered a form of dynamic optimization problem (DOO). Therefore, machine learning techniques designed for DOO, such as Reinforcement Learning, Deep Reinforcement Learning [214], and Evolutionary Dynamic Optimization (EDO) techniques [241], present promising avenues for exploration.

In addition, the security of ML-based solutions should be taken into account. Adversarial attacks, especially in mobile environments, require further investigation. The dynamic and lossy nature of UAV networks makes distinguishing between normal and abnormal behavior challenging, leaving room for exploitation by attackers. Moreover, new types of physical adversarial attacks are emerging, such as physical adversarial patches [242], which can effectively mislead UAV visual sensors. These evolving threats underscore the need for continued research and development of effective defense mechanisms to improve UAV resilience against a wider range of adversarial tactics.

As pointed above, secure communication among agents running ML algorithms locally can be achieved through blockchain or federated learning. Blockchain-assisted federated learning [243] facilitates decentralized model aggregation, eliminating the vulnerability of a central aggregator. Moreover, involving only authorized UAVs' local models in updating the system's model [244] could mitigate poisoning attacks. Designing lightweight blockchains for UAVs presents a promising area of investigation due to its inherent advantages [245].

**Resource Constraints:**

Resource consumption is another constraint and critical issue that needs to be addressed when designing security solutions for UAVs and FANETs. As some missions necessitate the deployment of small-sized UAVs having constraints in processing power, memory, and energy, such nodes may be more susceptible to attacks. It is essential to develop secure protocols or security solutions tailored for UAVs and FANETs that operate within stringent resource constraints. This involves developing protocols that ensure both security and efficiency, implementing appropriate access control and key management mechanisms, optimizing cryptographic techniques, authentication methods, and IDSs to operate effectively while minimizing resource usage within UAVs' limited capabilities.

In order to protect small-sized UAVs, lightweight solutions should be designed, and certain proposals in the literature have already targeted this aim [214], [215], [230]. These proposals have generally claimed their approaches to be lightweight due to the use of a known lightweight approach such as deep learning [215], or where it is considered to generate only a low overhead [230]. However, differences in the trade-offs between effectiveness and resource consumption have not been discussed in these proposals. Different trade-offs may be more suited to certain tasks or missions, hence this consideration requires considerable exploration.

To effectively discover these trade-offs, this multi-objective optimization (MOO) problem can be addressed using Pareto-based approaches, which focus on identifying a set of Pareto-optimal solutions. In these approaches, solutions are evaluated based on their trade-offs between conflicting objectives, allowing for the selection of solutions where no single objective can be improved without degrading at least one other objective. Techniques such as Multi-Objective Evolutionary Algorithms (MOEAs), Pareto Simulated Annealing, and Pareto Ant Colony Optimization are commonly employed to explore the solution space and converge toward the Pareto front. These techniques are particularly suitable for MANETs [246] or FANETs, as they can effectively manage the dynamic and complex nature of aerial networks, optimizing parameters such as energy efficiency, detection accuracy, and network coverage simultaneously.

In addition, the effectiveness of lightweight solutions with the inclusion of other nodes and a ground central system could be improved, and is also worthy of further investigation. GBS typically possess superior computation capabilities and energy resources compared to small-sized UAVs, which are constrained by limited battery power. Additionally, GBS could be deployable in either static or mobile configurations based on specific application needs.These central nodes can be utilized for tasks such as updating models or databases, incorporating signatures and rules during UAV battery charging at these stations [214].

**Strengthening Security with Emerging Technologies:**

In real-world scenarios, UAVs have the capability to interact with IoT devices, MANETs, VANETs or other systems. This capability not only enables UAVs to increase connectivity by working as a relay node in such systems but also bases the groundwork for UAV-assisted security solutions. These solutions play a crucial role in averting potential security vulnerabilities that may arise when diverse devices are integrated within hybrid environments. For instance, in [247], how UAVs can aid in reducing security risks in IoT systems against eavesdropping attacks is explored. A UAV acting as a relay receives packages to provide transmission to a specific destination and aims for their successful delivery. Another approach [248] uses UAVs as friendly jammers in order to deter unidentified eavesdroppers using artificial noise. When an eavesdropper is detected, they serve as relays between vehicles to prevent information leaks.

We expect to see more applications of UAV-assisted networks or IoT, since such systems make it possible to access locations where humans cannot access in applications such as emergency search and rescue and military. Exploring security solutions for such hybrid systems that cover all components with diverse characteristics is a new area of investigation. Additionally, the utilization of UAVs to enhance security in IoT or other network types is worth further exploration. Emerging areas such as UAV-assisted blockchain [249], trust management [250], and intrusion detection [247] are all areas that deserve deeper examination.

UAVs could also assist in Mobile Edge Computing (MEC), which involves offloading tasks to mobile edges to meet the requirements such as decreased latency, real-time processing, and enhanced Quality of Experience of next-wave applications like augmented reality and ultra-high definition video streaming [251]. Security presents a significant challenge in these systems, particularly for real-time applications where security and latency are competing factors. Future research directions for securing UAV-assisted MEC highlight areas such as PLS, machine learning, blockchain, and authentication protocols [251].

Enhancing the security of UAVs using emerging technologies such as digital twins or the Internet of Digital Twins represents a highly promising area for research. Beyond security improvements, digital twins offer capabilities for anomaly detection, early failure detection, and more. These systems can complement traditional security solutions, particularly when certain UAV agents face connectivity issues or device failures. Exploring methods to model UAVs and their communication within digital twin frameworks, as well as identifying meaningful semantic data for security purposes, are key areas that demand further study.

**Other Aspects- Diversity of Applications and Standards:**

UAVs can be used in various applications with various security requirements, ranging from military applications to civilian tasks such as infrastructure inspection, agricultural monitoring, environmental surveys, disaster management, and aerial photography, among others. Hence, when designing security solutions, the specific needs and diverse communication patterns inherent in different applications should be taken into account. As pointed in above, such mission-specific or scenario-based simulations is a notable gap in the literature. This gap also underscores the critical necessity for advancements in protocols and standards. Notably, existing standards provide varying levels of support for UAV communication, traffic management and flight operation among UAVs [252]–[254]. Many standards such as UASSC [254], PODIUM [253] focus solely on communication among homogeneous UAV, while others, like JAUS [252], extend their scope to accommodate heterogeneous UAV systems. Enhancing these protocols and standards is a pivotal step in addressing the growing challenges faced by UAV system security.

To summarize, UAVs bring about new challenges from the security perspective. Whilst there have already been studies published in this area, the research is still at an early stage. The development of suitable solutions are still needed for such dynamic and resource-constrained systems, and the deployment and evaluation of these solutions is an important area in which further exploration is necessary.

## IX. Conclusion

The increasing numbers and expanding applications of UAVs in both military and civilian domains have made them vulnerable targets for various attacks. This review study aims to comprehensively explore the security issues concerning UAVs and their communications facilitated by FANETs in various operational tasks. While existing research has extensively covered MANETs and VANETs in the literature, it is crucial to analyze FANETs due to their distinct characteristics, strengths, and vulnerabilities.

Initially, we evaluate the specific characteristics of UAVs and FANETs, considering their unique requirements and discussing their implications for security. Then, we present the attack surface analysis, which is one of the important contributions of this study. This analysis not only illuminates vulnerabilities within UAVs and FANETs but also serves as a foundation for identifying novel threats. The survey strategically aligns a taxonomy of attacks targeting UAVs and FANETs based on the attack surface analysis.

This study transcends a standard review by integrating an attack analysis based on extensive simulations. Four attacks against FANETs, including blackhole, sinkhole, drooping, and flooding attacks, are simulated using realistic real-world scenarios to illustrate the varying implications and potential results of each attack within the network. Then, the proposed solutions based on prevention and detection are presented and discussed. Finally, we have thoroughly explored open issues and research directions, ensuring a comprehensive understanding of the latest developments and their interdependence with related areas.

We believe this study offers a comprehensive survey covering security issues in UAVs and FANETs. It presents a taxonomy of attacks based on the attack surface analysis, conducts simulations and analysis of attacks in real-world scenarios, and discusses proposed security solutions in the literature, along with detailed research directions. The utilization of attack surface analysis allows for a more comprehensive understanding of the security landscape of UAVs. By thoroughly exploring open issues and delineating promising research directions, our study aims to pave new pathways within this research domain. Considering the accelerating focus on UAV security in recent years, we believe this study is timely and beneficial for researchers.

## References

[1] I. Bekmezci, O. K. Sahingoz, and Ş. Temel, "Flying ad-hoc networks (fanets): A survey," *Ad Hoc Networks*, vol. 11, no. 3, pp. 1254–1270, 2013.

[2] "Commercial drone market size, share &; covid-19 impact analysis, by weight," [Accessed 17-July-2022]. [Online]. Available: https://www.fortunebusinessinsights.com/commercial-drone-market-102171

[3] (2020) Why flying drones could disrupt mobility and transportation beyond COVID-19. Gartner. [Online]. Available: https://www.gartner.com/smarterwithgartner/why-flying-drones-could-disrupt-mobility-and-transportation-beyond-covid-19

[4] D. Orfanus, E. P. De Freitas, and F. Eliassen, "Self-organization as a supporting paradigm for military uav relay networks," *IEEE Communications letters*, vol. 20, no. 4, pp. 804–807, 2016.

[5] D. Erdos, A. Erdos, and S. E. Watkins, "An experimental uav system for search and rescue challenge," *IEEE Aerospace and Electronic Systems Magazine*, vol. 28, no. 5, pp. 32–37, 2013.

[6] J. Scherer, S. Yahyanejad, S. Hayat, E. Yanmaz, T. Andre, A. Khan, V. Vukadinovic, C. Bettstetter, H. Hellwagner, and B. Rinner, "An autonomous multi-uav system for search and rescue," ser. DroNet '15. New York, NY, USA: Association for Computing Machinery, 2015, p. 3338. [Online]. Available: https://doi.org/10.1145/2750675.2750683

[7] B. Li and Y. Wu, "Path planning for uav ground target tracking via deep reinforcement learning," *IEEE Access*, vol. 8, pp. 29 064–29 074, 2020.

[8] I. Bor-Yaliniz, S. S. Szyszkowicz, and H. Yanikomeroglu, "Environment-aware drone-base-station placements in modern metropolitans," *IEEE Wireless Communications Letters*, vol. 7, no. 3, pp. 372–375, 2018.

[9] P. Radoglou-Grammatikis, P. Sarigiannidis, T. Lagkas, and I. Moscholios, "A compilation of uav applications for precision agriculture," *Computer Networks*, vol. 172, p. 107148, 2020.

[10] R. Fu, X. Ren, Y. Li, Y. Wu, H. Sun, and M. A. Al-Absi, "Machine-learning-based uav-assisted agricultural information security architecture and intrusion detection," *IEEE Internet of Things Journal*, vol. 10, no. 21, pp. 18 589–18 598, 2023.

[11] I. Mahmud and Y. Z. Cho, "Adaptive Hello Interval in FANET Routing Protocols for Green UAVs," *IEEE Access*, vol. 7, pp. 63 004–63 015, 2019.

[12] J. George, S. PB, and J. B. Sousa, "Search strategies for multiple uav search and destroy missions," *Journal of Intelligent & Robotic Systems*, vol. 61, pp. 355–367, 2011.

[13] F. Al Fayez, M. Hammoudeh, B. Adebisi, and K. N. Abdul Sattar, "Assessing the effectiveness of flying ad hoc networks for international border surveillance," *International Journal of Distributed Sensor Networks*, vol. 15, no. 7, p. 1550147719860406, 2019.

[14] V. Hassija, V. Chamola, V. Saxena, D. Jain, P. Goyal, and B. Sikdar, "A survey on iot security: Application areas, security threats, and solution architectures," *IEEE Access*, vol. 7, pp. 82 721–82 743, 2019.

[15] C. Barrado, R. Messeguer, J. López, E. Pastor, E. Santamaria, and P. Royo, "Wildfire monitoring using a mixed air-ground mobile network," *IEEE Pervasive Computing*, vol. 9, no. 4, pp. 24–32, 2010.

[16] G. Faraci, S. A. Rizzo, and G. Schembra, "Green edge intelligence for smart management of a fanet in disaster-recovery scenarios," *IEEE Transactions on Vehicular Technology*, vol. 72, no. 3, pp. 3819–3831, 2023.

[17] H. Xiang and L. Tian, "Development of a low-cost agricultural remote sensing system based on an autonomous unmanned aerial vehicle (uav)," *Biosystems engineering*, vol. 108, no. 2, pp. 174–190, 2011.

[18] G. A. Kakamoukas, P. G. Sarigiannidis, and A. A. Economides, "Fanets in agriculture-a routing protocol survey," *Internet of Things*, vol. 18, p. 100183, 2022.

[19] A. Press, "Computer virus infects drone plane command centre in us," https://www.theguardian.com/technology/2011/oct/09/virus-infects-drone-plane-command, [Accessed 06-Jul-2022].

[20] J. Keller, "Iran–us rq-170 incident has defense industry saying never againto unmanned vehicle hacking," *Military & Aerospace Electronics*, vol. 3, 2016.

[21] İ. Bekmezci, E. Şentürk, and T. Türker, "Security issues in flying ad-hoc networks (fanets)," *Journal of Aeronautics and Space Technologies*, vol. 9, no. 2, pp. 13–21, 2016.

[22] J.-A. Maxa, M.-S. B. Mahmoud, and N. Larrieu, "Survey on uaanet routing protocols and network security challenges," *Ad Hoc & Sensor Wireless Networks*, 2017.

[23] G. K. Pandey, D. S. Gurjar, H. H. Nguyen, and S. Yadav, "Security threats and mitigation techniques in uav communications: A comprehensive survey," *IEEE Access*, 2022.

[24] A. Chriki, H. Touati, H. Snoussi, and F. Kamoun, "Fanet: Communication, mobility models and security issueszhi2020," *Computer Networks*, vol. 163, p. 106877, 2019.

[25] F. Tlili, L. C. Fourati, S. Ayed, and B. Ouni, "Investigation on vulnerabilities, threats and attacks prohibiting UAVs charging and depleting UAVs batteries: Assessments & countermeasures," *Ad Hoc Networks*, vol. 129, no. January, 2022.

[26] Y. Zhi, Z. Fu, X. Sun, and J. Yu, "Security and privacy issues of uav: A survey," *Mobile Networks and Applications*, vol. 25, no. 1, pp. 95–101, 2020.

[27] M. Yahuza, M. Y. I. Idris, I. B. Ahmedy, A. W. A. Wahab, T. Nandy, N. M. Noor, and A. Bala, "Internet of drones security and privacy issues: Taxonomy and open challenges," *IEEE Access*, vol. 9, pp. 57 243–57 270, 2021.

[28] V. Hassija, V. Chamola, A. Agrawal, A. Goyal, N. C. Luong, D. Niyato, F. R. Yu, and M. Guizani, "Fast, reliable, and secure drone communication: A comprehensive survey," *IEEE Communications Surveys & Tutorials*, vol. 23, no. 4, pp. 2802–2832, 2021.

[29] J.-P. Yaacoub, H. Noura, O. Salman, and A. Chehab, "Security analysis of drones systems: Attacks, limitations, and recommendations," *Internet of Things*, vol. 11, p. 100218, 2020.

[30] R. Altawy and A. M. Youssef, "Security, privacy, and safety aspects of civilian drones: A survey," *ACM Transactions on Cyber-Physical Systems*, vol. 1, no. 2, pp. 1–25, 2016.

[31] J. Sharma and P. S. Mehra, "Secure communication in iot-based uav networks: A systematic survey," *Internet of Things*, p. 100883, 2023.

[32] A. Rugo, C. A. Ardagna, and N. E. Ioini, "A security review in the uavnet era: threats, countermeasures, and gap analysis," *ACM Computing Surveys (CSUR)*, vol. 55, no. 1, pp. 1–35, 2022.

[33] N. Kumar and A. Chaudhary, "Surveying cybersecurity vulnerabilities and countermeasures for enhancing uav security," *Computer Networks*, vol. 252, p. 110695, 2024.

[34] Y. Mekdad, A. Aris, L. Babun, A. El Fergougui, M. Conti, R. Lazzeretti, and A. S. Uluagac, "A survey on security and privacy issues of uavs," *Computer Networks*, vol. 224, p. 109626, 2023.

[35] H. J. Hadi, Y. Cao, K. U. Nisa, A. M. Jamil, and Q. Ni, "A comprehensive survey on security, privacy issues and emerging defence technologies for uavs," *Journal of Network and Computer Applications*, vol. 213, p. 103607, 2023.

[36] Z. Wang, Y. Li, S. Wu, Y. Zhou, L. Yang, Y. Xu, T. Zhang, and Q. Pan, "A survey on cybersecurity attacks and defenses for unmanned aerial systems," *Journal of Systems Architecture*, vol. 138, p. 102870, 2023.

[37] K.-Y. Tsao, T. Girdler, and V. G. Vassilakis, "A survey of cyber security threats and solutions for uav communications and flying ad-hoc networks," *Ad Hoc Networks*, vol. 133, p. 102894, 2022.

[38] O. Ceviz, P. Sadioglu, and S. Sen, "Analysis of routing attacks in fanets," in *Ad Hoc Networks and Tools for IT*, W. Bao, X. Yuan, L. Gao, T. H. Luan, and D. B. J. Choi, Eds. Cham: Springer International Publishing, 2022, pp. 3–17.

[39] O. S. Oubbati, M. Atiquzzaman, P. Lorenz, M. H. Tareque, and M. S. Hossain, "Routing in flying Ad Hoc networks: Survey, constraints, and future challenge perspectives," *IEEE Access*, 2019.

[40] M. B. Yassein and . A. Damer, "Flying ad-hoc networks: Routing protocols, mobility models, issues," *International Journal of Advanced Computer Science and Applications*, vol. 7, no. 6, 2016. [Online]. Available: http://dx.doi.org/10.14569/IJACSA.2016.070621

[41] B. Van Der Bergh, A. Chiumento, and S. Pollin, "Lte in the sky: Trading off propagation benefits with interference costs for aerial nodes," *IEEE Communications Magazine*, vol. 54, no. 5, pp. 44–50, 2016.

[42] S. Şen and J. A. Clark, "Intrusion detection in mobile ad hoc networks," in *Guide to wireless ad hoc networks*. Springer, 2009, pp. 427–454.

[43] A. Purohit, F. Mokaya, and P. Zhang, "Collaborative indoor sensing with the sensorfly aerial sensor network," in *Proceedings of the 11th international conference on Information Processing in Sensor Networks*, 2012, pp. 145–146.

[44] S. Javed, A. Hassan, R. Ahmad, W. Ahmed, R. Ahmed, A. Saadat, and M. Guizani, "State-of-the-art and future research challenges in uav swarms," *IEEE Internet of Things Journal*, 2024.

[45] R. M. Fouda, "Security vulnerabilities of cyberphysical unmanned aircraft systems," *IEEE Aerospace and Electronic Systems Magazine*, vol. 33, no. 9, pp. 4–17, 2018.

[46] M. Bacco, E. Ferro, and A. Gotta, "Radio propagation models for uavs: what is missing?" in *Proceedings of the 11th International Conference on Mobile and Ubiquitous Systems: Computing, Networking and Services*, 2014, pp. 391–392.

[47] W. Khawaja, I. Guvenc, D. W. Matolak, U.-C. Fiebig, and N. Schneckenburger, "A survey of air-to-ground propagation channel modeling for unmanned aerial vehicles," *IEEE Communications Surveys & Tutorials*, vol. 21, no. 3, pp. 2361–2391, 2019.

[48] M. Bacco, P. Cassará, M. Colucci, A. Gotta, M. Marchese, and F. Patrone, "A survey on network architectures and applications for nanosat and uav swarms," in *Wireless and Satellite Systems: 9th International Conference, WiSATS 2017, Oxford, UK, September 14-15, 2017, Proceedings 9*. Springer, 2018, pp. 75–85.

[49] M. Banafaa, Ö. Pepeoğlu, I. Shayea, A. Alhammadi, Z. Shamsan, M. A. Razaz, M. Alsagabi, and S. Al-Sowayan, "A comprehensive survey on 5g-and-beyond networks with uavs: Applications, emerging technologies, regulatory aspects, research trends and challenges," *IEEE Access*, 2024.

[50] W. Zheng, Z. Zheng, X. Chen, K. Dai, P. Li, and R. Chen, "Nutbaas: A blockchain-as-a-service platform," *IEEE Access*, vol. 7, pp. 134 422–134 433, 2019.

[51] M. Kuzlu, M. Pipattanasomporn, L. Gurses, and S. Rahman, "Performance analysis of a hyperledger fabric blockchain framework: Throughput, latency and scalability," in *2019 IEEE International Conference on Blockchain (Blockchain)*, 2019, pp. 536–540.

[52] D. Shumeye Lakew, U. Sa'Ad, N. N. Dao, W. Na, and S. Cho, "Routing in Flying Ad Hoc Networks: A Comprehensive Survey," *IEEE Commun. Surv. Tutorials*, vol. 22, no. 2, pp. 1071–1120, 2020.

[53] D. Cerri and A. Ghioni, "Securing aodv: the a-saodv secure routing prototype," *IEEE Communications Magazine*, vol. 46, no. 2, pp. 120–125, 2008.

[54] M. A. Khan, A. Safi, I. M. Qureshi, and I. U. Khan, "Flying ad-hoc networks (fanets): A review of communication architectures, and routing protocols," in *2017 First international conference on latest trends in electrical engineering and computing technologies (INTELLECT)*. IEEE, 2017, pp. 1–9.

[55] N. Mansoor, M. I. Hossain, A. Rozario, M. Zareei, and A. R. Arreola, "A fresh look at routing protocols in unmanned aerial vehicular networks: a survey," *IEEE Access*, 2023.

[56] "attack surface - Glossary — CSRC — csrc.nist.gov," https://csrc.nist.gov/glossary/term/attack_surface, [Accessed 15-12-2023].

[57] C. Jansi Sophia Mary, K. Mahalakshmi, and B. Senthilkumar, "Deep dive on various security challenges, threats and attacks over the cloud security," in *2023 9th International Conference on Advanced Computing and Communication Systems (ICACCS)*, vol. 1, 2023, pp. 2089–2094.

[58] T. Li, X. He, S. Jiang, and J. Liu, "A survey of privacy-preserving offloading methods in mobile-edge computing," *Journal of Network and Computer Applications*, vol. 203, p. 103395, 2022. [Online]. Available: https://www.sciencedirect.com/science/article/pii/S1084804522000546

[59] D. Swessi and H. Idoudi, "A survey on internet-of-things security: Threats and emerging countermeasures," *Wireless Personal Communications*, vol. 124, no. 2, p. 15571592, Jan. 2022. [Online]. Available: http://dx.doi.org/10.1007/s11277-021-09420-0

[60] N. Mustari, M. A. Karabulut, A. S. Shah, and U. Tureli, "Cooperative thz communication for uavs in 6g and beyond," *Green Energy and Intelligent Transportation*, vol. 3, no. 1, p. 100131, 2024.

[61] L. D. Xu, Y. Lu, and L. Li, "Embedding blockchain technology into iot for security: A survey," *IEEE Internet of Things Journal*, vol. 8, no. 13, pp. 10 452–10 473, 2021.

[62] L. U. Khan, Z. Han, W. Saad, E. Hossain, M. Guizani, and C. S. Hong, "Digital twin of wireless systems: Overview, taxonomy, challenges, and opportunities," *IEEE Communications Surveys & Tutorials*, vol. 24, no. 4, pp. 2230–2254, 2022.

[63] S.-Y. Chang, K. Park, J. Kim, and J. Kim, "Securing uav flying base station for mobile networking: A review," *Future Internet*, vol. 15, no. 5, p. 176, 2023.

[64] N. Pitropakis, E. Panaousis, T. Giannetsos, E. Anastasiadis, and G. Loukas, "A taxonomy and survey of attacks against machine learning," *Computer Science Review*, vol. 34, p. 100199, 2019. [Online]. Available: https://www.sciencedirect.com/science/article/pii/S1574013718303289

[65] C. L. Krishna and R. R. Murphy, "A review on cybersecurity vulnerabilities for unmanned aerial vehicles," in *2017 IEEE international symposium on safety, security and rescue robotics (SSRR)*. IEEE, 2017, pp. 194–199.

[66] G. Cornelius, P. Caire, N. Hochgeschwender, M. A. Olivares-Mendez, P. Esteves-Verissimo, M. Völp, and H. Voos, "A perspective of security for mobile service robots," in *ROBOT 2017: Third Iberian Robotics Conference: Volume 1*. Springer, 2018, pp. 88–100.

[67] J. Marcus, "Belarus-made armed drone shot down in ukraine." [Online]. Available: https://defence-blog.com/belarus-made-armed-drone-shot-down-in-ukraine/

[68] "Russian electronic warfare drone shot down in ukraine: Report." [Online]. Available: https://www.thedefensepost.com/2022/06/21/russia-electronic-warfare-drone-ukraine/

[69] "Saudi oil attacks: Who's using drones in the middle east?" [Online]. Available: https://www.bbc.com/news/world-middle-east-49718828

[70] A. E. Omolara, M. Alawida, and O. I. Abiodun, "Drone cybersecurity issues, solutions, trend insights and future perspectives: a survey," *Neural computing and applications*, vol. 35, no. 31, pp. 23 063–23 101, 2023.

[71] S. Bhunia and M. M. Tehranipoor, *The hardware trojan war: Attacks, myths, and defenses*. Springer International Publishing, 2018.

[72] C. Gorman, "Counterfeit chips on the rise," *IEEE Spectrum*, vol. 49, no. 6, pp. 16–17, 2012.

[73] R. Spreitzer, V. Moonsamy, T. Korak, and S. Mangard, "Systematic classification of side-channel attacks: A case study for mobile devices," *IEEE Communications Surveys & Tutorials*, vol. 20, no. 1, pp. 465–488, 2017.

[74] F.-X. Standaert, "Introduction to side-channel attacks," in *Secure integrated circuits and systems*. Springer, 2010, pp. 27–42.

[75] F. Koeune and F.-X. Standaert, "A tutorial on physical security and side-channel attacks," *International School on Foundations of Security Analysis and Design*, pp. 78–108, 2004.

[76] Y. Park, O. C. Onar, and B. Ozpineci, "Potential cybersecurity issues of fast charging stations with quantitative severity analysis," in *2019 IEEE CyberPELS (CyberPELS)*. IEEE, 2019, pp. 1–7.

[77] V. Desnitsky and I. Kotenko, "Simulation and assessment of battery depletion attacks on unmanned aerial vehicles for crisis management infrastructures," *Simulation Modelling Practice and Theory*, vol. 107, p. 102244, 2021.

[78] V. Shakhov and I. Koo, "Depletion-of-battery attack: Specificity, modelling and analysis," *Sensors*, vol. 18, no. 6, p. 1849, 2018.

[79] A. B. Lopez, K. Vatanparvar, A. P. Deb Nath, S. Yang, S. Bhunia, and M. A. Al Faruque, "A security perspective on battery systems of the internet of things," *Journal of Hardware and Systems Security*, vol. 1, pp. 188–199, 2017.

[80] S. Belikovetsky, M. Yampolskiy, J. Toh, J. Gatlin, and Y. Elovici, "dr0wned–{Cyber-Physical} attack with additive manufacturing," in *11th USENIX workshop on offensive technologies (WOOT 17)*, 2017.

[81] U. Guin, K. Huang, D. DiMase, J. M. Carulli, M. Tehranipoor, and Y. Makris, "Counterfeit integrated circuits: A rising threat in the global semiconductor supply chain," *Proceedings of the IEEE*, vol. 102, no. 8, pp. 1207–1228, 2014.

[82] Y. Son, H. Shin, D. Kim, Y. Park, J. Noh, K. Choi, J. Choi, and Y. Kim, "Rocking drones with intentional sound noise on gyroscopic sensors," in *24th USENIX security symposium (USENIX Security 15)*, 2015, pp. 881–896.

[83] V. Lisa, "Drone hijacked by hackers from texas college with $1,000 spoofer," 2012. [Online]. Available: https://nakedsecurity.sophos.com/2012/07/02/drone-hackedwith-1000-spoofer/

[84] Z. Feng, N. Guan, M. Lv, W. Liu, Q. Deng, X. Liu, and W. Yi, "Efficient drone hijacking detection using two-step ga-xgboost," *Journal of Systems Architecture*, vol. 103, p. 101694, 2020.

[85] N. O. Tippenhauer, C. Pöpper, K. B. Rasmussen, and S. Capkun, "On the requirements for successful gps spoofing attacks," in *Proceedings of the 18th ACM conference on Computer and communications security*, 2011, pp. 75–86.

[86] M. Pasternak, N. Kahani, M. Bagherzadeh, J. Dingel, and J. R. Cordy, "Simgen: A tool for generating simulations and visualizations of embedded systems on the unity game engine," in *Proceedings of the 21st ACM/IEEE International Conference on Model Driven Engineering Languages and Systems: Companion Proceedings*, 2018, pp. 42–46.

[87] J. A. Saputro, E. E. Hartadi, and M. Syahral, "Implementation of gps attacks on dji phantom 3 standard drone as a security vulnerability test," in *2020 1st international conference on information technology, advanced mechanical and electrical engineering (ICITAMEE)*. IEEE, 2020, pp. 95–100.

[88] K. Mansfield, T. Eveleigh, T. H. Holzer, and S. Sarkani, "Unmanned aerial vehicle smart device ground control station cyber security threat model," in *2013 IEEE International Conference on Technologies for Homeland Security (HST)*. IEEE, 2013, pp. 722–728.

[89] J. Crook, "Infamous hacker creates skyjack to hunt, hack, and control other drones," *TechCrunch*, 2013.

[90] N. Shachtman, "Computer virus hits US drone fleet," *CNN. com, Oct.*, 2011. [Online]. Available: http://www.cs.clemson.edu/course/cpsc420/material/Papers/ComputerVirusHitsUSDroneFleet.pdf

[91] S. Rahul, "Maldrone the first backdoor for drones," 2015. [Online]. Available: http://http://garage4hackers.com/entry.php?b=3105

[92] C. Kolias, G. Kambourakis, A. Stavrou, and S. Gritzalis, "Intrusion detection in 802.11 networks: Empirical evaluation of threats and a public dataset," *IEEE Commun. Surv. Tutorials*, vol. 18, no. 1, pp. 184–208, 2016.

[93] H. Shin, K. Choi, Y. Park, J. Choi, and Y. Kim, "Security analysis of fhss-type drone controller," in *Information Security Applications: 16th International Workshop, WISA 2015, Jeju Island, Korea, August 20–22, 2015, Revised Selected Papers 16*. Springer, 2016, pp. 240–253.

[94] B. Wu, J. Chen, J. Wu, and M. Cardei, "A survey of attacks and countermeasures in mobile ad hoc networks," *Wireless network security*, pp. 103–135, 2007.

[95] X. Sun, D. W. K. Ng, Z. Ding, Y. Xu, and Z. Zhong, "Physical layer security in uav systems: Challenges and opportunities," *IEEE Wireless Communications*, vol. 26, no. 5, pp. 40–47, 2019.

[96] K. Li, S. S. Kanhere, W. Ni, E. Tovar, and M. Guizani, "Proactive eavesdropping via jamming for trajectory tracking of uavs," in *2019 15th International Wireless Communications & Mobile Computing Conference (IWCMC)*. IEEE, 2019, pp. 477–482.

[97] X. Wang, D. Li, C. Guo, X. Zhang, S. S. Kanhere, K. Li, and E. Tovar, "Eavesdropping and jamming selection policy for suspicious uavs based on low power consumption over fading channels," *Sensors*, vol. 19, no. 5, p. 1126, 2019.

[98] V. Chamola, P. Kotesh, A. Agarwal, N. Gupta, M. Guizani *et al.*, "A comprehensive review of unmanned aerial vehicle attacks and neutralization techniques," *Ad hoc networks*, vol. 111, p. 102324, 2021.

[99] A. Mpitziopoulos, D. Gavalas, C. Konstantopoulos, and G. Pantziou, "A survey on jamming attacks and countermeasures in wsns," *IEEE Communications Surveys & Tutorials*, vol. 11, no. 4, pp. 42–56, 2009.

[100] L. Gupta, R. Jain, and G. Vaszkun, "Survey of important issues in uav communication networks," *IEEE communications surveys & tutorials*, vol. 18, no. 2, pp. 1123–1152, 2015.

[101] Y.-C. Hu, A. Perrig, and D. B. Johnson, "Packet leashes: a defense against wormhole attacks in wireless networks," in *IEEE INFOCOM 2003. Twenty-second Annual Joint Conference of the IEEE Computer and Communications Societies (IEEE Cat. No. 03CH37428)*, vol. 3. IEEE, 2003, pp. 1976–1986.

[102] ——, "Rushing attacks and defense in wireless ad hoc network routing protocols," in *Proceedings of the 2nd ACM workshop on Wireless security*, 2003, pp. 30–40.

[103] C. Perkins, E. Belding-Royer, and S. Das, "Ad hoc on-demand distance vector (aodv) routing," Tech. Rep., 2003.

[104] Y.-C. Hu, D. A. Maltz, and D. B. Johnson, "The Dynamic Source Routing Protocol (DSR) for Mobile Ad Hoc Networks for IPv4," RFC 4728, Feb. 2007. [Online]. Available: https://www.rfc-editor.org/info/rfc4728

[105] M. G. Zapata, "Secure ad hoc on-demand distance vector routing," *ACM SIGMOBILE Mobile Computing and Communications Review*, vol. 6, no. 3, pp. 106–107, 2002.

[106] G. Montenegro and C. Castelluccia, "Statistically unique and cryptographically verifiable (sucv) identifiers and addresses," in *In Proceedings of the 9th Annual Network and Distributed System Security Symposium (NDSS*. Citeseer, 2002.

[107] S. Sen, J. A. Clark, and J. E. Tapiador, "Security threats in mobile ad hoc networks," *Security of Self-Organizing Networks: MANET, WSN, WMN, VANET, Auerbach Publications*, pp. 127–147, 2010.

[108] J.-A. Maxa, M. S. B. Mahmoud, and N. Larrieu, "Extended verification of secure uaanet routing protocol," in *2016 IEEE/AIAA 35th Digital Avionics Systems Conference (DASC)*. IEEE, 2016, pp. 1–16.

[109] C. Ge, X. Ma, and Z. Liu, "A semi-autonomous distributed blockchain-based framework for uavs system," *Journal of Systems Architecture*, vol. 107, p. 101728, 2020.

[110] G. Vasconcelos, G. Carrijo, R. Miani, J. Souza, and V. Guizilini, "The impact of dos attacks on the ar. drone 2.0," in *2016 XIII Latin American Robotics Symposium and IV Brazilian Robotics Symposium (LARS/SBR)*. IEEE, 2016, pp. 127–132.

[111] "hping3(8) - Linux man page — linux.die.net," https://linux.die.net/man/8/hping3, [Accessed 16-Jun-2023].

[112] "GitHub - NewEraCracker/LOIC: Low Orbit Ion Cannon - An open source network stress tool, written in C#. Based on Praetox's LOIC project. USE ON YOUR OWN RISK. WITHOUT ANY EXPRESS OR IMPLIED WARRANTIES. — github.com," https://github.com/NewEraCracker/LOIC, [Accessed 16-Jun-2023].

[113] "netwox(1) - Linux man page — linux.die.net," https://linux.die.net/man/1/netwox, [Accessed 16-Jun-2023].

[114] M. Conti, N. Dragoni, and V. Lesyk, "A survey of man in the middle attacks," *IEEE Communications Surveys & Tutorials*, vol. 18, no. 3, pp. 2027–2051, 2016.

[115] F. Ahmad, F. Kurugollu, A. Adnane, R. Hussain, and F. Hussain, "Marine: Man-in-the-middle attack resistant trust model in connected vehicles," *IEEE Internet of Things Journal*, vol. 7, no. 4, pp. 3310–3322, 2020.

[116] X. Tan, Z. Zuo, S. Su, X. Guo, and X. Sun, "Research of security routing protocol for UAV communication network based on AODV," *Electronics*, vol. 9, no. 8, pp. 1–18, 2020.

[117] J. P. Rohrer, E. K. Cetinkaya, H. Narra, D. Broyles, K. Peters, and J. P. Sterbenz, "Aerorp performance in highly-dynamic airborne networks using 3d gauss-markov mobility model," in *2011-MILCOM 2011 Military Communications Conference*. IEEE, 2011, pp. 834–841.

[118] D. Broyles and A. Jabbar, "Design and analysis of a 3-d gauss-markov model for highly dynamic airborne networks." International Foundation for Telemetering, 2010.

[119] "The ns-3 network simulator," http://www.nsnam.org/.

[120] S. Bhunia and M. Tehranipoor, "The hardware trojan war," *Cham,, Switzerland: Springer*, 2018.

[121] T. M. Supon and R. Rashidzadeh, "Hardware trojan prevention using memristor technology," *Microprocessors and Microsystems*, vol. 102, p. 104915, 2023. [Online]. Available: https://www.sciencedirect.com/science/article/pii/S014193312300159X

[122] Y. Wang, P. Chen, J. Hu, G. Li, and J. Rajendran, "The cat and mouse in split manufacturing," *IEEE Transactions on Very Large Scale Integration (VLSI) Systems*, vol. 26, no. 5, pp. 805–817, 2018.

[123] T. M. Supon and R. Rashidzadeh, "On-chip magnetic probes for hardware trojan prevention and detection," *IEEE Transactions on Electromagnetic Compatibility*, vol. 63, no. 2, pp. 353–364, 2021.

[124] H. Al-Aqrabi, A. P. Johnson, R. Hill, P. Lane, and T. Alsboui, "Hardware-intrinsic multi-layer security: A new frontier for 5g enabled iiot," *Sensors*, vol. 20, no. 7, 2020. [Online]. Available: https://www.mdpi.com/1424-8220/20/7/1963

[125] A. Pillai, M. Sindhu, and K. Lakshmy, "Securing firmware in internet of things using blockchain," in *2019 5th International Conference on Advanced Computing & Communication Systems (ICACCS)*, 2019, pp. 329–334.

[126] D. Davidson, H. Wu, R. Jellinek, V. Singh, and T. Ristenpart, "Controlling {UAVs} with sensor input spoofing attacks," in *10th USENIX workshop on offensive technologies (WOOT 16)*, 2016.

[127] Y. Zhao, Y. Liu, A. Tian, Y. Yu, and X. Du, "Blockchain based privacy-preserving software updates with proof-of-delivery for internet of things," *Journal of Parallel and Distributed Computing*, vol. 132, pp. 141–149, 2019.

[128] E. Dushku, M. M. Rabbani, M. Conti, L. V. Mancini, and S. Ranise, "Sara: Secure asynchronous remote attestation for iot systems," *IEEE Transactions on Information Forensics and Security*, vol. 15, pp. 3123–3136, 2020.

[129] M. Bloch, O. Gnl, A. Yener, F. Oggier, H. V. Poor, L. Sankar, and R. F. Schaefer, "An overview of information-theoretic security and privacy: Metrics, limits and applications," *IEEE Journal on Selected Areas in Information Theory*, vol. 2, no. 1, pp. 5–22, 2021.

[130] X. Sun, D. W. K. Ng, Z. Ding, Y. Xu, and Z. Zhong, "Physical layer security in uav systems: Challenges and opportunities," *IEEE Wireless Communications*, vol. 26, no. 5, pp. 40–47, 2019.

[131] A. Li, W. Zhang, and S. Dou, "Uav-enabled secure data dissemination via artificial noise: Joint trajectory and communication optimization," *IEEE Access*, vol. 8, pp. 102 348–102 356, 2020.

[132] S. J. Maeng, Y. Yapc, . Gven, A. Bhuyan, and H. Dai, "Precoder design for physical-layer security and authentication in massive mimo uav communications," *IEEE Transactions on Vehicular Technology*, vol. 71, no. 3, pp. 2949–2964, 2022.

[133] C. Zhong, J. Yao, and J. Xu, "Secure uav communication with cooperative jamming and trajectory control," *IEEE Communications Letters*, vol. 23, no. 2, pp. 286–289, 2019.

[134] R. Ma, W. Yang, Y. Zhang, J. Liu, and H. Shi, "Secure mmwave communication using uav-enabled relay and cooperative jammer," *IEEE Access*, vol. 7, pp. 119 729–119 741, 2019.

[135] H. Fu, Z. Sheng, A. A. Nasir, A. H. Muqaibel, and L. Hanzo, "Securing the uav-aided non-orthogonal downlink in the face of colluding eavesdroppers," *IEEE Transactions on Vehicular Technology*, vol. 71, no. 6, pp. 6837–6842, 2022.

[136] X. Pang, M. Liu, N. Zhao, Y. Chen, Y. Li, and F. R. Yu, "Secrecy analysis of uav-based mmwave relaying networks," *IEEE Transactions on Wireless Communications*, vol. 20, no. 8, pp. 4990–5002, 2021.

[137] W. Liang, J. Shi, Z. Tie, and F. Yang, "Performance analysis for uav-jammer aided covert communication," *IEEE Access*, vol. 8, pp. 111 394–111 400, 2020.

[138] J. Tang, G. Chen, and J. P. Coon, "Secrecy performance analysis of wireless communications in the presence of uav jammer and randomly located uav eavesdroppers," *IEEE Transactions on Information Forensics and Security*, vol. 14, no. 11, pp. 3026–3041, 2019.

[139] M. Zhang, Y. Chen, X. Tao, and I. Darwazeh, "Power allocation for proactive eavesdropping with spoofing relay in uav systems," in *2019 26th International Conference on Telecommunications (ICT)*, 2019, pp. 102–107.

[140] H. Hashida, Y. Kawamoto, and N. Kato, "Intelligent reflecting surface placement optimization in air-ground communication networks toward 6g," *IEEE Wireless Communications*, vol. 27, no. 6, pp. 146–151, 2020.

[141] H. Yang, Z. Xiong, J. Zhao, D. Niyato, Q. Wu, H. V. Poor, and M. Tornatore, "Intelligent reflecting surface assisted anti-jamming communications: A fast reinforcement learning approach," *IEEE Transactions on Wireless Communications*, vol. 20, no. 3, pp. 1963–1974, 2021.

[142] Y. Ye, M. Lei, and M. Zhao, "A new frequency hopping strategy based on federated reinforcement learning for fanet," in *2021 IEEE 94th Vehicular Technology Conference (VTC2021-Fall)*, 2021, pp. 1–5.

[143] K. Pärlin, M. M. Alam, and Y. Le Moullec, "Jamming of uav remote control systems using software defined radio," in *2018 International Conference on Military Communications and Information Systems (ICMCIS)*. IEEE, 2018, pp. 1–6.

[144] C. Liu, J. Lee, and T. Q. Quek, "Safeguarding uav communications against full-duplex active eavesdropper," *IEEE Transactions on Wireless Communications*, vol. 18, no. 6, pp. 2919–2931, 2019.

[145] Y. Chen, G. Liu, Z. Zhang, L. He, and S. He, "Improving physical layer security for multi-uav systems against hybrid wireless attacks," *IEEE Transactions on Vehicular Technology*, vol. 73, no. 5, pp. 7034–7048, 2024.

[146] G. L. Stber and G. L. Steuber, *Principles of Mobile Communication*, 2nd ed. Berlin, Germany: Springer, 1996.

[147] Y. Teng, P. Zhang, X. Chen, X. Jiang, and F. Xiao, "Phy-layer authentication exploiting channel sparsity in mmwave mimo uav-ground systems," *IEEE Transactions on Information Forensics and Security*, vol. 19, pp. 4642–4657, 2024.

[148] J.-A. Maxa, M. S. B. Mahmoud, and N. Larrieu, "Secure routing protocol design for uav ad hoc networks," in *2015 IEEE/AIAA 34th Digital Avionics Systems Conference (DASC)*. IEEE, 2015, pp. 4A5–1.

[149] ——, "Performance evaluation of a new secure routing protocol for uav ad hoc network," in *2019 IEEE/AIAA 38th Digital Avionics Systems Conference (DASC)*. IEEE, 2019, pp. 1–10.

[150] V. Bhardwaj, N. Kaur, S. Vashisht, and S. Jain, "Secrip: Secure and reliable intercluster routing protocol for efficient data transmission in flying ad hoc networks," *Transactions on Emerging Telecommunications Technologies*, vol. 32, no. 6, p. e4068, 2021.

[151] V. Bhardwaj and N. Kaur, "Seedrp: a secure energy efficient dynamic routing protocol in fanets," *Wireless Personal Communications*, vol. 120, no. 2, pp. 1251–1277, 2021.

[152] M. Hosseinzadeh, S. Ali, H. J. Ahmad, F. Alanazi, M. S. Yousefpoor, E. Yousefpoor, O. H. Ahmed, A. M. Rahmani, and S.-W. Lee, "A novel q-learning-based secure routing scheme with a robust defensive

system against wormhole attacks in flying ad hoc networks," *Vehicular Communications*, vol. 49, p. 100826, 2024.

[153] W. Buksh, Y. Guo, S. Iqbal, K. N. Qureshi, and J. Lloret, "Trust-oriented peered customized mechanism for malicious nodes isolation for flying ad hoc networks," *Transactions on Emerging Telecommunications Technologies*, vol. 35, no. 4, p. e4489, 2024.

[154] D. Muruganandam and J. Martin Leo Manickam, "Retracted article: an efficient technique for mitigating stealthy attacks using mnda in manet," *Neural Computing and Applications*, vol. 31, no. Suppl 1, pp. 15–22, 2019.

[155] X. Zhu, L. Lin, Y. Huang, X. Wang, Y. Que, B. Jedari, and M. Jalil Piran, "Secure data transmission based on reinforcement learning and position confusion for internet of uavs," *IEEE Internet of Things Journal*, vol. 11, no. 12, pp. 21 010–21 020, 2024.

[156] R. A. Nazib and S. Moh, "Reinforcement learning-based routing protocols for vehicular ad hoc networks: A comparative survey," *IEEE Access*, vol. 9, pp. 27 552–27 587, 2021.

[157] M. Ficco, D. Granata, F. Palmieri, and M. Rak, "A systematic approach for threat and vulnerability analysis of unmanned aerial vehicles," *Internet of Things*, vol. 26, p. 101180, 2024.

[158] "Introduction · MAVLink Developer Guide — mavlink.io," https://mavlink.io/en/, [Accessed 29-09-2024].

[159] C. Pu and Y. Li, "Lightweight authentication protocol for unmanned aerial vehicles using physical unclonable function and chaotic system," in *2020 IEEE International Symposium on Local and Metropolitan Area Networks (LANMAN*, 2020, pp. 1–6.

[160] A. Varga, "Omnet++," in *Modeling and tools for network simulation*. Springer, 2010, pp. 35–59.

[161] D. M'Raïhi, S. Machani, M. Pei, and J. Rydell, "TOTP: time-based one-time password algorithm," *RFC*, vol. 6238, pp. 1–16, 2011. [Online]. Available: https://doi.org/10.17487/RFC6238

[162] D. Mallikarachchi, K. Wong, and J. M.-Y. Lim, "An authentication scheme for fanet packet payload using data hiding," *Journal of Information Security and Applications*, vol. 77, p. 103559, 2023.

[163] S. Yu, J. Lee, A. K. Sutrala, A. K. Das, and Y. Park, "Laka-uav: Lightweight authentication and key agreement scheme for cloud-assisted unmanned aerial vehicle using blockchain in flying ad-hoc networks," *Computer Networks*, vol. 224, p. 109612, 2023. [Online]. Available: https://www.sciencedirect.com/science/article/pii/S1389128623000579

[164] T. Wu, X. Guo, Y. Chen, S. Kumari, and C. Chen, "Amassing the security: An enhanced authentication protocol for drone communications over 5g networks," *Drones*, vol. 6, no. 1, 2022. [Online]. Available: https://www.mdpi.com/2504-446X/6/1/10

[165] S. Hussain, K. Mahmood, M. K. Khan, C.-M. Chen, B. A. Alzahrani, and S. A. Chaudhry, "Designing secure and lightweight user access to drone for smart city surveillance," *Computer Standards & Interfaces*, vol. 80, p. 103566, 2022.

[166] Z. Ali, S. A. Chaudhry, M. S. Ramzan, and F. Al-Turjman, "Securing smart city surveillance: A lightweight authentication mechanism for unmanned vehicles," *IEEE Access*, vol. 8, pp. 43 711–43 724, 2020.

[167] Z. Zhang, C. Hsu, M. H. Au, L. Harn, J. Cui, Z. Xia, and Z. Zhao, "Prlap-iod: A puf-based robust and lightweight authentication protocol for internet of drones," *Computer Networks*, vol. 238, p. 110118, 2024. [Online]. Available: https://www.sciencedirect.com/science/article/pii/S1389128623005637

[168] A. Armando, D. Basin, Y. Boichut, Y. Chevalier, L. Compagna, J. Cuéllar, P. H. Drielsma, P.-C. Héam, O. Kouchnarenko, J. Mantovani *et al.*, "The avispa tool for the automated validation of internet security protocols and applications," in *Computer Aided Verification: 17th International Conference, CAV 2005, Edinburgh, Scotland, UK, July 6-10, 2005. Proceedings 17*. Springer, 2005, pp. 281–285.

[169] T. Alladi, V. Chamola, N. Kumar *et al.*, "Parth: A two-stage lightweight mutual authentication protocol for uav surveillance networks," *Computer Communications*, vol. 160, pp. 81–90, 2020.

[170] Y. Li and C. Pu, "Lightweight digital signature solution to defend micro aerial vehicles against man-in-the-middle attack," in *2020 IEEE 23rd International Conference on Computational Science and Engineering (CSE)*, 2020, pp. 92–97.

[171] S. A. Chaudhry, K. Yahya, M. Karuppiah, R. Kharel, A. K. Bashir, and Y. B. Zikria, "Gcacs-iod: A certificate based generic access control scheme for internet of drones," *Computer Networks*, vol. 191, p. 107999, 2021. [Online]. Available: https://www.sciencedirect.com/science/article/pii/S1389128621001195

[172] L. Mendiboure, M. Chalouf, and F. Krief, "Survey on blockchain-based applications in internet of vehicles," *Computers & Electrical Engineering*, vol. 84, p. 106646, 06 2020.

[173] B. Mikavica and A. Kostić-Ljubisavljević, "Blockchain-based solutions for security, privacy, and trust management in vehicular networks: a survey," *The Journal of Supercomputing*, vol. 77, no. 9, pp. 9520–9575, Sep 2021. [Online]. Available: https://doi.org/10.1007/s11227-021-03659-x

[174] D. Saraswat, A. Verma, P. Bhattacharya, S. Tanwar, G. Sharma, P. N. Bokoro, and R. Sharma, "Blockchain-based federated learning in uavs beyond 5g networks: A solution taxonomy and future directions," *IEEE Access*, vol. 10, pp. 33 154–33 182, 2022.

[175] X. Liang, J. Zhao, S. Shetty, and D. Li, "Towards data assurance and resilience in iot using blockchain," in *MILCOM 2017 - 2017 IEEE Military Communications Conference (MILCOM)*, 2017, pp. 261–266.

[176] T. Rana, A. Shankar, M. K. Sultan, R. Patan, and B. Balusamy, "An intelligent approach for uav and drone privacy security using blockchain methodology," in *2019 9th International Conference on Cloud Computing, Data Science & Engineering (Confluence)*, 2019, pp. 162–167.

[177] B. Bera, S. Saha, A. K. Das, N. Kumar, P. Lorenz, and M. Alazab, "Blockchain-envisioned secure data delivery and collection scheme for 5g-based iot-enabled internet of drones environment," *IEEE Transactions on Vehicular Technology*, vol. 69, no. 8, pp. 9097–9111, 2020.

[178] Y. Tan, J. Liu, and N. Kato, "Blockchain-based key management for heterogeneous flying ad hoc network," *IEEE Transactions on Industrial Informatics*, vol. 17, no. 11, pp. 7629–7638, 2020.

[179] R. Xiong, Q. Xiao, Z. Wang, Z. Xu, and F. Shan, "Leveraging lightweight blockchain for secure collaborative computing in uav ad-hoc networks," *Computer Networks*, vol. 251, p. 110612, 2024.

[180] M. Burrows, M. Abadi, and R. M. Needham, "A logic of authentication," *Proceedings of the Royal Society of London. A. Mathematical and Physical Sciences*, vol. 426, no. 1871, pp. 233–271, 1989. [Online]. Available: https://royalsocietypublishing.org/doi/abs/10.1098/rspa.1989.0125

[181] M. Hooper, Y. Tian, R. Zhou, B. Cao, A. P. Lauf, L. Watkins, W. H. Robinson, and W. Alexis, "Securing commercial WiFi-based UAVs from common security attacks," *Proc. - IEEE Mil. Commun. Conf. MILCOM*, pp. 1213–1218, 2016.

[182] J. Cui, J. Wen, S. Han, and H. Zhong, "Efficient privacy-preserving scheme for real-time location data in vehicular ad-hoc network," *IEEE Internet of Things Journal*, vol. 5, no. 5, pp. 3491–3498, 2018.

[183] R. Küsters and T. Truderung, "Using proverif to analyze protocols with diffie-hellman exponentiation," in *2009 22nd IEEE Computer Security Foundations Symposium*. IEEE, 2009, pp. 157–171.

[184] W. Mao and C. Boyd, "Towards formal analysis of security protocols," in *[1993] Proceedings Computer Security Foundations Workshop VI*. IEEE, 1993, pp. 147–158.

[185] R. Canetti, O. Goldreich, and S. Halevi, "The random oracle methodology, revisited," *J. ACM*, vol. 51, no. 4, p. 557594, jul 2004. [Online]. Available: https://doi.org/10.1145/1008731.1008734

[186] M. Abdalla, P.-A. Fouque, and D. Pointcheval, "Password-based authenticated key exchange in the three-party setting," in *Public Key Cryptography - PKC 2005*, S. Vaudenay, Ed. Berlin, Heidelberg: Springer Berlin Heidelberg, 2005, pp. 65–84.

[187] R. Fotohi, "Securing of unmanned aerial systems (uas) against security threats using human immune system," *Reliability Engineering & System Safety*, vol. 193, p. 106675, 2020.

[188] R. Fotohi, E. Nazemi, and F. S. Aliee, "An agent-based self-protective method to secure communication between uavs in unmanned aerial vehicle networks," *Vehicular Communications*, vol. 26, p. 100267, 2020.

[189] R. Fotohi and M. Abdan, "Self-adaptive intrusion detection system for uav-to-uav communications in uav networks," 2021.

[190] M. Faraji-Biregani and R. Fotohi, "Secure communication between uavs using a method based on smart agents in unmanned aerial vehicles," *The journal of supercomputing*, vol. 77, no. 5, pp. 5076–5103, 2021.

[191] H. Sedjelmaci and S. M. Senouci, "Cyber security methods for aerial vehicle networks: taxonomy, challenges and solution," *The Journal of Supercomputing*, vol. 74, no. 10, pp. 4928–4944, 2018.

[192] C. Pu and P. Zhu, "Defending against flooding attacks in the internet of drones environment," in *2021 IEEE Global Communications Conference (GLOBECOM)*. IEEE, 2021, pp. 1–6.

[193] Z. Baig, N. Syed, and N. Mohammad, "Securing the smart city airspace: Drone cyber attack detection through machine learning," *Future Internet*, vol. 14, no. 7, p. 205, 2022.

[194] Labs, "V. drone forensics," 2020. [Online]. Available: https://www.vtolabs.com/drone-forensics(accessedon15April2022)

[195] J. Whelan, T. Sangarapillai, O. Minawi, A. Almehmadi, and K. El-Khatib, "Novelty-based intrusion detection of sensor attacks on unmanned aerial vehicles," in *Proceedings of the 16th ACM symposium on QoS and security for wireless and mobile networks*, 2020, pp. 23–28.

[196] L. M. da Silva, H. B. d. B. Menezes, M. d. S. Luccas, C. Mailer, A. S. R. Pinto, A. Boava, M. Rodrigues, I. G. Ferrão, J. C. Estrella, and K. R. L. J. C. Branco, "Development of an efficiency platform based on mqtt for uav controlling and dos attack detection," *Sensors*, vol. 22, no. 17, p. 6567, 2022.

[197] C. Kolias, G. Kambourakis, A. Stavrou, and S. Gritzalis, "Intrusion detection in 802.11 networks: Empirical evaluation of threats and a public dataset," *IEEE Communications Surveys & Tutorials*, vol. 18, no. 1, pp. 184–208, 2015.

[198] R. Shrestha, A. Omidkar, S. A. Roudi, R. Abbas, and S. Kim, "Machine-learning-enabled intrusion detection system for cellular connected uav networks," *Electronics*, vol. 10, no. 13, p. 1549, 2021.

[199] I. Sharafaldin, A. H. Lashkari, and A. A. Ghorbani, "Toward generating a new intrusion detection dataset and intrusion traffic characterization." *ICISSp*, vol. 1, pp. 108–116, 2018.

[200] M. Nayfeh, Y. Li, K. Al Shamaileh, V. Devabhaktuni, and N. Kaabouch, "Machine learning modeling of gps features with applications to uav location spoofing detection and classification," *Computers & Security*, vol. 126, p. 103085, 2023.

[201] D. Chulerttiyawong and A. Jamalipour, "Sybil attack detection in internet of flying things-ioft: A machine learning approach," *IEEE Internet of Things Journal*, 2023.

[202] W. Zhai, L. Liu, Y. Ding, S. Sun, and Y. Gu, "Etd: An efficient time delay attack detection framework for uav networks," *IEEE Transactions on Information Forensics and Security*, 2023.

[203] A. Keränen, J. Ott, and T. Kärkkäinen, "The one simulator for dtn protocol evaluation," in *Proceedings of the 2nd international conference on simulation tools and techniques*, 2009, pp. 1–10.

[204] C. Titouna and F. Nat-Abdesselam, "Securing unmanned aerial systems using mobile agents and artificial neural networks," in *2021 International Wireless Communications and Mobile Computing (IWCMC)*, 2021, pp. 825–830.

[205] B. Taylor, "Thor flight 111. retrieved from the university of minnesota digital conservancy," 2014.

[206] C. Rani, H. Modares, R. Sriram, D. Mikulski, and F. L. Lewis, "Security of unmanned aerial vehicle systems against cyber-physical attacks," *The Journal of Defense Modeling and Simulation*, vol. 13, no. 3, pp. 331–342, 2016.

[207] K. Singh and A. K. Verma, "Tbcs: A trust based clustering scheme for secure communication in flying ad-hoc networks," *Wireless Personal Communications*, vol. 114, pp. 3173–3196, 2020.

[208] K. Cengiz, S. Lipsa, R. K. Dash, N. Ivković, and M. Konecki, "A novel intrusion detection system based on artificial neural network and genetic algorithm with a new dimensionality reduction technique for uav communication," *IEEE access*, 2024.

[209] Q. Abu Al-Haija and A. Al Badawi, "High-performance intrusion detection system for networked uavs via deep learning," *Neural Computing and Applications*, vol. 34, no. 13, pp. 10 885–10 900, 2022.

[210] N. Pitropakis, E. Panaousis, T. Giannetsos, E. Anastasiadis, and G. Loukas, "A taxonomy and survey of attacks against machine learning," *Computer Science Review*, vol. 34, p. 100199, 2019.

[211] R. A. Ramadan, A.-H. Emara, M. Al-Sarem, and M. Elhamahmy, "Internet of drones intrusion detection using deep learning," *Electronics*, vol. 10, no. 21, p. 2633, 2021.

[212] KDD Cup 1999, "Kdd cup 1999," http://kdd.ics.uci.edu/databases/kddcup99/kddcup99.html, October 2007.

[213] M. Tavallaee, E. Bagheri, W. Lu, and A. A. Ghorbani, "NSL-KDD: A new approach to improve the KDD cup 1999 dataset," *Journal of Information Systems*, vol. 32, no. 3, pp. 586–596, 2009.

[214] O. Bouhamed, O. Bouachir, M. Aloqaily, and I. Al Ridhawi, "Lightweight ids for uav networks: A periodic deep reinforcement learning-based approach," in *2021 IFIP/IEEE International Symposium on Integrated Network Management (IM)*. IEEE, 2021, pp. 1032–1037.

[215] M. P. Arthur, "Detecting signal spoofing and jamming attacks in uav networks using a lightweight ids," in *2019 international conference on computer, information and telecommunication systems (CITS)*. IEEE, 2019, pp. 1–5.

[216] Y. Wu, L. Yang, L. Zhang, L. Nie, and L. Zheng, "Intrusion detection for unmanned aerial vehicles security: A tiny machine learning model," *IEEE Internet of Things Journal*, 2024.

[217] S. Miao, Q. Pan, D. Zheng *et al.*, "Unmanned aerial vehicle intrusion detection: Deep-meta-heuristic system," *Vehicular Communications*, vol. 46, p. 100726, 2024.

[218] F. Tlili, S. Ayed, and L. C. Fourati, "Exhaustive distributed intrusion detection system for uavs attacks detection and security enforcement (e-dids)," *Computers & Security*, vol. 142, p. 103878, 2024.

[219] H. J. Hadi, Y. Cao, S. Li, Y. Hu, J. Wang, and S. Wang, "Real-time collaborative intrusion detection system in uav networks using deep learning," *IEEE Internet of Things Journal*, 2024.

[220] N. I. Mowla, N. H. Tran, I. Doh, and K. Chae, "Federated learning-based cognitive detection of jamming attack in flying ad-hoc network," *IEEE Access*, vol. 8, pp. 4338–4350, 2019.

[221] O. Puñal, C. Pereira, A. Aguiar, and J. Gross, "Crawdad dataset uportorwthaachen/vanetjamming2014 (v. 2014-05-12)," 2014.

[222] N. I. Mowla, N. H. Tran, I. Doh, and K. Chae, "Afrl: Adaptive federated reinforcement learning for intelligent jamming defense in fanet," *Journal of Communications and Networks*, vol. 22, no. 3, pp. 244–258, 2020.

[223] L. M. Da Silva, I. G. Ferrão, C. Dezan, D. Espes, and K. R. Branco, "Anomaly-based intrusion detection system for in-flight and network security in uav swarm," in *2023 International Conference on Unmanned Aircraft Systems (ICUAS)*. IEEE, 2023, pp. 812–819.

[224] J. Whelan, T. Sangarapillai, O. Minawi, A. Almehmadi, and K. El-Khatib, "Uav attack dataset," 2020. [Online]. Available: https://dx.doi.org/10.21227/00dg-0d12

[225] B. McMahan, E. Moore, D. Ramage, S. Hampson, and B. A. y Arcas, "Communication-efficient learning of deep networks from decentralized data," in *Artificial intelligence and statistics*. PMLR, 2017, pp. 1273–1282.

[226] S. Reddi, Z. Charles, M. Zaheer, Z. Garrett, K. Rush, J. Konečnỳ, S. Kumar, and H. B. McMahan, "Adaptive federated optimization," *arXiv preprint arXiv:2003.00295*, 2020.

[227] O. Ceviz, P. Sadioglu, S. Sen, and V. G. Vassilakis, "A novel federated learning-based intrusion detection system for flying ad hoc networks," *arXiv preprint arXiv:2312.04135*, 2023.

[228] O. Ceviz, S. Sen, and P. Sadioglu, "Distributed intrusion detection in dynamic networks of uavs using few-shot federated learning," in *Proceedings of the 20th EAI International Conference on Security and Privacy in Communication Networks*. EAI, 2024, to appear.

[229] J. P. Condomines, R. Zhang, and N. Larrieu, "Network intrusion detection system for UAV ad-hoc communication: From methodology design to real test validation," *Ad Hoc Networks*, vol. 90, p. 101759, 2019. [Online]. Available: https://doi.org/10.1016/j.adhoc.2018.09.004

[230] H. Sedjelmaci, S. M. Senouci, and N. Ansari, "A hierarchical detection and response system to enhance security against lethal cyber-attacks in uav networks," *IEEE Transactions on Systems, Man, and Cybernetics: Systems*, vol. 48, no. 9, pp. 1594–1606, 2017.

[231] H. Sedjelmaci, S. M. Senouci, and M. A. Messous, "How to detect cyber-attacks in unmanned aerial vehicles network?" *2016 IEEE Glob. Commun. Conf. GLOBECOM 2016 - Proc.*, 2016.

[232] L. Zomlot, S. C. Sundaramurthy, K. Luo, X. Ou, and S. R. Rajagopalan, "Prioritizing intrusion analysis using dempster-shafer theory," in *Proceedings of the 4th ACM workshop on Security and artificial intelligence*, 2011, pp. 59–70.

[233] D. Kim, Y. Song, S. Kwon, H. Kim, J. D. Yoo, and H. K. Kim, "Uavcan dataset description," *arXiv preprint arXiv:2212.09268*, 2022.

[234] Y. Mirsky, T. Doitshman, Y. Elovici, and A. Shabtai, "Kitsune: an ensemble of autoencoders for online network intrusion detection," *arXiv preprint arXiv:1802.09089*, 2018.

[235] A. Alipour-Fanid, M. Dabaghchian, N. Wang, P. Wang, L. Zhao, and K. Zeng, "Machine learning-based delay-aware uav detection and operation mode identification over encrypted wi-fi traffic," *IEEE Transactions on Information Forensics and Security*, vol. 15, pp. 2346–2360, 2019.

[236] I. Sharafaldin, A. H. Lashkari, S. Hakak, and A. A. Ghorbani, "Developing realistic distributed denial of service (ddos) attack dataset and taxonomy," in *2019 international carnahan conference on security technology (ICCST)*. IEEE, 2019, pp. 1–8.

[237] F. Sakiz and S. Sen, "A survey of attacks and detection mechanisms on intelligent transportation systems: Vanets and iov," *Ad Hoc Networks*, vol. 61, pp. 33–50, 2017.

[238] M. Aloqaily, O. Bouachir, A. Boukerche, and I. A. Ridhawi, "Design guidelines for blockchain-assisted 5g-uav networks," *IEEE Network*, vol. 35, no. 1, pp. 64–71, 2021.

[239] A. Jain, S. Barke, M. Garg, A. Gupta, B. Narwal, A. K. Mohapatra, D. K. Sharma, and G. Srivastava, "A walkthrough of blockchain-based

internet of drones architectures," *IEEE Internet of Things Journal*, 2024.

[240] Y. Wang, Z. Su, S. Guo, M. Dai, T. H. Luan, and Y. Liu, "A survey on digital twins: architecture, enabling technologies, security and privacy, and future prospects," *IEEE Internet of Things Journal*, 2023.

[241] M. Aslan and S. Sen, "A dynamic trust management model for vehicular ad hoc networks," *Vehicular Communications*, vol. 41, p. 100608, 2023.

[242] T. Liu, C. Yang, X. Liu, R. Han, and J. Ma, "Rpau: Fooling the eyes of uavs via physical adversarial patches," *IEEE Transactions on Intelligent Transportation Systems*, vol. 25, no. 3, pp. 2586–2598, 2024.

[243] S. H. Alsamhi, F. A. Almalki, F. Afghah, A. Hawbani, A. V. Shvetsov, B. Lee, and H. Song, "Drones edge intelligence over smart environments in b5g: Blockchain and federated learning synergy," *IEEE Transactions on Green Communications and Networking*, vol. 6, no. 1, pp. 295–312, 2021.

[244] C. Feng, B. Liu, K. Yu, S. K. Goudos, and S. Wan, "Blockchain-empowered decentralized horizontal federated learning for 5g-enabled uavs," *IEEE Transactions on Industrial Informatics*, vol. 18, no. 5, pp. 3582–3592, 2021.

[245] D. Saraswat, A. Verma, P. Bhattacharya, S. Tanwar, G. Sharma, P. N. Bokoro, and R. Sharma, "Blockchain-based federated learning in uavs beyond 5g networks: A solution taxonomy and future directions," *IEEE Access*, vol. 10, pp. 33 154–33 182, 2022.

[246] S. Sen and J. A. Clark, "Evolutionary computation techniques for intrusion detection in mobile ad hoc networks," *Computer Networks*, vol. 55, no. 15, pp. 3441–3457, 2011.

[247] A. S. Abdalla, B. Shang, V. Marojevic, and L. Liu, "Securing mobile iot with unmanned aerial systems," in *2020 IEEE 6th World Forum on Internet of Things (WF-IoT)*. IEEE, 2020, pp. 1–6.

[248] B. Shang, L. Liu, J. Ma, and P. Fan, "Unmanned aerial vehicle meets vehicle-to-everything in secure communications," *IEEE Communications Magazine*, vol. 57, no. 10, pp. 98–103, 2019.

[249] M. Golam, J.-M. Lee, and D.-S. Kim, "A uav-assisted blockchain based secure device-to-device communication in internet of military things," in *2020 International Conference on Information and Communication Technology Convergence (ICTC)*. IEEE, 2020, pp. 1896–1898.

[250] M. Huang, A. Liu, N. N. Xiong, and J. Wu, "A uav-assisted ubiquitous trust communication system in 5g and beyond networks," *IEEE Journal on Selected Areas in Communications*, vol. 39, no. 11, pp. 3444–3458, 2021.

[251] E. T. Michailidis, K. Maliatsos, D. N. Skoutas, D. Vouyioukas, and C. Skianis, "Secure uav-aided mobile edge computing for iot: A review," *IEEE Access*, vol. 10, pp. 86 353–86 383, 2022.

[252] S. Rowe and C. R. Wagner, "An introduction to the joint architecture for unmanned systems (jaus)," *Ann Arbor*, vol. 1001, p. 48108, 2008.

[253] EUROCONTROL, "Proving operations of drones with initial uas traffic management," https://www.eurocontrol.int/project/proving-operations-drones-initial-uas-traffic-management.

[254] A. N. S. I. (ANSI), "Unmanned aircraft systems standardization collaborative (uaccs)," https://www.ansi.org/standards-coordination/collaboratives-activities/unmanned-aircraft-systems-collaborative.